\documentclass[12pt]{article}
\advance\voffset by -2.0cm \advance\hoffset by -1.25cm
\usepackage[utf8x]{inputenc}
\usepackage[english]{babel}
\usepackage{refcheck}
\usepackage{amsmath}
\usepackage{amssymb}
\usepackage{young}
\usepackage[vcentermath]{youngtab}
\usepackage{amsfonts,amssymb, amsmath, amsthm, stmaryrd, mdframed}
\usepackage{pifont, indentfirst, dsfont, relsize}
\usepackage{fancyhdr}
\usepackage{color}

\usepackage{lastpage}

\newcommand{\mi}{\mathrm{i}} 
\def\nonu{\nonumber}
\def\a{\alpha}
\def\b{\beta}
\allowdisplaybreaks

\numberwithin{equation}{section}

\def\be{\begin{equation}}
\def\ee{\end{equation}}

\newcommand{\bea}{\begin{eqnarray}}
\newcommand{\eea}{\end{eqnarray}}

\begin{document}
\thispagestyle{empty} \addtocounter{page}{-1}
   \begin{flushright}
\end{flushright}

\vspace*{1.3cm}
  
\centerline{ \Large \bf
The ${\cal N}=4$
Higher Spin Algebra for Generic $\mu$ Parameter
}
\vspace*{1.5cm}
\centerline{{\bf  Changhyun Ahn$^\dagger$},
  and {\bf Man Hea Kim$^{\ast}$
}} 
\vspace*{1.0cm} 
\centerline{\it 
$\dagger$ Department of Physics, Kyungpook National University, Taegu
41566, Korea} 
\vspace*{0.3cm}
\centerline{\it 
  $\ast$
  Asia Pacific Center for Theoretical Physics, Pohang 37673, Korea }
\vspace*{0.8cm} 
\centerline{\tt ahn@knu.ac.kr, manhea.kim@apctp.org 
} 
\vskip2cm

\centerline{\bf Abstract}
\vspace*{0.5cm}

The ${\cal N}=4$ higher spin generators
for general superspin $s$ in terms of oscillators
in the matrix generalization of $AdS_3$ Vasiliev higher spin theory
at nonzero $\mu$ (which is equivalent to the 't Hooft-like coupling
constant $\lambda$) 
were found previously.
In this paper, by computing the (anti)commutators between these
${\cal N}=4$ higher spin generators for low spins $s_1$ and $s_2$
($s_1+s_2 \leq 11$) explicitly,
we determine 
the complete ${\cal N}=4$ higher spin algebra for generic $\mu$.
The three kinds of
structure constants contain the linear combination of two different
generalized hypergeometric functions. These structure constants
remain the same under the transformation
$\mu \leftrightarrow (1-\mu)$ up to signs.
We have checked that the above ${\cal N}=4$ higher spin algebra
contains the ${\cal N}=2$ higher spin algebra, as a subalgebra,
found by Fradkin and Linetsky
some time ago.

\baselineskip=18pt
\newpage
\renewcommand{\theequation}
{\arabic{section}\mbox{.}\arabic{equation}}

\tableofcontents

\section{Introduction}

It is known that there exists a large ${\cal N}=4$
holography \cite{GG1305} which
connects the matrix generalization of the Vasiliev higher spin theory
\cite{PV9806,PV9812}
on $AdS_3$ with 
the two dimensional minimal model conformal field theories
having the large ${\cal N}=4$ superconformal symmetry.
One of the connections can be seen from the fact that
the oscillator deformation parameter in the $AdS_3$ bulk theory
corresponds to the free parameter of the large ${\cal N}=4$ superconformal
algebra.
The former is related to the mass of the scalar field while the latter
is given by a particular combination of
$N$ and a level $k$ in the ${\cal N}=4$ unitary coset model.
There are seven spin $1$ fields and eight fields of spin  
$s =\frac{3}{2}, 2, \frac{5}{2}, 3, \cdots$ in the (matrix valued)
Vasiliev higher
spin theory.
They correspond to the ${\cal N}=4$ higher spin multiplets
in the boundary theory
which generate the nonlinear ${\cal W}_{\infty}^{{\cal N}=4}[\lambda]$
algebra where $\lambda$ is the 't Hooft-like coupling constant
(above free parameter).
By imposing the wedge condition and taking the infinity limit of
central charge (or infinity limit of $N$)
into the ${\cal W}_{\infty}^{{\cal N}=4}[\lambda]$
algebra, we are left with the linear higher spin algebra
$shs_2[\mu]$
generated by
above higher spin fields in the $AdS_3$ bulk theory, via the
large ${\cal N}=4$ holography. 
See also relevant works in \cite{GGH,FGJ,EGL,GHKPR,GG1803} for
the emergence of higher spin symmetry.

So far,   the nonlinear ${\cal W}_{\infty}^{{\cal N}=4}[\lambda]$
algebra in the ${\cal N}=4$ unitary coset model is known,
for generic $\lambda$,
from both the OPE between the first ${\cal N}=4$ higher spin
multiplet and the OPE between the first and second
${\cal N}=4$ higher spin multiplets \cite{AK1509,AKK1703,AKK}.
Beyond these OPEs, it is rather nontrivial to obtain more OPEs. 
See also relevant works in \cite{EGR,Ahn1311,GP,BCG,AGK,Ahn1408,AK1411,Ahn1504}
for the large ${\cal N}=4$ holography.
On the other hand,
at the vanishing $\lambda$ after taking infinity limit of $N$,
the linear 
 ${\cal W}_{\infty}^{{\cal N}=4}[0]$
algebra is realized by $2N$ free complex bosons and
$2N$ free complex fermions through bilinear construction \cite{EGR,CHR1306}
and the (anti)commutators
with explicit $SO(4)$ symmetry
between the ${\cal N}=4$
higher spin currents in component approach
are completely determined
for generic spins $s_1$ and $s_2$ \cite{AKK}. 
See also the relevant works \cite{GJL,GH,BKnpb,Prochazka}
for the construction of
its bosonic subsector.
In this case, the corresponding ${\cal N}=4$ higher spin algebra
$shs_2[0]$ 
with nondeformed parameter
can be determined by restricting the mode indices of higher spin currents
to the wedge indices. See also
previous works
\cite{Odake,BK,OS,PRS,PRS1990,PRS1990-1,BPRSS}
 on ${\cal N}=2$ higher spin algebra
$shs[\mu]$
some time ago.

In this paper, we would like to construct the ${\cal N}=4$ higher spin
algebra $shs_2[\mu]$ for generic $\mu$ (or $\lambda$) parameter.  
All we have is 
the ${\cal N}=4$ higher spin generators for superspin $s$ in terms of
the multiple products of
oscillators in the matrix generalization of $AdS_3$ Vasiliev higher spin
theory for nonzero $\mu$ \cite{AKK}.
We can compute the various (anti)commutators
between these ${\cal N}=4$ higher spin generators for low spins
$s_1$ and $s_2$ by using the fundamental (anti)commutator relations
between the oscillators.
We have considered the total spin $s=s_1+s_2$ which is less than or equal to
$11$.
Therefore, all the structure constants depend on $\lambda$ (or $\mu$)
for these fixed $s_1$ and $s_2$
explicitly.
We want to generalize these (anti)commutators in terms of closed
analytic form for any
arbitrary $s_1$ and $s_2$. 
We expect to have that the $SO(4)$ index structure and mode dependent parts
in the (anti)commutators for nonzero $\lambda$ (or $\mu$)
in $shs_2[\mu]$
remain the same as the
ones for vanishing $\lambda$ in $shs_2[0]$.
The structure constants are function of $s_1,s_2$ and $\lambda$
as well as modes.
The question is how we allocate these variables appropriately
in the generalized
hypergeometric functions in which there are four upper arguments
and three lower arguments in addition to one variable.

In section $2$,
after introducing the fundamental (anti)commutator relations
for the oscillators, the realizations for
the ${\cal N}=4$ (higher spin)
generators are given explicitly. The aim of this paper is sketched
briefly.

In section $3$,
after introducing the various (anti)commutators for the oscillators
with vanishing $\lambda$, they for nonzero $\lambda$ are determined
by considering the several $s_1$ and $s_2$ values.

In section $4$,
based on the results of section $3$, the three (or six) kinds of
structure constants are obtained and  the closed forms are presented
in terms of generalized hypergeometric functions explicitly.
Moreover, the final (anti)commutator relations for the
${\cal N}=4$ higher spin generators, $shs_2[\mu]$,
are determined completely. 

In Appendices, some various detailed results in sections $3,4$ and $5$
are given explicitly.
In Appendix $F$,
the ${\cal N}=2$ higher spin algebra $shs[\mu]$, as a subalgebra
of $shs_2[\mu]$,
is obtained from the
results of section $4$. 
See also
relevant works
\cite{FL,Korybut-1,Korybut,BBB,CG,CHR1111,CG1203,HLPR,HP,Ahn1206,Ahn1208,CFO,Romans}
on this direction.

See also recent works in \cite{CH1812,CH1906,CHU}.
The mathematica
\cite{mathematica} is used.

\section{Review}

\subsection{The fundamental relations}

The generators of ${\cal N}=4$ higher spin algebra
denoted by $shs_2[\mu]$
are given by the tensor product between
the generators of ${\cal N}=2$ higher spin algebra
denoted by $shs[\mu]$ and the generators of $2 \times 2$ matrices.
The generators of ${\cal N}=2$ higher spin algebra $shs[\mu]$
can be constructed by the multiple products between
the oscillators $\hat{y}_{\a}$ ($\a=1,2$) and the operator $K$ and
they satisfy the following fundamental relations
\cite{Vasiliev89,Vasiliev91,Vasiliev1804}
\footnote{ We denote $\mi$ by an imaginary number.
}
\begin{eqnarray}
[\hat{y}_{\a}, \hat{y}_{\b}]_\star= 2 \, \mi \, \epsilon_{\a \b} (1+ \nu \,K), \qquad  \{K ,\, \hat{y}_{\a}\}_\star =0, \qquad K^2 =1.
\label{fundcomm}
\end{eqnarray}
Without a deformation ($\nu$=0), the usual oscillator commutation
relation in $\hat{y}_{\a}$ is satisfied in the first relation of
(\ref{fundcomm}). Here $\nu = 2\mu -1$.
The $\epsilon_{1 2}= - \epsilon_{2 1}=1$ tensor is antisymmetric. 
The product between the oscillators and the operator $K$
is a Moyal product \cite{AKP} by using a $\star$ notation.
For example,
the (anti)commutators between the oscillators are given by
\footnote{For simplicity, we will not introduce a $\star$ notation
  explicitly
between the oscillators and the operator $K$.}
\begin{eqnarray}
[\hat{y}_{\a}, \hat{y}_{\b}\}_\star&=& \hat{y}_{\a}\star  \hat{y}_{\b} \pm \hat{y}_{\b}\star  \hat{y}_{\a}\,.
\label{anticomm}
\end{eqnarray}
The plus sign in (\ref{anticomm}) corresponds to the anticommutator while
the minus sign corresponds to the commutator.

We introduce an underbrace notation with the number of
oscillators
as
the completely symmetrized product between them \cite{BBB}
\footnote{
In \cite{AKK}, 
there is no prefactor $\frac{1}{n!}$.
}
\bea
\underbrace{
\hat{y}_{(\b_1 }\,\ldots\,\hat{y}_{\b_n)}
}_{n}
\,
&\equiv&
\frac{1}{n!}\sum_{\sigma \in \mathfrak{S}_n} \hat{y}_{\b_{\sigma(1)}} \star  \hat{y}_{\b_{\sigma(2)}} \star \cdots \star  \hat{y}_{\b_{\sigma(n)}}
\,,
\label{underbrace}
\eea
where $\mathfrak{S}_n$ in (\ref{underbrace})
stands for the group of permutations of
$n$ elements.

The precise relation between the parameter $\nu$ (or $\mu$)
in the ${\cal N}=4$ higher spin algebra and the parameter $\lambda$
in the ${\cal N}=4$ coset model in two dimensions
is as follows \cite{GG1305}:
\bea
\nu =2 \lambda -1, \qquad \mbox{or} \qquad \lambda = \mu =\frac{(\nu+1)}{2}.
\label{lamunu}
\eea
See also the relevant works in \cite{AKP,CHR1211,AP1902}
for the matrix generalization of the $AdS_3$ Vasiliev higher
spin theory.

From now on we will use the $\mu$ parameter rather than
$\lambda$ parameter all the time.

\subsection{The ${\cal N}=4$ generators}

By using the $SO(4)$ symmetry of the ${\cal N}=4$
superconformal algebra \cite{STV,STVS,Schoutens},
we rearrange the $SO(4)$ vector
and adjoint indices  in order to match them with those 
of Pauli matrices $\sigma^i$ ($i=1,2,3$) \cite{GG1305,Ferreira}.
\bea
L_{m} &=& \frac{1}{4 \mi}\,  \underbrace{
\hat{y}_{(1}\ldots\,\hat{y}_{1}
}_{1+m} \,
\underbrace{
\hat{y}_{2}\ldots\,\hat{y}_{2)}
}_{1-m}
\otimes 
\left(\begin{array}{cc}
1 & 0 \\
0 & 1 \\
\end{array}\right), \quad |m|<2,
\nonu\\
G^{1}_{\pm \frac{1}{2}} & = &
\frac{1}{2}\,e^{\mi\frac{\pi}{4}}\,\,\hat{y}_{\frac{3}{2}\mp \frac{1}{2}}K \otimes 
\sigma^1,
\qquad
G^{2}_{\pm \frac{1}{2}}  = 
-\frac{1}{2}\,e^{\mi\frac{\pi}{4}}\,\,\hat{y}_{\frac{3}{2}\mp \frac{1}{2}}K \otimes 
\sigma^2,
\nonu\\
G^{3}_{\pm \frac{1}{2}} & = &
\frac{1}{2}\,e^{\mi\frac{\pi}{4}}\,\,\hat{y}_{\frac{3}{2}\mp \frac{1}{2}}K \otimes 
\sigma^3,
\qquad
G^{4}_{\pm \frac{1}{2}}  = 
\frac{\mi}{2}\,e^{\mi\frac{\pi}{4}}\,\,\hat{y}_{\frac{3}{2}\mp \frac{1}{2}} \otimes 
\left(\begin{array}{cc}
1 & 0 \\
0 & 1 \\
\end{array}\right)\,,
\nonu\\
T^{12}_0 & = & \frac{1}{2} \otimes  \sigma^3,  \qquad 
T^{13}_0  =  \frac{1}{2} \otimes  \sigma^2,  \qquad 
T^{23}_0  =  \frac{1}{2} \otimes  \sigma^1,  \qquad 
\label{9plus8}
\\
T^{14}_0  &=& - \frac{1}{2} K \otimes  \sigma^1,  \qquad
T^{24}_0  =  \frac{1}{2} K \otimes  \sigma^2,  \qquad
T^{34}_0  =- \frac{1}{2} K \otimes  \sigma^3,  \qquad
(T^{ij}=-T^{ji}).
\nonu
 \eea
The $\epsilon^{1 2 3}=1$ tensor is antisymmetric.
The ${\cal N}=4$ wedge algebra generated by
nine bosonic ($L_m, T_0^{ij}, T_0^{i4}$) and eight fermionic
($G^{i}_{\pm \frac{1}{2}}, G^{4}_{\pm \frac{1}{2}}$) generators in
(\ref{9plus8}) can be obtained
and denoted by the global exceptional superalgebra $D(2,1|\alpha)$
with $\alpha \equiv \frac{(1+\nu)}{(1-\nu)}$ \cite{GG1305}.
We can obtain this ${\cal N}=4$ wedge algebra
by restricting to the wedge mode
from the standard (anti)commutators of the large ${\cal N}=4$
superconformal algebra after performing the large $(N,k)$ 't Hooft limit
\footnote{
\label{refereepoint}
  By identifying
$A^{\pm 3}= \frac{1}{2} (T^{12} \mp T^{34})$ and $G^{\pm}=
\frac{1}{\sqrt{2}}(G^1\mp \mi\, G^2)$ with the corresponding
generators of ${\cal N}=2$ superconformal algebra where the
$U(1)$ spin-$1$ current is given by $J=2 (\gamma \, A^{+3} +(1-\gamma) \,
A^{-3})$ with $\gamma \equiv \frac{\alpha}{(1+\alpha)}$, the ${\cal N}=2$
wedge algebra is generated by
four bosonic ($L_{\pm1}, L_0, J_0$) and four fermionic
$G^{\pm}_{\pm \frac{1}{2}}$ generators \cite{GG1305}. We thank the
referee for pointing this embedding and
also the embedding of the ${\cal N}=2$ higher spin algebra
$shs[\mu]$ inside the ${\cal N}=4$ higher spin algebra
$shs_2[\mu]$ out.}.

\subsection{The $s$-th ${\cal N}=4$ higher spin generators}

In \cite{AKK}, the $s$-th ${\cal N}=4$ higher spin generators
written in terms of the component field in ${\cal N}=4$ superspace
are described by the oscillators
\begin{eqnarray}
\Phi^{(s)}_{0,m} & = &
\underbrace{
\hat{y}_{(1 }\,\ldots\,\hat{y}_{1}
}_{s-1+m}
\,
\underbrace{
\hat{y}_{2}\,\ldots \ \hat{y}_{2)}
}_{s-1-m}
\big(K (2s-1)+\nu \big)
\otimes 
\left(\begin{array}{cc}
1 & 0 \\
0 & 1 \\
\end{array}\right),
 -(s-1) \leq m \leq (s-1),
\nonu \\
\Phi^{(s),1}_{\frac{1}{2},\rho}& = &
-(2s-1)
\, e^{\mi\frac{\pi}{4}}
\,
\underbrace{
\hat{y}_{(1 }\,\ldots\,\hat{y}_{1}
}_{s-\frac{1}{2}+\rho}
\,
\underbrace{
\hat{y}_{2 }\,\ldots\ \hat{y}_{2)}
}_{s-\frac{1}{2}-\rho}
\otimes \,
\sigma^1 ,
\qquad -(s-\frac{1}{2}) \leq \rho \leq (s-\frac{1}{2}),
\nonu \\
\Phi^{(s),2}_{\frac{1}{2},\rho}& = &
(2s-1)
\, e^{\mi\frac{\pi}{4}}
\,
\underbrace{
\hat{y}_{(1 }\,\ldots\,\hat{y}_{1}
}_{s-\frac{1}{2}+\rho}
\,
\underbrace{
\hat{y}_{2 }\,\ldots\ \hat{y}_{2)}
}_{s-\frac{1}{2}-\rho}
\otimes \,
\sigma^2 ,\qquad -(s-\frac{1}{2}) \leq \rho \leq (s-\frac{1}{2}),
\nonu \\
\Phi^{(s),3}_{\frac{1}{2},\rho}& = &
-(2s-1)
\, e^{\mi\frac{\pi}{4}}
\,
\underbrace{
\hat{y}_{(1 }\,\ldots\,\hat{y}_{1}
}_{s-\frac{1}{2}+\rho}
\,
\underbrace{
\hat{y}_{2 }\,\ldots\ \hat{y}_{2)}
}_{s-\frac{1}{2}-\rho}
\otimes \,
\sigma^3 ,\qquad -(s-\frac{1}{2}) \leq \rho \leq (s-\frac{1}{2}),
\nonu\\
\Phi^{(s),4}_{\frac{1}{2},\rho} & = &
-\mi
(2s-1)
\, e^{\mi\frac{\pi}{4}}
\,
\underbrace{
\hat{y}_{(1 }\,\ldots\,\hat{y}_{1}
}_{s-\frac{1}{2}+\rho}
\,
\underbrace{
\hat{y}_{2 }\,\ldots\ \hat{y}_{2)}
}_{s-\frac{1}{2}-\rho}
K
\otimes 
\left(\begin{array}{cc}
1 & 0 \\
0 & 1 \\
\end{array}\right),
-(s-\frac{1}{2}) \leq \rho \leq (s-\frac{1}{2}),
\nonu\\
\Phi^{(s),12}_{1,m} & = & 
-(2s-1)
\,
\underbrace{
\hat{y}_{(1} \,\ldots\,\hat{y}_{1}
}_{s+m}
\,
\underbrace{
\hat{y}_{2 }\,\ldots\ \hat{y}_{2)}
}_{s-m}
\otimes \, 
\sigma^3,
\qquad -s \leq m \leq s,
\nonu\\
\Phi^{(s),13}_{1,m} & = & 
-(2s-1)
\,
\underbrace{
\hat{y}_{(1} \,\ldots\,\hat{y}_{1}
}_{s+m}
\,
\underbrace{
\hat{y}_{2 }\,\ldots\ \hat{y}_{2)}
}_{s-m}
\otimes \, 
\sigma^2,
\qquad -s \leq m \leq s,
\nonu\\
\Phi^{(s),23}_{1,m} & = & 
-(2s-1)
\,
\underbrace{
\hat{y}_{(1} \,\ldots\,\hat{y}_{1}
}_{s+m}
\,
\underbrace{
\hat{y}_{2 }\,\ldots\ \hat{y}_{2)}
}_{s-m}
\otimes \, \sigma^1,
\qquad -s \leq m \leq s,
\nonu\\
\Phi^{(s),14}_{1,m} & = & 
(2s-1)
\,
\underbrace{
\hat{y}_{(1} \,\ldots\,\hat{y}_{1}
}_{s+m}
\,
\underbrace{
\hat{y}_{2 }\,\ldots\ \hat{y}_{2)}
}_{s-m}K
\otimes \, 
\sigma^1
 \,,
\qquad -s \leq m \leq s,
\nonu\\
\Phi^{(s),24}_{1,m} & = & 
-(2s-1)
\,
\underbrace{
\hat{y}_{(1} \,\ldots\,\hat{y}_{1}
}_{s+m}
\,
\underbrace{
\hat{y}_{2 }\,\ldots\ \hat{y}_{2)}
}_{s-m}K
\otimes \, 
\sigma^2
\,,
\qquad -s \leq m \leq s,
\nonu\\
\Phi^{(s),34}_{1,m} & = & 
(2s-1)
\,
\underbrace{
\hat{y}_{(1} \,\ldots\,\hat{y}_{1}
}_{s+m}
\,
\underbrace{
\hat{y}_{2 }\,\ldots\ \hat{y}_{2)}
}_{s-m}K
\otimes \, 
\sigma^3
 \,,
\qquad -s \leq m \leq s, \qquad  (\Phi^{(s),ij}_{1,m}=-\Phi^{(s),ji}_{1,m})
\nonu\\
\tilde{\Phi}^{(s),1}_{\frac{3}{2},\rho}& = &
\mi (2s-1)
\,e^{\mi\frac{\pi}{4}}
\,
\underbrace{
\hat{y}_{(1 }\,\ldots\,\hat{y}_{1}
}_{s+\frac{1}{2}+\rho}
\,
\underbrace{
\hat{y}_{2 }\,\ldots\ \hat{y}_{2)}
}_{s+\frac{1}{2}-\rho}
K
\otimes 
\sigma^1,
\qquad -(s+\frac{1}{2}) \leq \rho \leq (s+\frac{1}{2}),
\nonu\\
\tilde{\Phi}^{(s),2}_{\frac{3}{2},\rho}& = &
-\mi
(2s-1)
\,e^{\mi\frac{\pi}{4}}
\,
\underbrace{
\hat{y}_{(1 }\,\ldots\,\hat{y}_{1}
}_{s+\frac{1}{2}+\rho}
\,
\underbrace{
\hat{y}_{2 }\,\ldots\ \hat{y}_{2)}
}_{s+\frac{1}{2}-\rho}
K
\otimes 
\sigma^2,
\qquad -(s+\frac{1}{2}) \leq \rho \leq (s+\frac{1}{2}),
\nonu\\
\tilde{\Phi}^{(s),3}_{\frac{3}{2},\rho}& = &
\mi
(2s-1)
\,e^{\mi\frac{\pi}{4}}
\,
\underbrace{
\hat{y}_{(1 }\,\ldots\,\hat{y}_{1}
}_{s+\frac{1}{2}+\rho}
\,
\underbrace{
\hat{y}_{2 }\,\ldots\ \hat{y}_{2)}
}_{s+\frac{1}{2}-\rho}
K
\otimes 
\sigma^3,
\qquad -(s+\frac{1}{2}) \leq \rho \leq (s+\frac{1}{2}),
\nonu\\
\tilde{\Phi}^{(s),4}_{\frac{3}{2},\rho} & = &
-(2s-1)
\,e^{\mi\frac{\pi}{4}}
\,
\underbrace{
\hat{y}_{(1 }\,\ldots\,\hat{y}_{1}
}_{s+\frac{1}{2}+\rho}
\,
\underbrace{
\hat{y}_{2 }\,\ldots\ \hat{y}_{2)}
}_{s+\frac{1}{2}-\rho}
\otimes 
\left(\begin{array}{cc}
1 & 0 \\
0 & 1 \\
\end{array}\right),
\qquad -(s+\frac{1}{2}) \leq \rho \leq (s+\frac{1}{2}),
\nonu \\
\tilde{\Phi}^{(s)}_{2,m} & = &
-(2s-1)
\underbrace{
\hat{y}_{(1} \,\ldots\,\hat{y}_{1}
}_{s+1+m}
\,
\underbrace{
\hat{y}_{2} \,\ldots\ \hat{y}_{2)}
}_{s+1-m}
\otimes 
\left(\begin{array}{cc}
1 & 0 \\
0 & 1 \\
\end{array}\right),
-(s+1) \leq m \leq (s+1).
\label{HSbasis}
\end{eqnarray}
The spin of each operator is given by
the half of the number of oscillators plus one.
We can check that
the total number of higher spin generators
is given by $16(1+2s)$.
The difference between the number of $\hat{y}_1$
and
the number of $\hat{y}_2$ is two times of each mode.
Note that if we rewrite the above higher spin generators by ignoring
the symmetric combinations and moving all the $\hat{y}_1$ to the left,
then there exist various types of oscillators having different lengths  
in their expressions \footnote{
For example, 
there are following five bosonic generators of spin $3$  for $s=1$ 
\bea
\tilde{  \Phi}^{(1)}_{2,\,-2}& = &
-\hat{y}_2 
\hat{y}_2
\hat{y}_2
\hat{y}_2
\otimes \,I_{2\times2},
\qquad
\tilde{\Phi}^{(1)}_{2,\,-1} = 
\Big(
-\hat{y}_1 
\hat{y}_2
\hat{y}_2
\hat{y}_2
+\mi\,\nu\,
\hat{y}_2 
\hat{y}_2
K
+3\,\mi\,
\hat{y}_2
\hat{y}_2
\Big)
\otimes \,I_{2\times2},
\nonu\\
\tilde{\Phi}^{(1)}_{2,\,0}& = &
\Big(
-\hat{y}_1 
\hat{y}_1
\hat{y}_2
\hat{y}_2
+4 \,\mi\,
\hat{y}_1 
\hat{y}_2
+\frac{2}{3}
(3+4\,\nu \,K+\nu^2)
\Big)
\otimes \,I_{2\times2},
\nonu \\
\tilde{\Phi}^{(1)}_{2,\,+1} & = & 
\Big(
-\hat{y}_1 
\hat{y}_1
\hat{y}_1
\hat{y}_2
+\mi\,\nu\,
\hat{y}_1
\hat{y}_1
K
+3\,\mi\,
\hat{y}_1
\hat{y}_1
\Big)
\otimes \,I_{2\times2},
\qquad
\tilde{\Phi}^{(1)}_{2,\,+2} = 
-\hat{y}_1 
\hat{y}_1
\hat{y}_1
\hat{y}_1
\otimes \,I_{2\times2}.
\label{examples}
\eea
Among these in (\ref{examples})
we observe that by moving the oscillator $\hat{y}_1$ appearing in
(\ref{HSbasis})
to the left some of the generators have different kinds of
oscillators in the sense that each term in the middle three
generators in (\ref{examples}) has different number of
oscillators. }.

Recall that the large ${\cal N}=4$
superconformal algebra contains
two $su(2)$ algebra
and the higher spin generators from the $D(2,1|\alpha)$ multiplets \cite{GG1305}
\bea
\begin{array}{lllc}
& s: & ({\bf 1},{\bf 1})  &\!\! \longrightarrow \quad \Phi^{(s)}_{0,m} \\
& s+\tfrac{1}{2}: & ({\bf 2},{\bf 2})   &\longrightarrow  \quad \Phi^{(s),i}_{\frac{1}{2},\rho} \\
  R^{(s)}: \qquad & s+1: & ({\bf 3},{\bf 1}) \oplus ({\bf 1},{\bf 3}) &\longrightarrow
  \quad  \Phi^{(s),ij}_{1,m}  \\
  & s+\tfrac{3}{2}: & ({\bf 2},{\bf 2}) &\longrightarrow \quad
  \tilde{\Phi}^{(s),i}_{\frac{3}{2},\rho} \\
& s+2: & ({\bf 1},{\bf 1}) &\!\! \longrightarrow \quad \tilde{\Phi}^{(s)}_{2,m}
\end{array}
\label{R}
\eea
generate the ${\cal N}=4$ higher spin algebra $shs_2[\mu]$
and  two bold-face numbers in (\ref{R}) denote the representations
with respect to the two $su(2)$ algebra and $s=1,2, \cdots$.

\subsection{The goal of this paper}

The question we have raised in \cite{AKK}
is how we obtain the ${\cal N}=4$ higher spin algebra
$shs_2[\mu]$
for nonzero $\mu$ 
by using the explicit realization of (\ref{HSbasis}) via direct
calculation with the help of (\ref{fundcomm}).

Schematically  the generator of spin $h$ with mode $m$
has the following product of oscillators (the analysis for the
oscillators with the operator $K$ can be done) 
\begin{eqnarray}
\Phi^{h}_m &\propto&
 \underbrace{
\hat{y}_{(1}\ldots\,\hat{y}_{1}
}_{h-1+m}
 \underbrace{
\hat{y}_{2}\ldots\,\hat{y}_{2)}
}_{h-1-m}\,.
\label{field}
\end{eqnarray}
Then we can construct the (anti)commutator between them in (\ref{field})
as follows:
\begin{eqnarray}
\big[ \Phi^{h_1}_{m_1},\, \Phi^{h_2}_{m_2} \big\} \!\!\!
&\propto&
 \underbrace{
\hat{y}_{(1}\ldots\,\hat{y}_{1}
}_{h_1-1+m_1}
 \underbrace{
\hat{y}_{2}\ldots\,\hat{y}_{2)}
}_{h_1-1-m_1}
 \underbrace{
\hat{y}_{(1}\ldots\,\hat{y}_{1}
}_{h_2-1+m_2}
 \underbrace{
\hat{y}_{2}\ldots\,\hat{y}_{2)}
}_{h_2-1-m_2}
 \nonu \\
 & \pm &  
 \underbrace{
\hat{y}_{(1}\ldots\,\hat{y}_{1}
}_{h_2-1+m_2}
 \underbrace{
\hat{y}_{2}\ldots\,\hat{y}_{2)}
}_{h_2-1-m_2}
 \underbrace{
\hat{y}_{(1}\ldots\,\hat{y}_{1}
}_{h_1-1+m_1}
 \underbrace{
\hat{y}_{2}\ldots\,\hat{y}_{2)}
}_{h_1-1-m_1}
\nonumber\\
&=&
\sum_{h_3=1}^{h_1+h_2-1}\!\!\!
\Bigg[ c_{h_3}
(h_1,h_2,m_1,m_2) \underbrace{
\hat{y}_{(1}\ldots\ldots\,\hat{y}_{1}
}_{h_3-1+m_1+m_2}
\underbrace{
\hat{y}_{2}\ldots\ldots\,\hat{y}_{2)}
}_{h_3-1-m_1-m_2}
\nonumber\\
&+& d_{h_3}
(h_1,h_2,m_1,m_2) \underbrace{
\hat{y}_{(1}\ldots\ldots\,\hat{y}_{1}
}_{h_3-1+m_1+m_2}
\underbrace{
\hat{y}_{2}\ldots\ldots\,\hat{y}_{2)} 
}_{h_3-1-m_1-m_2} \, K \Bigg]
\nonu \\
&=&
\sum_{h_3=1}^{h_1+h_2-1}\!\!\!
e_{h_3}
(h_1,h_2,m_1,m_2)\,
\Phi^{h_3}_{m_1+m_2}
\,.
\label{Strategy}
\end{eqnarray}
The first expression of (\ref{Strategy})
is simply two terms from the (anti)commutator for given spins and modes.
The next expression is
that we would like to move the oscillator $\hat{y}_1$ to the left while the
operator $K$ should be moved to the right in order to classify
all the possible oscillators systematically. Due to the
fundamental relations (\ref{fundcomm}), the former will produce the
oscillator term whose number of oscillator is reduced by $2$ at one step
movement while
the latter will produce additional sign change. In this process
after  all the oscillators $\hat{y}_1$ are moved to the left, 
the right hand side  contains different kinds of oscillators
including the operator $K$.
This implies the summation over the dummy variable $h_3$ in the above.
At the final stage, we express the oscillators appearing in the
right hand side in terms of
${\cal N}=4$ higher spin generators.
The final goal is to obtain the structure constants
explicitly 
for given spins $h_1,_2$ and modes $m_1, m_2$.
We would like to express them
in terms of generalized hypergeometric functions.

\section{Direct calculation of (anti)commutators between
  the oscillators}

Let us start with the particular commutator 
of ${\cal N}=4$ higher spin algebra $shs_2[0]$ with vanishing
$\mu$.
The last commutator of Appendix $(J.1)$ in \cite{AKK}
 is
described by
\bea
&& \Big[\tilde{\Phi}^{(h_1)}_{2,\,m}, \tilde{\Phi}^{(h_2)}_{2,\,n}\Big]\!
=
-
\sum_{h= 0}^{[\frac{h_1+h_2}{2}]}
\,\frac{(2h_1-1)(2h_2-1)}{8(h_1+h_2-2h)+12}
 \Bigg[
\nonu\\
&& \times \,
\Big(
p_{\mathrm{F}, 2h}^{h_1+2 , h_2+2}
-p_{\mathrm{B}, 2h}^{h_1+2 , h_2+2}
\Big)\,
\Phi^{(h_1+h_2-2h+2)}_{0,\,m+n}
-
\frac{2}{2(h_1+h_2-2h)-1}
\nonu\\
&& \times \,
\Big(\!
(h_1\!+\!h_2\!-\!2h\!+\!1)\,p_{\mathrm{F}, 2h}^{h_1+2 ,h_2+2}
+ (h_1\!+\!h_2\!-\!2h\!+\!2)\,p_{\mathrm{B}, 2h}^{h_1+2 , h_2+2}
\!\Big)\,
\tilde{\Phi}^{(h_1+h_2-2h)}_{2,\,m+n}
\Bigg]
 \,.
 \label{lastcomm}
\eea
\footnote
{
The notation $[x]$ is the greatest integer less than or equal to x.
}
The mode dependent structure constants $p_F$ and $p_B$
are given by Appendix $A$.
In order to extract the commutator constructed from the oscillators
in the left hand side, we should change 
the higher spin generators into the corresponding oscillators
by using (\ref{HSbasis})
\bea
&&\big[\tilde{\Phi}^{(h_1)}_{2,\,m},\,\tilde{\Phi}^{(h_2)}_{2,\,n}\big]
=
\frac{\mi^{h_1+1}(2h_1-1)}{8}\, \frac{\mi^{h_2+1}(2h_2-1)}{8}
\nonu \\
&& \times \, \big[\underbrace{
\hat{y}_{(1 }\,\ldots\,\hat{y}_{1}
}_{h_1+1+m}
\,
\underbrace{
\hat{y}_{2}\,\ldots \, \hat{y}_{2)}
}_{h_1+1-m},\,
\underbrace{
\hat{y}_{(1 }\,\ldots\,\hat{y}_{1}
}_{h_2+1+n}
\,
\underbrace{
\hat{y}_{2}\,\ldots \, \hat{y}_{2)}
}_{h_2+1-n}\big] 
\otimes 
\left(\begin{array}{cc}
1 & 0 \\
0 & 1 \\
\end{array}\right).
\label{secondeq}
\eea
The prefactors come from the different normalization in this paper
compared to the one in \cite{AKK}.
Since
the commutator of two tensor products is given by
the tensor product between commutator of oscillators
and the $2 \times 2$ matrix product
$[A \otimes I,B \otimes I]= (A B -B A) \otimes I $
for any operators $A$ and $B$,
we can ignore the $2 \times 2 $ identity matrix in this particular case.
Then, the commutator from two bosonic oscillators in (\ref{secondeq})
can be written as 
\bea
&& \big[\underbrace{
\hat{y}_{(1 }\,\ldots\,\hat{y}_{1}
}_{h_1+1+m}
\,
\underbrace{
\hat{y}_{2}\,\ldots \, \hat{y}_{2)}
}_{h_1+1-m},\,
\underbrace{
\hat{y}_{(1 }\,\ldots\,\hat{y}_{1}
}_{h_2+1+n}
\,
\underbrace{
\hat{y}_{2}\,\ldots \, \hat{y}_{2)}
}_{h_2+1-n}\big] 
\otimes 
\left(\begin{array}{cc}
1 & 0 \\
0 & 1 \\
\end{array}\right)
=
\nonu \\
&&
\frac{64}{\mi^{h_1+h_2+2}(2h_1-1)(2h_2-1)} \,
\big[\tilde{\Phi}^{(h_1)}_{2,\,m},\,\tilde{\Phi}^{(h_2)}_{2,\,n}\big].
\label{case1example}
\eea

In this way, we can rewrite the various (anti)commutators
between the oscillators in terms of those between the
${\cal N}=4$ higher spin generators like as (\ref{case1example}).
We can classify all possible cases as follows:

\begin{itemize}
\item The case-one

We have a single commutator by constructing two bosonic oscillators.  
  
\item The case-two

  We have two kinds of commutators by considering one bosonic and one
  fermionic oscillators. 
  There are two cases for the bosonic oscillator depending on
  the presence of the operator $K$ according to (\ref{HSbasis}).
  
\item The case-three

   We have two kinds of anticommutators by considering two 
  fermionic oscillators. 
  There are two cases for the fermionic oscillator depending on
  the presence of the operator $K$.

\end{itemize}

We will construct the above five kinds of (anti)commutators
explicitly in this paper. Although there are other cases, but they will
belong to these three cases, in principle.
Once we determine these (anti)commutators
for the oscillators, we can obtain the ${\cal N}=4$ higher spin
algebra $shs_2[\mu]$ by multiplying the appropriate coefficients
appearing in (\ref{HSbasis}),
which depend on the numerical factors, spins and the
deformation parameter $\nu$, 
together with $2\times 2$ matrices.

\subsection{The (anti)commutators between the oscillators with vanishing
  $\mu$ }

In this subsection, we restrict to focus on the previous work
\cite{AKK} which holds for the vanishing $\mu$.
In next subsection, we will construct
the (anti)commutators between the oscillators with nonvanishing
$\mu$ \footnote{
  In this paper, the $h_1$, $h_2$ and $r$ for the spins
  are natural numbers.}.

 $\bullet$ The case-one with two bosonic oscillators 

We return to the relation (\ref{case1example}).
After substituting  the last expression of
(\ref{HSbasis}) into the right hand side,
we can compute this particular commutator. We can use the result of
previous work of \cite{AKK}: the last commutator of Appendix $(J.1)$
which is the right hand side of (\ref{lastcomm}).
We substitute the oscillator realization for the
higher spin generators with the condition $\mu=0$ (or $\nu=-1$
in (\ref{lamunu})).
By shifting to $h_i \rightarrow h_i-2$ and introducing a
dummy variable as $h \rightarrow r-1$, 
 and simplifying the right hand side,
 we eventually obtain  the summation of various
 commutators in the right hand side 
\bea
&&
\big[\underbrace{
\hat{y}_{(1 }\,\ldots\,\hat{y}_{1}
}_{h_1-1+m}
\,
\underbrace{
\hat{y}_{2}\,\ldots \, \hat{y}_{2)}
}_{h_1-1-m},\,
\underbrace{
\hat{y}_{(1 }\,\ldots\,\hat{y}_{1}
}_{h_2-1+n}
\,
\underbrace{
\hat{y}_{2}\,\ldots \, \hat{y}_{2)}
}_{h_2-1-n}\big]  
=
\nonu \\
&& -\mi 
\sum_{r=1}^{[\frac{h_1+h_2-1}{2}]}
2\,(-1)^{r}
\Bigg[
\!
\Big(p_{\mathrm{F},\,2(r-1)}^{h_1,h_2}(m,n)+p_{\mathrm{B},\,2(r-1)}^{h_1,h_2}(m,n)\Big)
\underbrace{
\hat{y}_{(1 }\ldots\ldots\ldots. \,\hat{y}_{1}
}_{h_1+h_2-2r-1+m+n}
\underbrace{
\hat{y}_{2}\ldots\ldots\ldots.  \,\hat{y}_{2)}
}_{h_1+h_2-2r-1-m-n}
\nonu\\
&&
-
\Big(p_{\mathrm{F},\,2(r-1)}^{h_1,h_2}(m,n)-p_{\mathrm{B},\,2(r-1)}^{h_1,h_2}(m,n)\Big)
\underbrace{
\hat{y}_{(1 }\ldots\ldots\ldots.\,\hat{y}_{1}
}_{h_1+h_2-2r-1+m+n}
\underbrace{
\hat{y}_{2}\ldots\ldots\ldots. \,\hat{y}_{2)}
}_{h_1+h_2-2r-1-m-n}K
\Bigg]. 
\label{IdEvenEvenOdake}
\eea
We denote $m,n, \cdots$ by  the integer mode. 
The spin of right hand side is given by $(h_1+h_2-2r)$
from the previous analysis by counting
the number of oscillators.
Also the dummy variable $r$ starts with $1$.
The structure constants are given in Appendix $A$
\footnote{
   \label{othercasesforcase1}
   We observe that for the bosonic oscillators having
   the operator $K$, the relevant commutator is given by
   $
   \big[ 
  \underbrace{
\hat{y}_{(1}\ldots\,\hat{y}_{1}
}_{h_1-1+m} 
\underbrace{
\hat{y}_{2}\ldots\,\hat{y}_{2)}
}_{h_1-1-m} K,\,
 \underbrace{
\hat{y}_{(1}\ldots\,\hat{y}_{1}
}_{h_2-1+n} 
\underbrace{
\hat{y}_{2}\ldots\,\hat{y}_{2)}
}_{h_2-1-n}K
\big]$.
   We can move the first operator $K$ in the plus sign
   to the right  and then this meets with the second operator $K$.
   This becomes $1$ from (\ref{fundcomm}). Similarly,
    the first operator $K$ in the minus sign
    to the right  and this meets with the second operator $K$.
    This also gives $1$. Then this commutator is the same as the one in
    (\ref{IdEvenEvenOdake}).
    For the case where one of the oscillators contains the operator $K$,
    we consider the relevant commutator
%
$   \big[ 
  \underbrace{
\hat{y}_{(1}\ldots\,\hat{y}_{1}
}_{h_1-1+m} 
\underbrace{
\hat{y}_{2}\ldots\,\hat{y}_{2)}
}_{h_1-1-m},\,
 \underbrace{
\hat{y}_{(1}\ldots\,\hat{y}_{1}
}_{h_2-1+n} 
\underbrace{
\hat{y}_{2}\ldots\,\hat{y}_{2)}
}_{h_2-1-n}K
\big] K
    $ where the additional operator $K$ is multiplied to the right.
    We can easily check that this commutator is the same as the one in
    (\ref{IdEvenEvenOdake}) by moving the operator $K$ properly.
    Therefore,  when one of the oscillators contains the operator $K$,
    the commutator can be obtained by multiplying the operator $K$
  into (\ref{IdEvenEvenOdake}) from the right.}.

$\bullet$ The case-two with one bosonic and on fermionic
oscillators

Let us consider that the two kinds of oscillators do not have the
operator $K$ dependence. We can use the fourth relation of Appendix $(J.1)$
of \cite{AKK}. Together with the shifts,
$h_2 \rightarrow h_1-2$, $h_1 \rightarrow h_2$ and $h \rightarrow r-
\frac{3}{2}$, we substitute the corresponding higher spin generators
given in (\ref{HSbasis}) into the right hand side.
It turns out that 
\bea
&&\big[ 
  \underbrace{
\hat{y}_{(1}\ldots\,\hat{y}_{1}
}_{h_1-1+m} 
\underbrace{
\hat{y}_{2}\ldots\,\hat{y}_{2)}
}_{h_1-1-m},\,
 \underbrace{
\hat{y}_{(1}\ldots\,\hat{y}_{1}
}_{h_2-\frac{1}{2}+\rho} 
\underbrace{
\hat{y}_{2}\ldots\,\hat{y}_{2)}
}_{h_2-\frac{1}{2}-\rho}
\big]
=
\nonu \\
&& -4  \sum_{r=1}^{[\frac{h_1+h_2-1}{2}]}
(-1)^r 
\Bigg[
\mi
\Big(q_{\mathrm{F},\,2(r-1)}^{h_1,h_2+\frac{1}{2}}(m,\rho)+
q_{\mathrm{B},\,2(r-1)}^{h_1,h_2+\frac{1}{2}}(m,\rho)\Big)
\,  \underbrace{
\hat{y}_{(1}\ldots\ldots\ldots. \,\hat{y}_{1}
}_{h_1+h_2-2r-\frac{1}{2}+m+\rho} 
\underbrace{
\hat{y}_{2}\ldots\ldots\ldots. \,\hat{y}_{2)}
}_{h_1+h_2-2r-\frac{1}{2}-m-\rho}
\nonumber\\
&&
-\Big(q_{\mathrm{F},\,2r-3}^{h_1,h_2+\frac{1}{2}}(m,\rho)+
q_{\mathrm{B},\,2r-3}^{h_1,h_2+\frac{1}{2}}(m,\rho)\Big)
\,  \underbrace{
\hat{y}_{(1}\ldots\ldots\ldots. \,\hat{y}_{1}
}_{h_1+h_2-2r+\frac{1}{2}+m+\rho} 
\underbrace{
\hat{y}_{2}\ldots\ldots\ldots. \,\hat{y}_{2)}
}_{h_1+h_2-2r+\frac{1}{2}-m-\rho}K
\Bigg].
\label{fourthcomm}
\eea
We denote $\rho,\omega, \cdots$ by  the half-integer mode.
The spins of right hand side of (\ref{fourthcomm})
can be read off and are given by $(h_1+h_2-2r+
\frac{1}{2})$ and $(h_1+h_2-2r+\frac{3}{2})$ respectively.
The structure constants are given in Appendix $A$
\footnote{
\label{othercasesforcase2}
  For the case where the fermionic oscillators contain the operator
  $K$,
the commutator  $  \big[ 
  \underbrace{
\hat{y}_{(1}\ldots\,\hat{y}_{1}
}_{h_1-1+m} 
\underbrace{
\hat{y}_{2}\ldots\,\hat{y}_{2)}
}_{h_1-1-m},\,
 \underbrace{
\hat{y}_{(1}\ldots\,\hat{y}_{1}
}_{h_2-\frac{1}{2}+\rho} 
\underbrace{
\hat{y}_{2}\ldots\,\hat{y}_{2)}
}_{h_2-\frac{1}{2}-\rho} K
\big]   K
$ is the same as the one in (\ref{fourthcomm}).
Then by multiplying the operator $K$ into
(\ref{fourthcomm})
from the right, we obtain the corresponding commutator.
Similarly,
when we have the operator $K$ in both bosonic and fermionic
oscillators,
the commutator $
\big[ 
  \underbrace{
\hat{y}_{(1}\ldots\,\hat{y}_{1}
}_{h_1-1+m} 
\underbrace{
\hat{y}_{2}\ldots\,\hat{y}_{2)}
}_{h_1-1-m}K,\,
 \underbrace{
\hat{y}_{(1}\ldots\,\hat{y}_{1}
}_{h_2-\frac{1}{2}+\rho} 
\underbrace{
\hat{y}_{2}\ldots\,\hat{y}_{2)}
}_{h_2-\frac{1}{2}-\rho}K
\big] K$ with additional
operator $K$ is the same as the one in (\ref{IdEvenOddOdake}).
Again, by multiplying the operator $K$ into (\ref{IdEvenOddOdake})
from the right, we obtain the final result. 
}.

When the operator $K$ appears in the bosonic oscillator,
we can use the second equation of $(6.12)$ of \cite{AKK}.
By shifting the dummy variable $h \rightarrow r-\frac{3}{2}$,
we obtain the following result 
\bea
&&\big[ 
  \underbrace{
\hat{y}_{(1}\ldots\,\hat{y}_{1}
}_{h_1-1+m} 
\underbrace{
\hat{y}_{2}\ldots\,\hat{y}_{2)}
}_{h_1-1-m} K,\,
 \underbrace{
\hat{y}_{(1}\ldots\,\hat{y}_{1}
}_{h_2-\frac{1}{2}+\rho} 
\underbrace{
\hat{y}_{2}\ldots\,\hat{y}_{2)}
}_{h_2-\frac{1}{2}-\rho}
\big]
=
\nonu \\
&& -4  \sum_{r=1}^{[\frac{h_1+h_2}{2}]}
\Bigg[(-1)^r 
\Big(q_{\mathrm{F},\,2r-3}^{h_1,h_2+\frac{1}{2}}(m,\rho)-
q_{\mathrm{B},\,2r-3}^{h_1,h_2+\frac{1}{2}}(m,\rho)\Big)
\,  \underbrace{
\hat{y}_{(1}\ldots\ldots\ldots. \,\hat{y}_{1}
}_{ h_1+h_2-2r+\frac{1}{2}+m+\rho } 
\underbrace{
\hat{y}_{2}\ldots\ldots\ldots. \,\hat{y}_{2)}
}_{ h_1+h_2-2r+\frac{1}{2}-m-\rho }K  \Bigg]
\nonumber\\
&&
+4\,\mi  \sum_{r=1}^{[\frac{h_1+h_2-1}{2}]}
\Bigg[(-1)^r \Big(q_{\mathrm{F},\,2(r-1)}^{h_1,h_2+\frac{1}{2}}(m,\rho)-
q_{\mathrm{B},\,2(r-1)}^{h_1,h_2+\frac{1}{2}}(m,\rho)\Big)
\,  \underbrace{
\hat{y}_{(1}\ldots\ldots\ldots. \,\hat{y}_{1}
}_{ h_1+h_2-2r-\frac{1}{2}+m+\rho } 
\underbrace{
\hat{y}_{2}\ldots\ldots\ldots. \,\hat{y}_{2)}
}_{ h_1+h_2-2r-\frac{1}{2}-m-\rho }
\Bigg].
\nonu\\
\label{IdEvenOddOdake}
\eea
Again the spins of right hand side are given by $(h_1+h_2-2r+\frac{3}{2})$
and $(h_1+h_2-2r+\frac{1}{2})$ respectively.

$\bullet$ The case-three with two fermionic oscillators

We consider the last case.
For the case where there is no operator $K$ in the oscillator,
the first equation of Appendix $(J.1)$ of \cite{AKK} can be used.
We shift the dummy variable $h \rightarrow r-1$.
Then we obtain
the following anticommutator
\footnote{
  We can check that
  the anticommutator,
  $-\big\{ 
  \underbrace{
\hat{y}_{(1}\ldots\,\hat{y}_{1}
}_{h_1-\frac{1}{2}+\rho} 
\underbrace{
\hat{y}_{2}\ldots\,\hat{y}_{2)}
}_{h_1-\frac{1}{2}-\rho}K,\,
 \underbrace{
\hat{y}_{(1}\ldots\,\hat{y}_{1}
}_{h_2-\frac{1}{2}+\omega} 
\underbrace{
\hat{y}_{2}\ldots\,\hat{y}_{2)}
}_{h_2-\frac{1}{2}-\omega} K
\big\}$, is the same as the one in (\ref{casethreeone}) by similar
analysis.
}
\bea
&&
\big\{\underbrace{
\hat{y}_{(1 }\,\ldots\,\hat{y}_{1}
}_{h_1-\frac{1}{2}+\rho}
\,
\underbrace{
\hat{y}_{2}\,\ldots \, \hat{y}_{2)}
}_{h_1-\frac{1}{2}-\rho},\,
\underbrace{
\hat{y}_{(1 }\,\ldots\,\hat{y}_{1}
}_{h_2-\frac{1}{2}+\omega}
\,
\underbrace{
\hat{y}_{2}\,\ldots \, \hat{y}_{2)}
}_{h_2-\frac{1}{2}-\omega}\big\} 
=
\nonu \\
&& -\frac{1}{2}
\sum_{r=1}^{[\frac{h_1+h_2+1}{2}]}
(-1)^{r}
\Bigg[
\Big(o_{\mathrm{F},\,2(r-1)}^{h_1+\frac{1}{2},h_2+\frac{1}{2}}(\rho,\omega)+o_{\mathrm{B},\,2(r-1)}^{h_1+\frac{1}{2},h_2+\frac{1}{2}}(\rho,\omega)\Big)
\, \underbrace{
\hat{y}_{(1 }\ldots\ldots\ldots. \, \hat{y}_{1}
}_{ h_1+h_2-2r+1+\rho+\omega }\,
\underbrace{
\hat{y}_{2}\ldots\ldots\ldots. \, \hat{y}_{2)}
}_{ h_1+h_2-2r+1-\rho-\omega }
\nonu\\
&&
-
\Big(o_{\mathrm{F},\,2(r-1)}^{h_1+\frac{1}{2},h_2+\frac{1}{2}}(\rho,\omega)-o_{\mathrm{B},\,2(r-1)}^{h_1+\frac{1}{2},h_2+\frac{1}{2}}(\rho,\omega)\Big)
\,
\underbrace{
\hat{y}_{(1 }\ldots\ldots\ldots. \,\hat{y}_{1}
}_{ h_1+h_2-2r+1+\rho+\omega }\,
\underbrace{
\hat{y}_{2}\ldots\ldots\ldots. \,\hat{y}_{2)}
}_{ h_1+h_2-2r+1-\rho-\omega }K
\Bigg].
\label{casethreeone}
\eea
The spin of the right hand side is given by $(h_1+h_2-2r+2)$. 
The structure constants are given by Appendix $A$.

Let us consider the next case where the operator $K$ appears in one of the
oscillators.
It turns out that 
\bea
&&
\big\{\underbrace{
\hat{y}_{(1 }\,\ldots\,\hat{y}_{1}
}_{h_1-\frac{1}{2}+\rho}
\,
\underbrace{
\hat{y}_{2}\,\ldots \, \hat{y}_{2)}
}_{h_1-\frac{1}{2}-\rho},\,
\underbrace{
\hat{y}_{(1 }\,\ldots\,\hat{y}_{1}
}_{h_2-\frac{1}{2}+\omega}
\,
\underbrace{
\hat{y}_{2}\,\ldots \, \hat{y}_{2)}
}_{h_2-\frac{1}{2}-\omega} K \big\} 
=
\nonu \\
&& -\frac{\mi}{2}\,
\sum_{r=1}^{[\frac{h_1+h_2}{2}]}
(-1)^{r}
\Bigg[
\Big(o_{\mathrm{F},\,2r-1}^{h_1+\frac{1}{2},h_2+\frac{1}{2}}(\rho,\omega)+o_{\mathrm{B},\,2r-1}^{h_1+\frac{1}{2},h_2+\frac{1}{2}}(\rho,\omega)\Big)
\, \underbrace{
\hat{y}_{(1 }\ldots\ldots\ldots. \, \hat{y}_{1}
}_{h_1+h_2-2r+\rho+\omega}\,
\underbrace{
\hat{y}_{2}\ldots\ldots\ldots. \, \hat{y}_{2)}
}_{h_1+h_2-2r-\rho-\omega}
\nonu\\
&&
-
\Big(o_{\mathrm{F},\,2r-1}^{h_1+\frac{1}{2},h_2+\frac{1}{2}}(\rho,\omega)-o_{\mathrm{B},\,2r-1}^{h_1+\frac{1}{2},h_2+\frac{1}{2}}(\rho,\omega)\Big)
\,
\underbrace{
\hat{y}_{(1 }\ldots\ldots\ldots. \,\hat{y}_{1}
}_{h_1+h_2-2r+\rho+\omega}\,
\underbrace{
\hat{y}_{2}\ldots\ldots\ldots. \,\hat{y}_{2)}
}_{h_1+h_2-2r-\rho-\omega}K
\Bigg],
\label{IdOddOddOdake}
\eea
where the first equation of Appendix $(J.1)$ of \cite{AKK} is used
and dummy variable $h$ is replaced by $(r-1)$.

Therefore, there are nontrivial (anti)commutators (\ref{IdEvenEvenOdake}),
(\ref{fourthcomm}),(\ref{IdEvenOddOdake}),(\ref{casethreeone}),
and (\ref{IdOddOddOdake}).
See also Appendices $A$ and $B$
where the structure constants are written in terms of
generalized hypergeometric function.
As an intermediate step
for the ${\cal N}=4$ higher spin algebra $shs_2[\mu]$,
we should compute these for nonzero $\mu$.

\subsection{The (anti)commutators between the oscillators with nonzero
  $\mu$ }

The result of \cite{AKK} implies that 
although the structure constants contain
linear $\mu$ factor,
they will vanish under the wedge condition.
See also Appendices $(G.3)$ or  $(G.4)$ of \cite{AKK}.
We realize that the field contents for the vanishing $\mu$
are the same as the ones for nonvanishing $\mu$.
This implies that the mode dependent parts of any (anti)commutators
remain the same \footnote{We have checked that 
  this validity is true  up to total spin $11$ ($h_1+h_2 \leq 11$).
See also Appendix $C$.}.
This reflects that
in the OPE language of the ${\cal W}_{\infty}^{{\cal N}=4}[\mu]$
algebra,
the field contents in the
right hand side of OPE for the vanishing $\mu$
can be used for nonzero $\mu$.
By shifting $h \rightarrow r-2$, we introduce the mode dependent
function \cite{PRS}
\bea
&& N_r^{h_1, h_2}(m,n)  \equiv
 \sum_{k=0}^{r-1}(-1)^k
\left(\begin{array}{c}r-1 \\k\end{array}\right)
  \nonu \\
  && \times \,
        [h_1-1+m]_{r-1-k}[h_1-1-m]_k[h_2-1+n]_k[h_2-1-n]_{r-1-k}\,.
        \label{capitalN}
\eea
The falling Pochhammer symbol $[a]_n \equiv a(a-1) \cdots (a-n+1)$
is used in (\ref{capitalN}).

$\bullet$ The case-one with two bosonic oscillators

After we take the same mode dependent part as the one (coming from
$q_F$ or $q_B$) for
vanishing $\mu$ case in (\ref{fourthcomm}),
we introduce two $\mu$ dependent
parts, $a^{h_1,h_2}_{2r}(\nu)$ and $b^{h_1,h_2}_{2r} (\nu)$.
Note that there is a relation in (\ref{lamunu}). 
Once we have determined these two structure constants,
then we are done.
We can start with \footnote{ We take the maximum value of
  $r$ as an infinity which should appear in the upper bound
  of the dummy variable $r$ of the summation over $r$
  in this section and section $4$. For fixed $h_1,h_2$,
  most of the values for the $r$ which are greater than the sum of
  $(h_1+h_2)$ vanishes.}
\bea
&& \big[ 
  \underbrace{
\hat{y}_{(1}\ldots\,\hat{y}_{1}
}_{h_1-1+m} 
\underbrace{
\hat{y}_{2}\ldots\,\hat{y}_{2)}
}_{h_1-1-m},\,
 \underbrace{
\hat{y}_{(1}\ldots\,\hat{y}_{1}
}_{h_2-1+n} 
\underbrace{
\hat{y}_{2}\ldots\,\hat{y}_{2)}
}_{h_2-1-n}
\big]
=
\nonu \\
&& -\mi \sum_{r=1}^{[\frac{h_1+h_2-1}{2}]}
(-1)^r\, N^{h_1  h_2}_{2r}(m,n)\frac{1}{(2r-1)!}\,
\Bigg[\,
a^{h_1,h_2}_{2r}(\nu) \,
\,  \underbrace{
\hat{y}_{(1}\ldots\ldots\ldots\ldots\hat{y}_{1}
}_{h_1+h_2-1-2r+m+n} \,
\underbrace{
\hat{y}_{2}\ldots\ldots\ldots\ldots\hat{y}_{2)}
}_{h_1+h_2-1-2r-m-n}
\nonumber\\
&&
-b^{h_1,h_2}_{2r} (\nu)
\, \underbrace{
\hat{y}_{(1}\ldots\ldots\ldots\ldots\hat{y}_{1}
}_{h_1+h_2-1-2r+m+n} \,
\underbrace{
\hat{y}_{2}\ldots\ldots\ldots\ldots\hat{y}_{2)}
}_{h_1+h_2-1-2r-m-n}K
\,
\Bigg].
\label{first}
\eea
We can compute the above commutators starting with $(h_1,h_2)=(2,2)$
with appropriate dummy variable $r$. 
One kind of the structure constants leads to the following
$\mu$ (or $\nu$) dependent polynomials
\bea
&& a^{2, 2} _ {2} = 2\, , \qquad
a^{2, 3} _ {2} = 2\, , \qquad
a^{2, 4} _ {2} = 2\, , \qquad
a^{2, 5} _ {2} = 2\, , \qquad
a^{2, 6} _ {2} = 2\, , \qquad
a^{2, 7} _ {2} = 2\, ,  \nonu\\
&& a^{2, 8} _ {2} = 2\, , \qquad
a^{2, 9} _ {2} = 2\, , \qquad
a^{3, 2} _ {2} = 2\, , \qquad
a^{3, 3} _ {2} = 2\, , \qquad
a^{3, 3} _ {4} = -\frac {2} {15} (-15 + \nu^2)\, , \nonu\\
&& a^{3, 4} _ {2} = 2\, , \qquad
a^{3, 4} _ {4} = -\frac {2} {35} (-35 + \nu^2)\, , \qquad
a^{3, 5} _ {2} = 2\, , \qquad
a^{3, 5} _ {4} = -\frac {2} {63} (-63 + \nu^2)\, , \nonu\\
&& a^{3, 6} _ {2} = 2\, , \qquad
a^{3, 6} _ {4} = -\frac {2} {99} (-99 + \nu^2)\, , \qquad
a^{3, 7} _ {2} = 2\, , \qquad
a^{3, 7} _ {4} = -\frac {2} {143} (-143 + \nu^2)\, , \nonu\\
&& a^{3, 8} _ {2} = 2\, , \qquad
a^{3, 8} _ {4} = -\frac {2} {195} (-195 + \nu^2)\, , \qquad
a^{4, 2} _ {2} = 2\, , \qquad
a^{4, 3} _ {2} = 2\, , \nonu\\
&& a^{4, 3} _ {4} = -\frac {2} {35} (-35 + \nu^2)\, , \qquad
a^{4, 4} _ {2} = 2\, , \qquad
a^{4, 4} _ {4} = -\frac {2} {75} (-75 + \nu^2)\, , \nonu\\
&& a^{4, 4} _ {6} = \frac {2} {525} (-5 + \nu) (5 + \nu) (-21 + \nu^2)\, , \qquad
a^{4, 5} _ {2} = 2\, , \qquad
a^{4, 5} _ {4} = -\frac {2} {385} (-385 + 3 \nu^2)\, , \nonu\\
&& a^{4, 5} _ {6} = \frac {2} {2205} (-7 + \nu) (7 + \nu) (-45 + \nu^2)\, , \qquad
a^{4, 6} _ {2} = 2\, , \qquad
a^{4, 6} _ {4} = -\frac {2} {195} (-195 + \nu^2)\, ,\nonu \\
&& a^{4, 6} _ {6} = \frac {2} {6237} (-9 + \nu) (9 + \nu) (-77 + \nu^2)\, , \qquad
a^{4, 7} _ {2} = 2\, , \qquad
a^{4, 7} _ {4} = -\frac {2} {275} (-275 + \nu^2)\, ,\nonu \\
&& a^{4, 7} _ {6} = \frac {2} {14157} (-11 + \nu) (11 + \nu) (-117 + \nu^2)\, , \qquad
a^{5, 2} _ {2} = 2\, , \qquad
a^{5, 3} _ {2} = 2\, ,  \nonu\\
&& a^{5, 3} _ {4} = -\frac {2} {63} (-63 + \nu^2)\, , \qquad
a^{5, 4} _ {2} = 2\, , \qquad
a^{5, 4} _ {4} = -\frac {2} {385} (-385 + 3 \nu^2)\, ,\nonu \\
&& a^{5, 4} _ {6} = \frac {2} {2205} (-7 + \nu) (7 + \nu) (-45 + \nu^2)\, , \qquad
a^{5, 5} _ {2} = 2\, , 
a^{5, 5} _ {4} = -\frac {2} {637} (-637 + 3 \nu^2)\, , \nonu\\
&& a^{5, 5} _ {6} = \frac {2} {24255} (24255 - 562 \nu^2 +   3 \nu^4)\, ,
\nonu \\ 
&& a^{5, 5} _ {8} = -\frac {2} {33075} (-7 + \nu) (-5 + \nu) (5 + \nu) (7 + \nu) (-27 + \nu^2)\, , \qquad
a^{5, 6} _ {2} = 2\, ,
\nonu \\
&& a^{5, 6} _ {4} = -\frac {2} {315} (-315 + \nu^2)\, , \qquad
a^{5, 6} _ {6} = \frac {2} {63063} (63063 - 922 \nu^2 +      3 \nu^4)\, ,  \nonu\\
&& a^{5, 6} _ {8} = \frac {2} {218295} (-9 + \nu) (-7 + \nu) (7 + \nu) (9 + \nu) (-55 + \nu^2)\, , \qquad
a^{6, 2} _ {2} = 2\, , 
a^{6, 3} _ {2} = 2\, ,  \nonu\\
&& a^{6, 3} _ {4} = -\frac {2} {99} (-99 + \nu^2)\, , \qquad
a^{6, 4} _ {2} = 2\, , \qquad
a^{6, 4} _ {4} = -\frac {2} {195} (-195 + \nu^2)\, , \nonu\\
&& a^{6, 4} _ {6} = \frac {2} {6237} (-9 + \nu) (9 + \nu) (-77 + \nu^2)\, , \qquad
a^{6, 5} _ {2} = 2\, , \qquad
a^{6, 5} _ {4} = -\frac {2} {315} (-315 + \nu^2)\, , \nonu\\
&& a^{6, 5} _ {6} = \frac {2} {63063} (63063 - 922 \nu^2 +   3 \nu^4)\, , \nonu\\
&& a^{6, 5} _ {8} = -\frac {2} {218295} (-9 + \nu) (-7 + \nu) (7 + \nu) (9 + \nu) (-55 + \nu^2)\, , \qquad
a^{7, 2} _ {2} = 2\, , 
a^{7, 3} _ {2} = 2\, , \nonu\\
&& a^{7, 3} _ {4} = -\frac {2} {143} (-143 + \nu^2)\, , \qquad
a^{7, 4} _ {2} = 2\, , \qquad
a^{7, 4} _ {4} = -\frac {2} {275} (-275 + \nu^2)\, , \nonu\\
&& a^{7, 4} _ {6} = \frac {2} {14157} (-11 + \nu) (11 + \nu) (-117 + \nu^2)\, , \qquad
a^{8, 2} _ {2} = 2\, , \qquad
a^{8, 3} _ {2} = 2\, , \nonu\\
&& a^{8, 3} _ {4} = -\frac {2} {195} (-195 + \nu^2)\, , \qquad
a^{9, 2} _ {2} = 2\,. 
\label{avalue}
\eea
First of all, the $\nu$ independent constant coefficients
appear as $2$.
Let us keep track of the $\nu$ dependence.
The $\nu^2$ dependence appears in the coefficient $a^{3, 3} _ {4}$.
This behavior goes to $a^{3, 8} _ {4}$ by varying $h_2$
for fixed $h_1$ and $2r$.
The $\nu^4$ dependence appears in the coefficient $a^{4, 4} _ {6}$.
This behavior appears until the coefficient
$a^{4, 7} _ {6}$ for fixed $h_1$ and $2r$.
Moreover,
 $\nu^6$ dependence appears in the coefficient $a^{5, 5} _ {8}$.
This behavior appears until the coefficient
$a^{5, 6} _ {8}$ for fixed $h_1$ and $2r$.

By considering the constant piece of 
the coefficient $a^{3, 3} _ {4}$ further,
then $(\nu^2-15)$ can be written in terms of
$(\nu-3)(\nu+5)$ plus other term.
By changing as $\nu \rightarrow -\nu$, then we have
$(\nu-5)(\nu+3)$ which is equivalent to
change the signs in the constant terms in the original expression.
That is $-3$ goes to $+3$ and $+5$ goes to $-5$.
By adding these two, we obtain $2(\nu^2-15)$ which is proportional to
the previous expression.
We will observe that the factor $(\nu-3)(\nu+5)$
can be written in terms of the linear combination of
two generalized hypergeometric functions at the specific upper and lower
arguments.
We can analyze each coefficient
$a^{3, h_2} _ {4}$ term where $h_2=4,5,6,7,8$ by adding
the number $2$ in the constant terms at each step.
Then for  the coefficient $a^{3, 8} _ {4}$, we can consider the two factors,
$(\nu-13)(\nu+15)$ which can be seen from the above
$(\nu-3)(\nu+5)$ after five steps
and $(\nu-15)(\nu+13)$ which can be obtained by sign changes.
By adding these two factors, we
obtain the coefficient  $a^{3, 8} _ {4}$ up to an overall scale
\footnote{Explicitly we have $\phi_4^{3,3}(\mu,1)=
  -\frac{1}{15}(-3+\nu)(5+\nu)$ by using the definition of
  (\ref{spf}) with (\ref{lamunu}) and
  $\phi_4^{3,3}(1-\mu,1) =\phi_4^{3,3}(\mu,1)
  \big|_{\nu \rightarrow -\nu}$.}.

We can consider the numerical value $21$
as the product of odd numbers $3$ and $7$ in the coefficient
$a^{4, 4} _ {6}$.
This implies that we consider the factor
$(\nu-3)(\nu+5)(\nu-5)(\nu+7)$ which is the multiplication between
the previous factor $(\nu-3)(\nu+5)$ appeared in previous paragraph
and $(\nu-5)(\nu+7)$ which is
obtained by adding the number $2$ to each constant term
in the first factor (ignore the signs).
Furthermore, we consider the next quartic term
$(\nu+3)(\nu-5)(\nu+5)(\nu-7)$ where the first two factors can be seen from
the previous paragraph and the next two factors are obtained by
adding the number $2$ to the constant terms in the first two factors.
Then we realize that by adding these two quartic terms we have
the coefficient $a^{4, 4} _ {6}$ up to an overall scale.
The quartic behavior comes from the fact that
the spins $h_1, h_2$ and $r$ are increased. 
It is easy to observe that the relevant quartic terms
for the coefficient $a^{4, 7} _ {6}$ are given by the factor
$(\nu-9)(\nu+11)(\nu-11)(\nu+13)$ where $117=9 \times 13$
and the factor
$(\nu+9)(\nu-11)(\nu+11)(\nu-13)$
\footnote{In this case we have  $\phi_6^{4,4}(\mu,1)=
  \frac{1}{525} (-5+\nu)(-3+\nu)(5+\nu)(7+\nu)$
  with 
  (\ref{spf}) and (\ref{lamunu})
  and also we have $\phi_6^{4,4}(1-\mu,1)=
  \phi_6^{4,4}(\mu,1)\big|_{\nu \rightarrow -\nu}$.}.

Let us analyze the sextic term in the coefficient
$a^{5, 5} _ {8}$ where the spins $h_1, h_2$ and $r$ are further
increased.
In this case, we reconsider $27$ as the multiple of $3$ and $9$.
Then we have $(\nu-3)(\nu+5)(\nu-5)(\nu+7)(\nu-7)(\nu+9)$
where the first and last factors come from $(\nu^2-27)$.
We understand  that the above sextic term can be written in terms of
the first quartic term which appears in previous paragraph and the
factor $(\nu-7)(\nu+9)$ which is obtained by adding the number $2$
from the factor $(\nu-5)(\nu+7)$.
The other sextic term can be seen from the above by changing $\nu \rightarrow
-\nu$ and is $(\nu+3)(\nu-5)(\nu+5)(\nu-7)(\nu+7)(\nu-9)$ (or simply by
changing the signs in the constant terms).
Therefore, by adding these two sextic terms we obtain  the coefficient
$a^{5, 5} _ {8}$ up an overall scale. 
For the  coefficient $a^{5, 6} _ {8}$, by realizing that $55$ is
the product of $5$ and $11$, the relevant sextic terms are
$(\nu \pm 5)(\nu \mp 7)(\nu \pm 7)(\nu \mp 9)( \nu \pm 9)( \nu \mp 11)$
\footnote{Similarly, we have  $\phi_8^{5,5}(\mu,1)=
  \frac{1}{33075}(-7+\nu)(-5+\nu)(-3+\nu)(5+\nu)(7+\nu)(9+\nu)$
together with
  (\ref{spf}) and we have $\phi_8^{5,5}(1-\mu,1)=
  \phi_8^{5,5}(\mu,1)\big|_{\nu \rightarrow -\nu}$.}.

Although it is not obvious, at the moment, how we can rearrange for the coefficients
$a^{4, 4} _ {4}$ and $ a^{4, 5} _ {4}$ because the constant terms cannot be
written as the product of two neighboring odd numbers, once we determine
the $\mu$ dependence for the above descriptions where we have nice
factorized forms, then these exotic coefficients also behave nicely. 

The other kind of structure constants gives 
\bea
&&
b^{3,3}_{4}=-\frac{4 \nu}{15}\,,\qquad
b^{3,4}_{4}=-\frac{4 \nu}{35}\,,\qquad
b^{3,5}_{4}=-\frac{4 \nu}{63}\,,\qquad
b^{3,6}_{4}=-\frac{4 \nu}{99}\,,
b^{3,7}_{4}=-\frac{4 \nu}{143}\,,
\nonu\\
&&
b^{3,8}_{4}=-\frac{4 \nu}{195}\,,\qquad
b^{4,3}_{4}=-\frac{4 \nu}{35}\,,\qquad
b^{4,4}_{4}=-\frac{4 \nu}{75}\,,\qquad
b^{4,4}_{6}=\frac{8}{525} (-5+\nu) \nu (5+\nu)\,,
\nonu\\
&&
b^{4,5}_{4}=-\frac{12 \nu}{385}\,,\qquad
b^{4,5}_{6}=\frac{8}{2205} (-7+\nu) \nu (7+\nu)\,,\qquad
b^{4,6}_{4}=-\frac{4 \nu}{195}\,,
\nonu\\
&&
b^{4,6}_{6}=\frac{8}{6237} (-9+\nu) \nu (9+\nu)\,,\qquad
b^{4,7}_{4}=-\frac{4 \nu}{275}\,,
b^{4,7}_{6}=\frac{8}{14157} (-11+\nu) \nu (11+\nu)\,,
\nonu\\
&&
b^{5,3}_{4}=-\frac{4 \nu}{63}\,,\qquad
b^{5,4}_{4}=-\frac{12 \nu}{385}\,,\qquad
b^{5,4}_{6}=\frac{8}{2205} (-7+\nu) \nu (7+\nu)\,,\qquad
b^{5,5}_{4}=-\frac{12 \nu}{637}\,,
\nonu\\
&&
b^{5,5}_{6}=\frac{8}{24255} \nu (-287+3 \nu^2)\,,\qquad
b^{5,5}_{8}=-\frac{4}{11025} (-7+\nu) (-5+\nu) \nu (5+\nu) (7+\nu)\,,
\nonu\\
&&
b^{5,6}_{4}=-\frac{4 \nu}{315}\,,\qquad
b^{5,6}_{6}=\frac{8}{63063} \nu (-467+3 \nu^2)\,,
\nonu \\
&&
b^{5,6}_{8}=-\frac{4}{72765} (-9+\nu) (-7+\nu) \nu (7+\nu) (9+\nu)\,,\qquad
b^{6,3}_{4}=-\frac{4 \nu}{99}\,,\qquad
b^{6,4}_{4}=-\frac{4 \nu}{195}\,,
\nonu\\
&&
b^{6,4}_{6}=\frac{8}{6237} (-9+\nu) \nu (9+\nu)\,,\qquad
b^{6,5}_{4}=-\frac{4 \nu}{315}\,,\qquad
b^{6,5}_{6}=\frac{8}{63063} \nu (-467+3 \nu^2)\,,
\nonu\\
&&
b^{6,5}_{8}=-\frac{4}{72765} (-9+\nu) (-7+\nu) \nu (7+\nu) (9+\nu)\,,\qquad
b^{7,3}_{4}=-\frac{4 \nu}{143}\,,
b^{7,4}_{4}=-\frac{4 \nu}{275}\,,
\nonu\\
&&
b^{7,4}_{6}=\frac{8}{14157} (-11+\nu) \nu (11+\nu)\,,\qquad
b^{8,3}_{4}=-\frac{4 \nu}{195}\,.
\label{bvalue}
\eea
There are no numerical constants. All the coefficients
depend on $\nu$ (At $\nu=0$, these are vanishing).
The $\nu$ dependence appears in the coefficient $b^{3, 3} _ {4}$.
This behavior goes to $b^{3, 8} _ {4}$ by varying
the $h_2$ for fixed $h_1$ and $2r$.
It is not straightforward to analyze these cases because they do depend on
the linear $\nu$ with different numerical coefficients.

The $\nu^3$ dependence appears in the coefficient $b^{4, 4} _ {6}$.
This behavior appears until the coefficient
$b^{4, 7} _ {6}$ for fixed $h_1$ and $2r$.
Let us look at the coefficient $b^{4, 4} _ {6}$ which has
$(-5+\nu)\nu(5+\nu)$.
We saw the quartic terms are given by
$(\nu \mp 3)(\nu \pm 5)(\nu \mp 5)(\nu \pm 7)$
in the coefficient of $a_{6}^{4,4}$.
Note that the middle two factors are the same for each quartic term.
Then by subtracting these two, we obtain $8 \nu(\nu-5)(\nu+5)$ which is
related to the above coefficient.
Similarly,
from the quartic terms $(\nu \mp 9)(\nu \pm 11)(\nu \mp 11)(\nu \pm 13)$
in the  coefficient of $a_{6}^{4,7}$, by subtracting them,
we arrive at
$8 \nu(\nu-11)(\nu+11)$ which is
related to the coefficient  $b^{4, 7} _ {6}$.
This implies that the second kind of structure constants can be obtained
by taking the subtraction between two generalized hypergeometric functions.

Moreover,
 $\nu^5$ dependence appears in the coefficient $b^{5, 5} _ {8}$.
This behavior appears until the coefficient
$b^{5, 6} _ {8}$ for fixed $h_1$ and $2r$.
We can analyze similarly.
From the sextic terms 
$(\nu \pm 3)(\nu \mp 5)(\nu \pm 5)(\nu \mp 7)( \nu \pm 7)( \nu \mp 9)$
in   the coefficient of $a_{8}^{5,5}$,
we can take the difference between them and obtain
$12 \nu (\nu - 5)(\nu + 5)(\nu - 7)( \nu + 7)$
which is proportional to the coefficient $b^{5, 5} _ {8}$.
From the sextic terms 
$(\nu \pm 5)(\nu \mp 7)(\nu \pm 7)(\nu \mp 9)( \nu \pm 9)( \nu \mp 11)$
in   the coefficient of $a_{8}^{5,6}$,
we can take the difference between them and obtain
$12 \nu (\nu - 7)(\nu + 7)(\nu - 9)( \nu + 9)$
which is related to the coefficient $b^{5, 6} _ {8}$.

Then the question is how we generalize the above structure constants
for generic $h_1,h_2$ and $r$.
Recall that the form of $\mu$ dependence of (\ref{IdEvenEvenOdake}),
(\ref{fourthcomm}), (\ref{IdEvenOddOdake}),
(\ref{casethreeone}),  and (\ref{IdOddOddOdake})    is the combination
of two generalized hypergeometric functions.
The question is how the $\mu$ or $\nu$ dependence appears  
in the structure constant.
Remember that the upper four elements of generalized hypergeometric
function appear in the numerator
of the fractional coefficient while
the lower three elements  appear in the denominator
of the fractional coefficient.
The behaviors of (\ref{avalue})
and (\ref{bvalue}) imply that the deformation parameter $\nu$ (or $\mu$)
can appear in the  upper four elements of generalized hypergeometric
function. The simplest case we can
consider is  to put the deformation parameter into the first two
elements respectively while we do not touch the third and fourth elements
which are associated with the dummy variable $r$ (Note that we can
choose other two elements rather than first two because from the
definition of generalized hypergeometric function
Appendix (\ref{4F3function}) the assignments of four upper
elements are arbitrary.).

It turns out that the $\mu$ dependent structure constants
can be written in terms of generalized hypergeometric functions
as follows:
\bea
a^{h_1,h_ 2} _ {2r}(\mu)
& = & {}_4F_3\left[
\begin{array}{c|}
\frac{1}{2} + \mu \ ,  \frac{1}{2} - \mu  \ , \frac{2-2r}{ 2} \ , \frac{1-2r}{2}\\
\frac{3}{2}-h_1 \ , \frac{3}{2}-h_2\ , \frac{1}{2}+ h_1+h_2-2r
\end{array}  \ 1\right]
\nonu \\
& + & {}_4F_3\left[
\begin{array}{c|}
\frac{3}{2} - \mu \ ,  -\frac{1}{2} +\mu  \ , \frac{2-2r}{ 2} \ , \frac{1-2r}{2}\\
\frac{3}{2}-h_1 \ , \frac{3}{2}-h_2\ , \frac{1}{2}+ h_1+h_2-2r
\end{array}  \ 1\right]\,,
\nonu\\
b^{h_1,h_ 2} _ {2r}(\mu)
& = & {}_4F_3\left[
\begin{array}{c|}
\frac{1}{2} + \mu \ ,  \frac{1}{2} - \mu  \ , \frac{2-2r}{ 2} \ , \frac{1-2r}{2}\\
\frac{3}{2}-h_1 \ , \frac{3}{2}-h_2\ , \frac{1}{2}+ h_1+h_2-2r
\end{array}  \ 1\right]
\nonu \\
& - & {}_4F_3\left[
\begin{array}{c|}
\frac{3}{2} - \mu \ ,  -\frac{1}{2} +\mu  \ , \frac{2-2r}{ 2} \ , \frac{1-2r}{2}\\
\frac{3}{2}-h_1 \ , \frac{3}{2}-h_2\ , \frac{1}{2}+ h_1+h_2-2r
\end{array}  \ 1\right]\,.
\label{abconstphi}
\eea
Note that by taking $\mu \rightarrow (1-\mu)$
(or $\nu \rightarrow -\nu$) in the second
generalized hypergeometric function in each case,
we obtain the first one. They have common elements, the last two in the upper
and the three in the lower.
We will see that the above two structure constants are denoted by
the first one in (\ref{3struct}).
Of course, the previous explicit results in (\ref{avalue})
and (\ref{bvalue}) satisfy
the relation (\ref{abconstphi}).
Note that the infinite series of generalized hypergeometric function
terminates if any of upper elements is negative integer or zero.
It is obvious that the common upper third element of
generalized hypergeometric functions in (\ref{abconstphi}),
$\frac{(2-2r)}{2}$, becomes negative integer or zero for (\ref{avalue})
and (\ref{bvalue}). Furthermore, the sum of upper elements plus $1$
is equal to the sum of lower elements of generalized hypergeometric
functions which is called ``saalschutzian'' \cite{PRS}. In this
case, there exists some relation which connects two different
generalized hypergeometric functions.
We can analyze the other two cases which will be described in
Appendix $C$.
 
\section{ The ${\cal N}=4$ higher spin algebra  $shs_2[\mu]$}

\subsection{Mode dependent structure constants}

From the results of previous section,
we realize that
the third and fourth upper elements of
generalized hypergeometric function
for the case-two and case-three differ from the
one for the case-one. See also Appendix $C$.
By introducing the notation for the generalized hypergeometric
function \footnote{Note that there are relations in (\ref{lamunu}).
  We can present the structure constants by using $\mu$ or $\nu$.
    For the $\nu$, we replace $\mu$ by $\frac{(\nu+1)}{2}$.
  This leads to the changing $\nu \leftrightarrow -\nu$ for
  the shift $\mu \leftrightarrow 1-\mu$.
}
\bea
\phi_{r}^{h_1 ,h_2}(\Lambda,a)  \equiv \ _4F_3\left[
\begin{array}{c|}
\frac{1}{2} + \Lambda \ ,  \frac{1}{2} - \Lambda  \ , \frac{1+a-r}{ 2}\ , \frac{a-r}{2}\\
\frac{3}{2}-h_1 \ , \frac{3}{2} -h_2\ , \frac{1}{2}+ h_1+h_2-r
\end{array}  \ 1\right] \ ,
\label{spf}
\eea
where
the sum of upper four elements plus $1$ is not equal to
the sum of lower three elements unless $a = 1$ 
and mode dependent function in (\ref{capitalN}),
we can write down the mode dependent structure constants
which are polynomials of $\mu$ as follows:
\bea
\mathrm{BB}^{h_1,h_2}_{r,\,\pm}(m,n; \mu )
&\equiv&
 -\frac{1  }{ (r-1)!} N_r^{h_1, h_2}(m,n) \Bigg[
\phi_{r}^{h_1 ,h_2}(\mu,1)  \pm \phi_{r}^{h_1 ,h_2}(1-\mu,1)  
\Bigg],
\nonu\\
\mathrm{BF}^{h_1,h_2+\frac{1}{2}}_{r,\,\pm}(m,\rho; \mu )
&\equiv&
 -\frac{1  }{ (r-1)!} N_r^{h_1, h_2+\frac{1}{2}}(m,\rho) \Bigg[
 \phi_{r+1}^{h_1 ,h_2+1}(\mu,\tfrac{3\pm1}{2})
 \nonu \\
 & \pm & \phi_{r+1}^{h_1 ,h_2+1}(1-\mu,\tfrac{3\pm1}{2})  
\Bigg],
\nonu\\
\mathrm{FF}^{h_1+\frac{1}{2},h_2+\frac{1}{2}}_{r,\,\pm}(\rho,\omega; \mu )
&\equiv&
-\frac{1  }{ (r-1)!}N_{r}^{h_1+\frac{1}{2}, h_2+\frac{1}{2}}(\rho,\omega)
\Bigg[
 \phi_{r+1}^{h_1+1 ,h_2+1}(\mu,\tfrac{3\pm1}{2})  \nonu \\
 & \pm & \phi_{r+1}^{h_1+1 ,h_2+1}(1-\mu,\tfrac{3\pm1}{2})  
 \Bigg].
 \label{3struct}
\eea
From the expression of (\ref{spf}), although
by increasing the index $r$ as $r+1$ for fixed $h_1$ and $h_2$, 
the third lower element is changed, we can make the last two upper
elements remain the same by further change of $a$ as $a+1$.
In this case, the number of terms in the generalized
hypergeometric functions in Appendix (\ref{4F3function}) are the same while
its coefficients are different (from the different lower elements)
and the highest power of
$\mu$ is the same.

Moreover, by changing $a$ as $a+1$ for fixed $h_1, h_2$ and $r$,
we observe that the third upper element in the former
is the same as the the fourth upper element in the latter
because the third element is given by the four element plus
$\frac{1}{2}$.
For example, when $a=1$ with even $r$, we have negative integer
in the third upper element and this will terminate the generalized
hypergeometric function in Appendix (\ref{4F3function}).
Therefore, the two cases, $a=1$ and $a=2$ for the same
$h_1, h_2, r$ values, will have the same number of terms
(the higher power of $\mu$ is the same) 
with its different coefficients.

Let us emphasize that these structure constants appear
in the ${\cal N}=4$ higher spin generators of spins
(equal to the sum of upper indices  minus lower index) in the
right hand side of (anti)commutators. 
Note that
in the second and third structure constants of
(\ref{3struct}), the three indices (upper two and lower one)
in the generalized hypergeometric
functions appear differently: $h_2$ by $h_2+\frac{1}{2}$
and $r$ by $r+1$ for the former (and $h_i$ by $h_i+\frac{1}{2}$
and $r$ by $r+1$ for the latter).
For the lower signs of these structure constants, 
the second argument of (\ref{spf}) which is equal to $1$
is the same as the one of first structure constant of (\ref{3struct})
although the upper and lower indices are different from each other.

We have the following symmetry under the transformation
$\mu \leftrightarrow 1-\mu$ (corresponding to
$N \leftrightarrow k$ symmetry in the ${\cal N}=4$ coset model \cite{GG1305,AK1506})
\bea
\mathrm{BB}^{h_1,h_2}_{r,\,\pm}(m,n; \mu ) & = &
\pm
\mathrm{BB}^{h_1,h_2}_{r,\,\pm}(m,n; 1-\mu ),
\nonu \\
\mathrm{BF}^{h_1,h_2+\frac{1}{2}}_{r,\,\pm}(m,\rho; \mu ) &=&
\pm
\mathrm{BF}^{h_1,h_2+\frac{1}{2}}_{r,\,\pm}(m,\rho; 1-\mu ),
\nonu \\
\mathrm{FF}^{h_1+\frac{1}{2},h_2+\frac{1}{2}}_{r,\,\pm}(\rho,\omega; \mu )
& = &
\pm 
\mathrm{FF}^{h_1+\frac{1}{2},h_2+\frac{1}{2}}_{r,\,\pm}(\rho,\omega; 1-\mu ).
\label{symmetry}
\eea
We observe that the half of the structure constants from (\ref{symmetry})
vanish at $\mu=
\frac{1}{2}$ (or $\nu=0$ which is the undeformed case).
We can write down the mode dependent structure constants
for vanishing $\mu$ and they can be related to the ones in
Odake basis \cite{Odake} as follows:
\bea
\mathrm{BB}^{h_1,h_2}_{h+2,\,\pm}(m,n;0)
 &=&
-2\Big(p_{\mathrm{F},\, h}^{h_1,h_2}(m,n)\pm \,p_{\mathrm{B},\, h}^{h_1,h_2}(m,n)\Big)\,,
\nonu\\
\mathrm{BF}^{h_1,h_2+\frac{1}{2}}_{2(h+1),\,+}(m,\rho; 0 )
&=&
-4\Big(q_{\mathrm{F}, \,2h}^{h_1,h_2+\frac{1}{2}}(m,\rho)+q_{\mathrm{B},\, 2h}^{h_1,h_2+\frac{1}{2}}(m,\rho)\Big)\,,
\nonu\\
\mathrm{BF}^{h_1,h_2+\frac{1}{2}}_{2(h+1),\,-}(m,\rho; 0 )
&=&
\frac{8(h+1)}{(2h_1-2h-3)}\Big(q_{\mathrm{F},\, 2h}^{h_1,h_2+\frac{1}{2}}(m,\rho)-q_{\mathrm{B},\, 2h}^{h_1,h_2+\frac{1}{2}}(m,\rho)\Big)\,,
\nonu\\
\mathrm{BF}^{h_1,h_2+\frac{1}{2}}_{2h+3,\,+}(m,\rho; 0)
&=&
4\Big(q_{\mathrm{F},\, 2h+1}^{h_1,h_2+\frac{1}{2}}(m,\rho)-q_{\mathrm{B},\, 2h+1}^{h_1,h_2+\frac{1}{2}}(m,\rho)\Big)\,,
\nonu\\
\mathrm{BF}^{h_1,h_2+\frac{1}{2}}_{2h+3,\,-}(m,\rho; 0 )
&=& -\frac{2(2h+3)}{(h_1-h-2)} \,
\Big(q_{\mathrm{F},\, 2h+1}^{h_1,h_2+\frac{1}{2}}(m,\rho)+q_{\mathrm{B},\, 2h+1}^{h_1,h_2+\frac{1}{2}}(m,\rho)\Big)\,,
\nonu\\
\mathrm{FF}^{h_1+\frac{1}{2},h_2+\frac{1}{2}}_{2h+1,\,+}(\rho,\omega; 0 )
&=&
-\frac{1}{2}\Big(o_{\mathrm{F}, \,2h}^{h_1+\frac{1}{2},h_2+\frac{1}{2}}(\rho,\omega)+o_{\mathrm{B},\, 2h}^{h_1+\frac{1}{2},h_2+\frac{1}{2}}(\rho,\omega)\Big)\,,
\nonu\\
\mathrm{FF}^{h_1+\frac{1}{2},h_2+\frac{1}{2}}_{2h+1,\,-}(\rho,\omega; 0 )
&=&
-\frac{(2h+1)}{4(h_1+h_2-h)}\Big(o_{\mathrm{F}, \,2h}^{h_1+\frac{1}{2},h_2+\frac{1}{2}}(\rho,\omega)-o_{\mathrm{B},\, 2h}^{h_1+\frac{1}{2},h_2+\frac{1}{2}}(\rho,\omega)\Big)\,,
\nonu\\
\mathrm{FF}^{h_1+\frac{1}{2},h_2+\frac{1}{2}}_{2(h+1),\,+}(\rho,\omega; 0 )
&=&
\frac{1}{2}\Big(o_{\mathrm{F}, \,2h+1}^{h_1+\frac{1}{2},h_2+\frac{1}{2}}(\rho,\omega)-o_{\mathrm{B},\, 2h+1}^{h_1+\frac{1}{2},h_2+\frac{1}{2}}(\rho,\omega)\Big)\,,
\label{BBBFFFetal}
\\
\mathrm{FF}^{h_1+\frac{1}{2},h_2+\frac{1}{2}}_{2(h+1),\,-}(\rho,\omega; 0 )
 &=&
\frac{(h+1)}{2(h_1+h_2-h)-1}\Big(o_{\mathrm{F}, \,2h+1}^{h_1+\frac{1}{2},h_2+\frac{1}{2}}(\rho,\omega)+o_{\mathrm{B},\, 2h+1}^{h_1+\frac{1}{2},h_2+\frac{1}{2}}(\rho,\omega)\Big)\,.
\nonu
\eea
In other words, the six independent structure constants
can be written in terms of those in (\ref{3struct}). 
Note that there are shifts between the left hand side and
the right hand side in the lower spin indices in (\ref{BBBFFFetal}).

$\bullet$ The case-one with two bosonic oscillators

We can rewrite (\ref{first}) with the help of (\ref{abconstphi})
(or from Appendix (\ref{IdEvenEven4F3}))
as follows:
\bea
&&\big[ 
  \underbrace{
\hat{y}_{(1}\ldots\,\hat{y}_{1}
}_{h_1-1+m} 
\underbrace{
\hat{y}_{2}\ldots\,\hat{y}_{2)}
}_{h_1-1-m},\,
 \underbrace{
\hat{y}_{(1}\ldots\,\hat{y}_{1}
}_{h_2-1+n} 
\underbrace{
\hat{y}_{2}\ldots\,\hat{y}_{2)}
}_{h_2-1-n}
\big]
\nonu\\
&&=
\mi \, \sum_{r=1}^{[\frac{h_1+h_2-1}{2}]} \,(-1)^r \,
\Bigg[\,
\mathrm{BB}^{h_1,h_2}_{2r,\,+}(m,n; \mu)
\,  \underbrace{
\hat{y}_{(1}\ldots\ldots\ldots\ldots\hat{y}_{1}
}_{h_1+h_2-2r-1+m+n} \,
\underbrace{
\hat{y}_{2}\ldots\ldots\ldots\ldots\hat{y}_{2)}
}_{h_1+h_2-2r-1-m-n}
\nonumber\\
&&
\hspace{2.5cm}\,
-\mathrm{BB}^{h_1,h_2}_{2r,\,-}(m,n; \mu)
\,  \underbrace{
\hat{y}_{(1}\ldots\ldots\ldots\ldots\hat{y}_{1}
}_{h_1+h_2-2r-1+m+n} \,
\underbrace{
\hat{y}_{2}\ldots\ldots\ldots\ldots\hat{y}_{2)}
}_{h_1+h_2-2r-1-m-n}K
\Bigg]
\,.
\label{IdEvenEven}
\eea
The only difference between
the first structure constant and the second one is
the sign in the generalized hypergeometric function containing
the argument $(1-\mu)$ from (\ref{3struct}).

$\bullet$ The case-two with one bosonic and one fermionic oscillators

From Appendix (\ref{IdEvenOdd4F3-1}), we can
reexpress this commutator as follows:
\bea
&&\big[ 
  \underbrace{
\hat{y}_{(1}\ldots\,\hat{y}_{1}
}_{h_1-1+m} 
\underbrace{
\hat{y}_{2}\ldots\,\hat{y}_{2)}
}_{h_1-1-m},\,
 \underbrace{
\hat{y}_{(1}\ldots\,\hat{y}_{1}
}_{h_2-\frac{1}{2}+\rho} 
\underbrace{
\hat{y}_{2}\ldots\,\hat{y}_{2)}
}_{h_2-\frac{1}{2}-\rho}
\big]
\nonumber\\
&&=
 \sum_{r=1}^{[\frac{h_1+h_2-1}{2}]}\,
\Bigg[
\mi\,(-1)^r \,
\mathrm{BF}^{h_1,h_2+\frac{1}{2}}_{2r,\,+}(m,\rho; \mu)
\,  \underbrace{
\hat{y}_{(1}\ldots\ldots\ldots\ldots\hat{y}_{1}
}_{h_1+h_2-2r-\frac{1}{2}+m+\rho} \,
\underbrace{
\hat{y}_{2}\ldots\ldots\ldots\ldots\hat{y}_{2)}
}_{h_1+h_2-2r-\frac{1}{2}-m-\rho}\,\Bigg]
\nonumber\\
&&
-\sum_{r=1}^{[\frac{h_1+h_2}{2}]}\,
\Bigg[(-1)^r \, \frac{2(h_1-r)}{2r-1}\, \mathrm{BF}^{h_1,h_2+\frac{1}{2}}_{2r-1,\,-}(m,\rho; \mu)
\,  \underbrace{
\hat{y}_{(1}\ldots\ldots\ldots\ldots\hat{y}_{1}
}_{h_1+h_2-2r+\frac{1}{2}+m+\rho} \,
\underbrace{
\hat{y}_{2}\ldots\ldots\ldots\ldots\hat{y}_{2)}
}_{h_1+h_2-2r+\frac{1}{2}-m-\rho}K
\Bigg].
\nonu\\
\label{IdEvenOdd-1}
\eea

From Appendix (\ref{IdEvenOdd4F3}), we can
reexpress by using the above mode dependent structure constant
\bea
&&\big[ 
  \underbrace{
\hat{y}_{(1}\ldots\,\hat{y}_{1}
}_{h_1-1+m} 
\underbrace{
\hat{y}_{2}\ldots\,\hat{y}_{2)}
}_{h_1-1-m}K,\,
 \underbrace{
\hat{y}_{(1}\ldots\,\hat{y}_{1}
}_{h_2-\frac{1}{2}+\rho} 
\underbrace{
\hat{y}_{2}\ldots\,\hat{y}_{2)}
}_{h_2-\frac{1}{2}-\rho}
\big]
\nonumber\\
&&
=
-\sum_{r=1}^{[\frac{h_1+h_2}{2}]}
\,
\Bigg[ \!
\,(-1)^r\,\mathrm{BF}^{h_1,h_2+\frac{1}{2}}_{2r-1,\,+}(m,\rho; \mu)
\,  \underbrace{
\hat{y}_{(1}\ldots\ldots\ldots\ldots\hat{y}_{1}
}_{h_1+h_2-2r+\frac{1}{2}+m+\rho} \,
\underbrace{
\hat{y}_{2}\ldots\ldots\ldots\ldots\hat{y}_{2)}
}_{h_1+h_2-2r+\frac{1}{2}-m-\rho}K \Bigg]
\nonumber\\
&&
\,\,+
\sum_{r=1}^{[\frac{h_1+h_2-1}{2}]}
\Bigg[ \!(-1)^r\,
 \frac{\mi \,(h_1-r-\frac{1}{2})}{r}\, 
\mathrm{BF}^{h_1,h_2+\frac{1}{2}}_{2r,\,-}(m,\rho; \mu)
\,  \underbrace{
\hat{y}_{(1}\ldots\ldots\ldots\ldots\hat{y}_{1}
}_{h_1+h_2-2r-\frac{1}{2}+m+\rho} \,
\underbrace{
\hat{y}_{2}\ldots\ldots\ldots\ldots\hat{y}_{2)}
}_{h_1+h_2-2r-\frac{1}{2}-m-\rho}
\Bigg].\nonumber\\
\label{IdEvenOdd}
\eea

$\bullet$ The case-three with two fermionic oscillators

The following expression can be obtained by considering Appendix
(\ref{IdOddOdd4F3-1})
\bea
&&
\big\{ 
  \underbrace{
\hat{y}_{(1}\ldots\,\hat{y}_{1}
}_{h_1-\frac{1}{2}+\rho} 
\underbrace{
\hat{y}_{2}\ldots\,\hat{y}_{2)}
}_{h_1-\frac{1}{2}-\rho},\,
 \underbrace{
\hat{y}_{(1}\ldots\,\hat{y}_{1}
}_{h_2-\frac{1}{2}+\omega} 
\underbrace{
\hat{y}_{2}\ldots\,\hat{y}_{2)}
}_{h_2-\frac{1}{2}-\omega}
\big\}
\nonumber\\
&&=
 \sum_{r=1}^{[\frac{h_1+h_2+1}{2}]}\,(-1)^r\,
\Bigg[
\mathrm{FF}^{h_1+\frac{1}{2},h_2+\frac{1}{2}}_{2r-1,\,+}(\rho,\omega; \mu)
\,  \underbrace{
\hat{y}_{(1}\ldots\ldots\ldots.\hat{y}_{1}
}_{h_1+h_2-2r+1+\rho+\omega} \,
\underbrace{
\hat{y}_{2}\ldots\ldots\ldots.\hat{y}_{2)}
}_{h_1+h_2-2r+1-\rho-\omega}
\label{IdOddOdd-1}
\\
&&
-\frac{2(h_1+h_2-r+1)}{2r-1}\,
\mathrm{FF}^{h_1+\frac{1}{2},h_2+\frac{1}{2}}_{2r-1,\,-}(\rho,\omega; \mu)
\,  \underbrace{
\hat{y}_{(1}\ldots\ldots\ldots.\hat{y}_{1}
}_{h_1+h_2-2r+1+\rho+\omega} \,
\underbrace{
\hat{y}_{2}\ldots\ldots\ldots.\hat{y}_{2)}
}_{h_1+h_2-2r+1-\rho-\omega}K
\Bigg].
\nonu
\eea

We have the following expression by considering Appendix
(\ref{IdOddOdd4F3})
\bea
&&
\big\{ 
  \underbrace{
\hat{y}_{(1}\ldots\,\hat{y}_{1}
}_{h_1-\frac{1}{2}+\rho} 
\underbrace{
\hat{y}_{2}\ldots\,\hat{y}_{2)}
}_{h_1-\frac{1}{2}-\rho},\,
 \underbrace{
\hat{y}_{(1}\ldots\,\hat{y}_{1}
}_{h_2-\frac{1}{2}+\omega} 
\underbrace{
\hat{y}_{2}\ldots\,\hat{y}_{2)}
}_{h_2-\frac{1}{2}-\omega}K
\big\}
\nonumber\\
&&=
\mi\, \sum_{r=1}^{[\frac{h_1+h_2}{2}]}\,(-1)^r\,
\Bigg[
\mathrm{FF}^{h_1+\frac{1}{2},h_2+\frac{1}{2}}_{2r,\,+}(\rho,\omega; \mu)
\,  \underbrace{
\hat{y}_{(1}\ldots\ldots\ldots.\hat{y}_{1}
}_{h_1+h_2-2r+\rho+\omega}\,
\underbrace{
\hat{y}_{2}\ldots\ldots\ldots.\hat{y}_{2)}
}_{h_1+h_2-2r-\rho-\omega}K
\nonumber\\
&&
-\frac{2(h_1+h_2-r)+1}{2r}
\,\mathrm{FF}^{h_1+\frac{1}{2},h_2+\frac{1}{2}}_{2r,\,-}(\rho,\omega; \mu)
\,  \underbrace{
\hat{y}_{(1}\ldots\ldots\ldots.\hat{y}_{1}
}_{h_1+h_2-2r+\rho+\omega} \,
\underbrace{
\hat{y}_{2}\ldots\ldots\ldots.\hat{y}_{2)}
}_{h_1+h_2-2r-\rho-\omega}
\Bigg].
\label{IdOddOdd}
\eea
Based on the results of this subsection, we will obtain
the ${\cal N}=4$ higher spin algebra $shs_2[\mu]$ in next subsection
\footnote{Let us emphasize again that from the indices of the structure
  constants we can determine how they appear in the (anti)commutators
  of oscillators. That is, the sum of upper indices minus lower index
  provides the spin of the oscillator in the right hand side of
  (anti)commutators of oscillators. Of course, the upper two indices
  give the spins of the oscillators in the
  left hand side of the (anti)commutators respectively.
  In (\ref{IdEvenEven}), (\ref{IdOddOdd-1}) and (\ref{IdOddOdd}),
  there exists only one kind of spin for the oscillator in the right
  hand side for fixed $r$. Each two structure constants
  has the same indices for the spins.
  In (\ref{IdEvenOdd-1}) and (\ref{IdEvenOdd}),
  there are two kinds of spin for the oscillator for fixed $r$.
  In each case, the lower indices of two structure constants are different
from each other.}.

\subsection{The (anti)commutators between the ${\cal N}=4$ higher spin
  generators with nonzero
  $\mu$ }

We determine the (anti)commutators of the ${\cal N}=4$
higher spin generators by
using (\ref{IdEvenEven}), (\ref{IdEvenOdd-1}), (\ref{IdEvenOdd}),
(\ref{IdOddOdd-1}),
and (\ref{IdOddOdd}).

Let us start with the commutator
$\big[\tilde{\Phi}^{(s_1)}_{2,\,m},\,\tilde{\Phi}^{(s_2)}_{2,\,n}\big]$
in (\ref{lastcomm}). The bosonic oscillators do not have the operator
$K$ and the $2 \times 2$ identity matrix is included.
From the previous observation $ [A \otimes I,B \otimes I]= [A,B]\otimes I$  and the realization of 
higher spin generators in (\ref{HSbasis}), 
we have the following intermediate result
for the commutator
\bea
&&\big[\tilde{\Phi}^{(s_1)}_{2,\,m},\,\tilde{\Phi}^{(s_2)}_{2,\,n}\big]
=
(2s_1-1)(2s_2-1)\nonu \\
&& \times  \Bigg(\!\!
\big[\underbrace{
\hat{y}_{(1 }\,\cdots\,\hat{y}_{1}
}_{ s_1+1+m }
\,
\underbrace{
\hat{y}_{2}\,\cdots \ \hat{y}_{2)}
}_{ s_1+1-m },\,
\underbrace{
\hat{y}_{(1 }\,\cdots\,\hat{y}_{1}
}_{ s_2+1+n }
\,
\underbrace{
\hat{y}_{2}\,\cdots \ \hat{y}_{2)}
}_{ s_2+1-n }\big]
\Bigg)
\otimes 
I_{2\times2}.
\label{comm1}
\eea
From the relation (\ref{IdEvenEven}) we have obtained,
the above commutator (\ref{comm1})
becomes
\bea
&&
(2s_1-1)(2s_2-1) 
\nonu\\
&&
\times
\Bigg(\!
\sum_{r=1}^{[\frac{s_1+s_2+3}{2}]}\,\mi \, (-1)^r
\Bigg[\,
\mathrm{BB}_{2r,\,+}^{s_1+2, \,s_2+2}(m,n;\mu)
\,
\underbrace{
\hat{y}_{(1}\ldots\ldots\ldots.\,\hat{y}_{1}
}_{ s_1+s_2+3-2r+m+n } \,
\underbrace{
\hat{y}_{2}\ldots\ldots\ldots.\,\hat{y}_{2)}
}_{ s_1+s_2+3-2r-m-n }
\nonumber\\
&&
-\mathrm{BB}_{2r,\,-}^{s_1+2,\,s_2+2}(m,n;\mu)
\,
 \underbrace{
\hat{y}_{(1}\ldots\ldots\ldots.\,\hat{y}_{1}
}_{ s_1+s_2+3-2r+m+n } \,
\underbrace{
\hat{y}_{2}\ldots\ldots\ldots.\,\hat{y}_{2)}
}_{ s_1+s_2+3-2r-m-n } K
\Bigg]\!
\Bigg)
\otimes 
I_{2\times2}.
\label{inter}
\eea
Because the right hand side contains only $2 \times 2$ identity matrix
and the spin is given by $(s_1+s_2+4-2r)$ which is an integer,
the only  ${\cal N}=4$
higher spin generators, $\Phi^{(s)}_0 $ and
$\tilde{\Phi}^{(s)}_2 $, are allowed.
From the relation (\ref{HSbasis}), the operator $K$ independent
oscillator term in (\ref{inter}) can be written as
\bea
&& 
\underbrace{
\hat{y}_{(1}\ldots\ldots\ldots.\,\hat{y}_{1}
}_{ s_1+s_2+3-2r+m+n } \,
\underbrace{
\hat{y}_{2}\ldots\ldots\ldots.\,\hat{y}_{2)}
}_{ s_1+s_2+3-2r-m-n }
   \otimes I_{2\times 2}
=
\nonu \\
&& -\frac{1}{2(s_1+s_2-2r+2)-1} \, \tilde{\Phi}^{(s_1+s_2-2r+2)}_{2,\,m+n}\,.
\label{part1}
\eea
Moreover, the operator
$K$ dependent term in (\ref{inter}) can be described by
the following linear combination
\bea
&&
 \underbrace{
\hat{y}_{(1}\ldots\ldots\ldots.\,\hat{y}_{1}
}_{ s_1+s_2+3-2r+m+n } \,
\underbrace{
\hat{y}_{2}\ldots\ldots\ldots.\,\hat{y}_{2)}
}_{ s_1+s_2+3-2r-m-n }K
   \otimes I_{2\times 2}
=
\label{part2}
\\\
&& \frac{1}{2(s_1+s_2-2r+4)-1} \bigg(
\Phi^{(s_1+s_2-2r+4)}_{0,\,m+n}
+
\frac{\nu}{2(s_1+s_2-2r+2)-1}\,
\tilde{\Phi}^{(s_1+s_2-2r+2)}_{2,\,m+n}
\bigg)\,.
\nonu
\eea
Therefore,  by combining (\ref{part1}) with (\ref{part2}),
we obtain the final commutator relation
which is given by 
Appendix (\ref{22comm}).

Let us emphasize that
in order to apply the three cases (we have described before)
for the construction of (anti)commutators between the ${\cal N}=4$
higher spin generators,
we should include at least one higher spin generator
in the left hand side 
which contains $2 \times 2$ identity matrix.
If this is not the case, we should find the relevant (anti)commutators
separately.
For example, 
the case
$\delta_{ik}\,\big[ \Phi^{(s),ij}_{1,\,m}, \, \Phi^{(s),kl}_{1,\,n}\big] $
described by Appendix (\ref{Com11}) does not have
any combination of the commutative matrix.
Note that the following commutator
between the tensor products 
provides the tensor product of the anticommutator between the operators
and $2 \times 2$ matrix 
\footnote{
In this case we have 
\bea
&&\big\{ 
  \underbrace{
\hat{y}_{(1}\ldots\,\hat{y}_{1}
}_{h_1-1+m} 
\underbrace{
\hat{y}_{2}\ldots\,\hat{y}_{2)}
}_{h_1-1-m},\,
 \underbrace{
\hat{y}_{(1}\ldots\,\hat{y}_{1}
}_{h_2-1+n} 
\underbrace{
\hat{y}_{2}\ldots\,\hat{y}_{2)}
}_{h_2-1-n}
\big\}
\nonumber\\
&&=
 \sum_{r=1}^{[\frac{h_1+h_2}{2}]}\,(-1)^r\,
\Bigg[
\,
\mathrm{BB}^{h_1,h_2}_{2r-1,\,+}(m,n; \mu)
\,  \underbrace{
\hat{y}_{(1}\ldots\ldots\ldots .\,\hat{y}_{1}
}_{h_1+h_2-2r+m+n} \,
\underbrace{
\hat{y}_{2}\ldots\ldots\ldots . \,\hat{y}_{2)}
}_{h_1+h_2-2r-m-n}
\nonumber\\
&&
\hspace{2.2cm}\,
-
\mathrm{BB}^{h_1,h_2}_{2r-1,\,-}(m,n; \mu)
\,  \underbrace{
\hat{y}_{(1}\ldots\ldots\ldots .\,\hat{y}_{1}
}_{h_1+h_2-2r+m+n} \,
\underbrace{
\hat{y}_{2}\ldots\ldots\ldots .\,\hat{y}_{2)}
}_{h_1+h_2-2r-m-n}K
\Bigg].
\label{IdAntiEvenEven}
\eea
Note that there exists an anticommutator rather than commutator
in the left side.
}
\bea
    [A \otimes \sigma^i,B \otimes \sigma^j]= \{A, B\} \otimes
   \,  \mi \, \epsilon^{ijk}\sigma^k + [A,B]\otimes \delta^{ij} I_{2\times 2} .
\label{nontrivial}
    \eea
    According to this relation (\ref{nontrivial}),
    we should put the relevant ${\cal N}=4$ higher spin generators
    both sides.
    
Moreover, it is easy to see that the anticommutators,
$\big\{ \Phi^{(s),1}_{\frac{1}{2},\,\rho}, \,  \Phi^{(s),1}_{\frac{1}{2},\,\sigma}\big\}$  and
$\big\{ \Phi^{(s),1}_{\frac{1}{2},\,\rho}, \,  \Phi^{(s),2}_{\frac{1}{2},\,\sigma}\big\}$,  cannot be obtained directly by using the relation (\ref{IdOddOdd-1})
because we have nontrivial structure of Pauli matrices. 
Instead, we should determine
the anticommutators,
$\big\{ \Phi^{(s),4}_{\frac{1}{2},\,\rho}, \,  \Phi^{(s),4}_{\frac{1}{2},\,\sigma}\big\} 
$ and  $\big\{ \Phi^{(s),1}_{\frac{1}{2},\,\rho}, \,  \Phi^{(s),4}_{\frac{1}{2},\,\sigma}\big\} $ first where the $2\times 2$ identity matrix arises in these
anticommutators. 
Based on these anticommutators, we can determine the above anticommutators
by keeping track of $SO(4)$ indices coming from ${\cal N}=4$ superspace
description. 

We present the first five commutators in this section and
the remaining ones will be summarized by Appendix $E$.
From the case-one (\ref{IdEvenEven})
with the footnote \ref{othercasesforcase1},
the following commutator is obtained
\begin{eqnarray}
&&\big[\Phi^{(s_1)}_{0,\,m},\,\Phi^{(s_2)}_{0,\,n}\big]=
\nonumber\\
&&
\sum_{r=1}^{[\frac{s_1+s_2-1}{2}]}
\,
\mi\,(-1)^r \,
\Bigg[
\frac{1}{2(s_1+s_2-2r)-1}
\bigg(
2 \nu(s_1+s_2-1)
\mathrm{BB}_{2r,\,+}^{s_1, s_2}(m,n;\mu)
\nonumber\\
&&
\hspace{4cm}
-((2s_1-1)(2s_2-1)+\nu^2) \mathrm{BB}_{2r,\,-}^{s_1, s_2}(m,n;\mu) \bigg) \Phi^{(s_1+s_2-2r)}_{0,\,m+n}
\nonumber\\
&&
-\frac{1}{(2s_1+2s_2-4r-5)(2s_1+2s_2-4r-1)}\bigg(
\nonumber\\
&&
(
(2s_1-1)(2s_2-1)(2s_1+2s_2-4r-1)+(1-4r)\nu^2
) 
\mathrm{BB}_{2r,\,+}^{s_1, s_2}(m,n;\mu)
\nonumber\\
&&
+
\nu(4s_2-4(s_1^2-s_1+s_1s_2+s_2^2)   -   8r  +
8(s_1+s_2)r+\nu^2-1)
\mathrm{BB}_{2r,\,-}^{s_1, s_2}(m,n;\mu)
 \bigg) 
\nonumber\\
&&
\times \tilde{\Phi}^{(s_1+s_2-2r-2)}_{2,\,m+n}
\Bigg]\,.
\label{00}
\end{eqnarray}
There are two different structure constants having the
same indices for fixed $r$.

The following commutator can be determined by using (\ref{IdEvenOdd-1})
and (\ref{IdEvenOdd})
\begin{eqnarray}
&&
\big[\Phi^{(s_1)}_{0,\,m},\,\Phi^{(s_2),i}_{\frac{1}{2},\,\rho}\big]=
\mi\,(2s_2-1)
\sum_{r=1}^{[\frac{s_1+s_2-1}{2}]}\,\Bigg[
(-1)^r \,
\frac{1}{2(s_1+s_2-2r)-1}
\label{0half}
  \\
&&
 \bigg(
\nu\,\mathrm{BF}_{2r,\,+}^{s_1, s_2+\frac{1}{2}}(m,\rho;\mu)
+\frac{(2s_1-1)(2s_1-2r-1)}{2r}\,\mathrm{BF}_{2r,\,-}^{s_1, s_2+\frac{1}{2}}(m,\rho;\mu)
 \bigg) \Phi^{(s_1+s_2-2r),i}_{\frac{1}{2},\,m+\rho}
\Bigg]
\nonumber\\
&&
-\mi\,(2s_2-1)
\sum_{r=1}^{[\frac{s_1+s_2}{2}]}\,
\Bigg[(-1)^r \,\frac{1}{2(s_1+s_2-2r)-1}
\nonumber\\
&&
\bigg(
(2s_1-1)\,\mathrm{BF}_{2r-1,\,+}^{s_1, s_2+\frac{1}{2}}(m,\rho;\mu)
+\frac{2\nu(s_1-r)}{(2r-1)}\,\mathrm{BF}_{2r-1,\,-}^{s_1, s_2+\frac{1}{2}}(m,\rho;\mu)
 \bigg) \tilde{\Phi}^{(s_1+s_2-2r),i}_{\frac{3}{2},\,m+\rho}
 \Bigg]
\,.
\nonu
\end{eqnarray}
There are four different structure constants where each two of them
have same lower indices for fixed $r$.

As done for (\ref{00}), we obtain the following commutator
\begin{eqnarray}
&&
\big[\Phi^{(s_1)}_{0,\,m},\,\Phi^{(s_2),ij}_{1,\,n}\big]=
-\sum_{r=1}^{[\frac{s_1+s_2}{2}]}\,
 \frac{\mi\,(-1)^r (2s_2-1)}{2(s_1+s_2-2r)-1}
\Bigg[
\nonumber\\
&&
\qquad 
\bigg(
(2s_1-1)\,\mathrm{BB}_{2r,\,-}^{s_1, s_2+1}(m,n;\mu)
-\nu\,\mathrm{BB}_{2r,\,+}^{s_1, s_2+1}(m,n;\mu)
 \bigg) \Phi^{(s_1+s_2-2r),ij}_{1,\,m+n}
\nonumber\\
&&
\quad \,
+\bigg(
(2s_1-1)\,\mathrm{BB}_{2r,\,+}^{s_1, s_2+1}(m,n;\mu)
-\nu\,\mathrm{BB}_{2r,\,-}^{s_1, s_2+1}(m,n;\mu)
 \bigg) \Phi^{(s_1+s_2-2r),ij}_{1,\,m+n}\Bigg]\,.
\label{01}
\end{eqnarray}
The two different structure constants have the
same indices for fixed $r$.

From the analysis of footnote \ref{othercasesforcase2} and (\ref{IdEvenOdd}),
the following commutator satisfies
\begin{eqnarray}
&&
\big[\Phi^{(s_1)}_{0,\,m},\,\tilde{\Phi}^{(s_2),i}_{\frac{3}{2},\,\rho}\big]=
\sum_{r=1}^{[\frac{s_1+s_2+1}{2}]}
\Bigg[\,
\mi\,(-1)^r \,(2s_2-1)
\frac{1}{2(s_1+s_2-2r)+3}
\label{03half}
  \\
&&
\bigg(
(2s_1-1)\,\mathrm{BF}_{2r-1,\,+}^{s_1, s_2+\frac{3}{2}}(m,\rho;\mu)
+\frac{2\nu(s_1-r)}{(2r-1)}\,\mathrm{BF}_{2r-1,\,-}^{s_1, s_2+\frac{3}{2}}(m,\rho;\mu)
 \bigg) \Phi^{(s_1+s_2-2r+2),i}_{\frac{1}{2},\,m+\rho}\Bigg]
\nonumber\\
&&
+
\sum_{r=1}^{[\frac{s_1+s_2}{2}]}\Bigg[
\,
\mi\,(-1)^r \,(2s_2-1)\frac{1}{2(s_1+s_2-2r)-1}
\nonumber\\
&&
\bigg(
\nu\,\mathrm{BF}_{2r,\,+}^{s_1, s_2+\frac{3}{2}}(m,\rho;\mu)
+\frac{(2s_1-1)(2s_1-2r-1)}{2r}\,\mathrm{BF}_{2r,\,-}^{s_1, s_2+\frac{3}{2}}(m,\rho;\mu)
 \bigg) \tilde{\Phi}^{(s_1+s_2-2r),i}_{\frac{3}{2},\,m+\rho}\Bigg]
\,.
\nonu
\end{eqnarray}
Two of the structure constants 
have the lower indices $(2r-1)$ and $2r$
respectively for fixed $r$.

Finally, we have the following commutator relation 
\begin{eqnarray}
&&\big[\Phi^{(s_1)}_{0,\,m},\,\tilde{\Phi}^{(s_2)}_{2,\,n}\big]=
\mi\,(2s_2-1)
\Bigg[
\sum_{r=1}^{[\frac{s_1+s_2+1}{2}]}
\,\frac{(-1)^r }{2(s_1+s_2-2r)+3}
\bigg(
\nonumber\\
&&
\nu\,\mathrm{BB}_{2r,\,-}^{s_1, s_2+2}(m,n;\mu)
-(2s_1-1)\, \mathrm{BB}_{2r,\,+}^{s_1, s_2+2}(m,n;\mu) \bigg) \Phi^{(s_1+s_2-2r+2)}_{0,\,m+n}
\nonumber\\
&&
+\sum_{r=1}^{[\frac{s_1+s_2}{2}]}
\,\frac{(-1)^r }{2(s_1+s_2-2r)+3}
\frac{1}{(2s_1+2s_2-4r-1)}\bigg(
2\nu(s_2-2r+2)\, \mathrm{BB}_{2r,\,+}^{s_1, s_2+2}(m,n;\mu)
\nonumber\\
&&
-((2s_1-1)(2s_1+2s_2-4r+3)-\nu^2)\,\mathrm{BB}_{2r,\,-}^{s_1, s_2+2}(m,n;\mu)
 \bigg)
 \tilde{\Phi}^{(s_1+s_2-2r)}_{2,\,m+n}
\Bigg]\,.
\label{02}
\end{eqnarray}
Again 
the two different structure constants have the
same indices for fixed $r$.

Therefore, the five commutators, (\ref{00}), (\ref{0half}), (\ref{01}),
(\ref{03half}) and (\ref{02}) with common higher spin-$s_1$
generator $\Phi^{(s_1)}_{0,\,m}$, where the $2\times$2 identity matrix
arises in the left hand side, are determined.
The remaining ten (anti)commutators are presented in Appendix $E$.

\section{Conclusions and outlook}

We have obtained the ${\cal N}=4$ higher spin algebra
$shs_2[\mu]$ which is given by (\ref{00})-(\ref{02}) and 
Appendices (\ref{halfhalf})-(\ref{22comm}) for generic $\mu$.
The structure constants are given in (\ref{3struct}).
As a subalgebra, the ${\cal N}=2$ higher spin algebra
$shs[\mu]$ arises.

So far, we have considered $2 \times 2$ matrix generalization
of $AdS_3$ Vasiliev higher spin theory. It is natural to 
ask about its $M \times M$ matrix generalization where $M \geq 3$.
According to \cite{GG1305}, the field contents are given by
$(2M^2-1)$ spin $1$ fields and $2M^2$ fields of spin $s =\frac{3}{2}, 2,
\frac{5}{2}, 3, \cdots$.
Although it is known that there is no supersymmetry,
it would be interesting to understand the higher spin algebra
$hs_M[\mu]$ and its nonlinear ${\cal W}_{\infty}^M[l,\mu]$ algebra
where $l$ is the level of $SU(M)$ currents \cite{CH1906}.

Furthermore, it is an open problem to 
observe any relations between this higher spin algebra $hs_M[\mu]$
and the rectangular $W$ algebra studied in
\cite{CH1812,CH1906,CHU}.
In the higher spin square \cite{GG1501,GG1512} or
in the two dimensional SYK model \cite{AP1812},
there exists an additional parameter, like as the above $M$,
in the theory.
It would be interesting to understand the role of this additional parameter
in the various different models.

Other type of matrix generalization of $AdS_3$
Vasiliev higher spin theory can be seen from the work of \cite{CHR1406}
and corresponding coset model is studied in \cite{AK1607} further.
It would be interesting to obtain the ${\cal N}=3$ higher spin algebra
explicitly.
Finally, we can apply the present method to
the ${\cal N}=4$ orthogonal coset model \cite{AKP1904,AP1410} and want to
obtain the corresponding ${\cal N}=4$ higher spin algebra \cite{EGR,CHU}.


\vspace{.7cm}

\centerline{\bf Acknowledgments}

MHK was supported by an appointment to the YST Program at the APCTP through the Science and Technology Promotion Fund and Lottery Fund of the Korean Government. 
MHK was also supported by the Korean Local Governments - Gyeongsangbuk-do Province and Pohang City.
This work of CA was supported by  the
National Research Foundation of Korea(NRF)  grant funded
by the Korea government(MSIT)(No. 2020R1F1A1066893).
CA acknowledges warm hospitality from 
the School of  Liberal Arts (and Institute of Convergence Fundamental
Studies), Seoul National University of Science and Technology.

\newpage

\appendix

\renewcommand{\theequation}{\Alph{section}\mbox{.}\arabic{equation}}

\section{Structure constants for vanishing $\mu$}

The structure constants in subsection $3.1$
are given by \cite{AKK}
\bea
p_{\mathrm{F},h}^{h_1h_2}(m,n)
&
=&\frac{1}{2(h+1)!}\,\phi^{h_1,h_2}_{h}(0,\textstyle{-\frac{1}{2}})
\,N^{h_1,h_2}_{h}(m,n),
\nonu\\
p_{\mathrm{B},h}^{h_1h_2}(m,n)
&
=&\frac{1}{2(h+1)!}\,\phi^{h_1,h_2}_{h}(0,0)
\,N^{h_1,h_2}_{h}(m,n),
\nonu\\
q_{\mathrm{F},h}^{h_1h_2}(m,r) 
&
=&\frac{(-1)^h}{4(h+2)!}\Big(
(h_1-1)\,\phi^{h_1,h_2+\frac{1}{2}}_{h+1}(0,0)
\nonu \\
& - & (h_1-h-3)\,\phi^{h_1,h_2+\frac{1}{2}}_{h+1}(0,\textstyle{-\frac{1}{2}})
\Big)
\,N^{h_1,h_2}_{h}(m,n),
\nonu\\
q_{\mathrm{B},h}^{h_1h_2}(m,r) 
&
=&\frac{-1}{4(h+2)!}\Big(
(h_1-h-2)\,\phi^{h_1,h_2+\frac{1}{2}}_{h+1}(0,0)
\nonu \\
& - & (h_1)\,\phi^{h_1,h_2+\frac{1}{2}}_{h+1}(0,\textstyle{-\frac{1}{2}})
\Big)
\,N^{h_1,h_2}_{h}(m,n),
\nonu\\
o_{\mathrm{F},h}^{h_1h_2}(r,s) 
&
=&\frac{4(-1)^h}{h!}\Big(
(h_1+h_2-1-h)\,\phi^{h_1+\frac{1}{2},h_2+\frac{1}{2}}_{h}(
\textstyle{\frac{1}{2}},\textstyle{-\frac{1}{4}})
\nonu \\
& - & (h_1+h_2-\frac{3}{2}-h)\,\phi^{h_1+\frac{1}{2},h_2+\frac{1}{2}}_{h+1}(
\textstyle{\frac{1}{2}},\textstyle{-\frac{1}{4}})
\Big)\,  N^{h_1,h_2}_{h-1}(m,n),
\nonu\\
o_{\mathrm{B},h}^{h_1h_2}(r,s) 
&
=&-\frac{4}{h!}\Big(
(h_1+h_2-2-h)\,\phi^{h_1+\frac{1}{2},h_2+\frac{1}{2}}_{h}(
\textstyle{\frac{1}{2}},\textstyle{-\frac{1}{4}})
\nonu \\
& - &
(h_1+h_2-\frac{3}{2}-h)\,\phi^{h_1+\frac{1}{2},h_2+\frac{1}{2}}_{h+1}(\textstyle{\frac{1}{2}},\textstyle{-\frac{1}{4}})
\Big)
 \,  N^{h_1,h_2}_{h-1}(m,n),
\label{6struct}
 \eea
 where the mode dependent function in Appendix (\ref{6struct})
and the symbol $\phi_h^{h_1,h_2}(x,y)$
are given by 
\bea
N^{h_1,h_2}_{h}(m,n)
&
=&
\sum_{l=0 }^{h+1}(-1)^l
\left(\begin{array}{c}
h+1 \\  l \\
\end{array}\right)
[h_1-1+m]_{h+1-l}[h_1-1-m]_l
\nonu \\
      & \times & [h_2-1+n]_l [h_2-1-n]_{h+1-l},
\nonu\\
\phi^{h_1,h_2}_{h}(x,y)
&
= &
{}_4 F_3
 \Bigg[
\begin{array}{c}
  -\frac{1}{2}-x-2y, \frac{3}{2}-x+2y, -\frac{h+1}{2}+x,
  -\frac{h}{2} +x \\
-h_1+\frac{3}{2},-h_2+\frac{3}{2},h_1+h_2-h-\frac{3}{2}
\end{array} ; 1
  \Bigg].
\label{Nphi}
 \eea
 In Appendix (\ref{Nphi}),
 we introduce the generalized hypergeometric function
as follows:
\bea
 {}_4 F_3
 \Bigg[
\begin{array}{c}
a_1, a_2, a_3, a_4 \\
b_1,b_2,b_3
\end{array} ; z
  \Bigg] 
=
\sum_{n=0 }^{\infty}
\frac{(a_1)_n (a_2)_n (a_3)_n (a_4)_n}
{(b_1)_n (b_2)_n (b_3)_n}
\frac{z^n}{n!}\,.
\label{4F3function}
\eea
The falling and rising Pochhammer symbols in Appendix (\ref{Nphi}) and
Appendix (\ref{4F3function})
$[a]_n$ and $(a)_n$ are defined as
$[a]_n \equiv a(a-1) \cdots (a-n+1)$ and
                $(a)_n \equiv a(a+1) \cdots (a+n-1)$.

\section{The (anti)commutators between the oscillators with vanishing
  $\mu$ in Odake basis}

We present here the (anti)commutators between the oscillators
with vanishing
$\mu$ in Odake basis where the structure constants
are given in terms of generalized hypergeometric functions.

$\bullet$ The case-one with two bosonic oscillators

We have 
\bea
&&\big[ 
  \underbrace{
\hat{y}_{(1}\ldots\,\hat{y}_{1}
}_{h_1-1+m} 
\underbrace{
\hat{y}_{2}\ldots\,\hat{y}_{2)}
}_{h_1-1-m},\,
 \underbrace{
\hat{y}_{(1}\ldots\,\hat{y}_{1}
}_{h_2-1+n} 
\underbrace{
\hat{y}_{2}\ldots\,\hat{y}_{2)}
}_{h_2-1-n}
\big]
=
\nonu \\
&& -\mi \, \sum_{r=1}^{[\frac{h_1+h_2-1}{2}]} \,(-1)^r \,\frac{1}{(2r-1)!}\,N_{2r}^{h_1, h_2}(m,n) 
\Bigg[
\,
\Bigg( {}
_4F_3\left[
\begin{array}{c|}
\frac{1}{2}  , \frac{1}{2} , \frac{-2r+1}{2}  ,\frac{-2r+2}{2}\\
\frac{3}{2}-h_1 ,  \frac{3}{2}-h_2 ,  h_1+h_2-2r+\frac{1}{2}
\end{array}   1\right]
\nonu \\
&& +\,_4F_3\left[
\begin{array}{c|}
\frac{3}{2}  , -\frac{1}{2}  , \frac{-2r+1}{2}  ,\frac{-2r+2}{2}\\
\frac{3}{2}-h_1 ,  \frac{3}{2}-h_2 ,  h_1+h_2-2r+\frac{1}{2}
\end{array}   1\right]
\Bigg)
\underbrace{
\hat{y}_{(1}\ldots\ldots\ldots\ldots\hat{y}_{1}
}_{h_1+h_2-2r-1+m+n} \,
\underbrace{
\hat{y}_{2}\ldots\ldots\ldots\ldots\hat{y}_{2)}
}_{h_1+h_2-2r-1-m-n}
\nonumber\\
&&
-
\Bigg( {}
_4F_3\left[
\begin{array}{c|}
\frac{1}{2}  , \frac{1}{2}  , \frac{-2r+1}{2}  ,\frac{-2r+2}{2}\\
\frac{3}{2}-h_1 ,  \frac{3}{2}-h_2 ,  h_1+h_2-2r+\frac{1}{2}
\end{array}   1\right]
\nonu \\
&& -\,_4F_3\left[
\begin{array}{c|}
\frac{3}{2}  , -\frac{1}{2}  , \frac{-2r+1}{2}  ,\frac{-2r+2}{2}\\
\frac{3}{2}-h_1 ,  \frac{3}{2}-h_2 ,  h_1+h_2-2r+\frac{1}{2}
\end{array}   1\right]
\Bigg)
\nonu \\
&& \times \underbrace{
\hat{y}_{(1}\ldots\ldots\ldots\ldots\hat{y}_{1}
}_{h_1+h_2-2r-1+m+n} \,
\underbrace{
\hat{y}_{2}\ldots\ldots\ldots\ldots\hat{y}_{2)}
}_{h_1+h_2-2r-1-m-n}K
\Bigg]
\,.
\label{IdEvenEven4F3Odake}
\eea
The
above commutation relation
Appendix (\ref{IdEvenEven4F3Odake})
  is the same as the one in (\ref{IdEvenEvenOdake}).
Moreover, 
this commutation relation Appendix
(\ref{IdEvenEven4F3Odake}) can be obtained
by taking $\mu$ to zero  in Appendix (\ref{IdEvenEven4F3}).

$\bullet$ The case-two with one bosonic and one fermionic oscillators

By substituting the corresponding structure constants, we obtain 
\bea
&&\big[ 
  \underbrace{
\hat{y}_{(1}\ldots\,\hat{y}_{1}
}_{h_1-1+m} 
\underbrace{
\hat{y}_{2}\ldots\,\hat{y}_{2)}
}_{h_1-1-m},\,
 \underbrace{
\hat{y}_{(1}\ldots\,\hat{y}_{1}
}_{h_2-\frac{1}{2}+\rho} 
\underbrace{
\hat{y}_{2}\ldots\,\hat{y}_{2)}
}_{h_2-\frac{1}{2}-\rho}
\big]
=
\nonu \\
&& - \sum_{r=1}^{[\frac{h_1+h_2-1}{2}]}
\Bigg[ \,
\frac{\mi\,(-1)^r}{(2r)!}
N_{2r}^{h_1, h_2+\frac{1}{2}}(m,\rho) 
\Bigg( (2r+1)
_4F_3\left[
\begin{array}{c|}
\frac{1}{2}  , \frac{1}{2}  , \frac{-2r}{2}  ,\frac{-2r+1}{2}\\
\frac{3}{2}-h_1 ,  \frac{1}{2}-h_2 ,  h_1+h_2-2r+\frac{1}{2}
\end{array}   1 \right]
\nonu\\
&&
+(2r-1)\,
_4F_3\left[
\begin{array}{c|}
\frac{3}{2} \ , -\frac{1}{2} \ , \frac{-2r}{2} \ ,\frac{-2r+1}{2}\\
\frac{3}{2}-h_1\ ,  \frac{1}{2}-h_2\ ,  h_1+h_2-2r+\frac{1}{2}
\end{array}  \ 1\right]
\Bigg)
\nonu\\
&&
\times \underbrace{
\hat{y}_{(1}\ldots\ldots\ldots\ldots\hat{y}_{1}
}_{h_1+h_2-2r-\frac{1}{2}+m+\rho} \,
\underbrace{
\hat{y}_{2}\ldots\ldots\ldots\ldots\hat{y}_{2)}
}_{h_1+h_2-2r-\frac{1}{2}-m-\rho} \Bigg]
\nonumber\\
&&
+\sum_{r=1}^{[\frac{h_1+h_2}{2}]}
\Bigg[ (-1)^r\,
\frac{2(h_1-r)}{(2r-1)!}\,
N_{2r-1}^{h_1, h_2+\frac{1}{2}}(m,\rho) 
\times
\Bigg({}
_4F_3\left[
\begin{array}{c|}
\frac{1}{2} \ , \frac{1}{2} \ , \frac{-2r+1}{2} \ ,\frac{-2r+2}{2}\\
\frac{3}{2}-h_1\ ,  \frac{1}{2}-h_2\ ,  h_1+h_2-2r+\frac{3}{2}
\end{array}  \ 1\right]
\nonu \\
&& -\,_4F_3\left[
\begin{array}{c|}
\frac{3}{2} \ , -\frac{1}{2} \ , \frac{-2r+1}{2} \ ,\frac{-2r+2}{2}\\
\frac{3}{2}-h_1\ ,  \frac{1}{2}-h_2\ ,  h_1+h_2-2r+\frac{3}{2}
\end{array}  \ 1\right]
\Bigg)
\nonu\\
&&
\times \underbrace{
\hat{y}_{(1}\ldots\ldots\ldots\ldots\hat{y}_{1}
}_{h_1+h_2-2r+\frac{1}{2}+m+\rho} \,
\underbrace{
\hat{y}_{2}\ldots\ldots\ldots\ldots\hat{y}_{2)}
}_{h_1+h_2-2r+\frac{1}{2}-m-\rho}K
\Bigg].
\label{case2other}
\eea
This is equivalent to (\ref{fourthcomm})
and we observe that this is generalized to Appendix
(\ref{IdEvenOdd4F3-1}).

Similarly, we have
\bea
&&\big[ 
  \underbrace{
\hat{y}_{(1}\ldots\,\hat{y}_{1}
}_{h_1-1+m} 
\underbrace{
\hat{y}_{2}\ldots\,\hat{y}_{2)}
}_{h_1-1-m} K ,\,
 \underbrace{
\hat{y}_{(1}\ldots\,\hat{y}_{1}
}_{h_2-\frac{1}{2}+\rho} 
\underbrace{
\hat{y}_{2}\ldots\,\hat{y}_{2)}
}_{h_2-\frac{1}{2}-\rho}
\big]
=
\nonu \\
&& \sum_{r=1}^{[\frac{h_1+h_2}{2}]}
\Bigg[
\frac{(-1)^r }{(2r-1)!}
N_{2r-1}^{h_1, h_2+\frac{1}{2}}(m,\rho) 
\Bigg( (2r)
_4F_3\left[
\begin{array}{c|}
\frac{1}{2}  , \frac{1}{2}  , \frac{-2r+1}{2}  ,\frac{-2r+2}{2}\\
\frac{3}{2}-h_1 ,  \frac{1}{2}-h_2 ,  h_1+h_2-2r+\frac{3}{2}
\end{array}   1\right]
\nonu\\
&&
+(2r-2)\,
_4F_3\left[
\begin{array}{c|}
\frac{3}{2} \ , -\frac{1}{2} \ , \frac{-2r+1}{2} \ ,\frac{-2r+2}{2}\\
\frac{3}{2}-h_1\ ,  \frac{1}{2}-h_2\ ,  h_1+h_2-2r+\frac{3}{2}
\end{array}  \ 1\right]
\Bigg)
\nonu\\
&&
\times \underbrace{
\hat{y}_{(1}\ldots\ldots\ldots\ldots\hat{y}_{1}
}_{h_1+h_2-2r+\frac{1}{2}+m+\rho} \,
\underbrace{
\hat{y}_{2}\ldots\ldots\ldots\ldots\hat{y}_{2)}
}_{h_1+h_2-2r+\frac{1}{2}-m-\rho}K \,\Bigg]
\nonumber\\
&&
-\mi\sum_{r=1}^{[\frac{h_1+h_2-1}{2}]}\Bigg[(-1)^r 
\frac{2(h_1-r-\frac{1}{2})}{(2r)!}
\,
N_{2r}^{h_1, h_2+\frac{1}{2}}(m,\rho) 
\,
\Bigg( {}
_4F_3\left[
\begin{array}{c|}
\frac{1}{2} \ , \frac{1}{2} \ , \frac{-2r}{2} \ ,\frac{-2r+1}{2}\\
\frac{3}{2}-h_1\ ,  \frac{1}{2}-h_2\ ,  h_1+h_2-2r+\frac{1}{2}
\end{array}  \ 1\right]
\nonu \\
&& -\,_4F_3\left[
\begin{array}{c|}
\frac{3}{2} \ , -\frac{1}{2} \ , \frac{-2r}{2} \ ,\frac{-2r+1}{2}\\
\frac{3}{2}-h_1\ ,  \frac{1}{2}-h_2\ ,  h_1+h_2-2r+\frac{1}{2}
\end{array}  \ 1\right]
\Bigg)
\nonu \\
&& \underbrace{
\hat{y}_{(1}\ldots\ldots\ldots\ldots\hat{y}_{1}
}_{h_1+h_2-2r-\frac{1}{2}+m+\rho} \,
\underbrace{
\hat{y}_{2}\ldots\ldots\ldots\ldots\hat{y}_{2)}
}_{h_1+h_2-2r-\frac{1}{2}-m-\rho}
\Bigg],
\label{IdEvenOdd4F3Odake}
\eea
which is equivalent to 
(\ref{IdEvenOddOdake}) and
is generalized to Appendix (\ref{IdEvenOdd4F3}). 

$\bullet$ The case-three with two fermionic oscillators

The next case can be written as
\bea
&&
\big\{ 
  \underbrace{
\hat{y}_{(1}\ldots\,\hat{y}_{1}
}_{h_1-\frac{1}{2}+\rho} 
\underbrace{
\hat{y}_{2}\ldots\,\hat{y}_{2)}
}_{h_1-\frac{1}{2}-\rho},\,
 \underbrace{
\hat{y}_{(1}\ldots\,\hat{y}_{1}
}_{h_2-\frac{1}{2}+\omega} 
\underbrace{
\hat{y}_{2}\ldots\,\hat{y}_{2)}
}_{h_2-\frac{1}{2}-\omega}
\big\}
=
\nonu \\
&& -2 \sum_{r=1}^{[\frac{h_1+h_2+1}{2}]}\,(-1)^r\,\frac{1}{(2r-2)!}\,
N_{2r-1}^{h_1+\frac{1}{2}, h_2+\frac{1}{2}}(\rho,\omega) \,
\Bigg[
\nonu\\
&&
\times
\Bigg({}
_4F_3\left[
\begin{array}{c|}
\frac{1}{2}  , -\frac{1}{2}  , \frac{-2r+2}{2}  ,\frac{-2r+3}{2}\\
\frac{1}{2}-h_1 ,  \frac{1}{2}-h_2 ,  h_1+h_2-2r+\frac{5}{2}
\end{array}   1\right]
\Bigg)
\,  \underbrace{
\hat{y}_{(1}\ldots\ldots\ldots.\hat{y}_{1}
}_{h_1+h_2-2r+1+\rho+\omega} \,
\underbrace{
\hat{y}_{2}\ldots\ldots\ldots.\hat{y}_{2)}
}_{h_1+h_2-2r+1-\rho-\omega}
\nonumber
\\
&&
+(2h_1\!+\!2h_2\!-\!4r\!+3)
\,
\Bigg({}
_4F_3\left[
\begin{array}{c|}
\frac{1}{2}  , -\frac{1}{2}  , \frac{-2r+1}{2}  ,\frac{-2r+2}{2}\\
\frac{1}{2}-h_1 ,  \frac{1}{2}-h_2 ,  h_1+h_2-2r+\frac{3}{2}
\end{array}   1\right]
\label{equivIdOddOdd4F3-1}
\\
&& -\,_4F_3\left[
\begin{array}{c|}
\frac{1}{2}  , -\frac{1}{2}  , \frac{-2r+2}{2} \ ,\frac{-2r+3}{2}\\
\frac{1}{2}-h_1 ,  \frac{1}{2}-h_2 ,  h_1+h_2-2r+\frac{5}{2}
\end{array}   1\right]
\Bigg)
  \underbrace{
\hat{y}_{(1}\ldots\ldots\ldots.\hat{y}_{1}
}_{h_1+h_2-2r+1+\rho+\omega} \,
\underbrace{
\hat{y}_{2}\ldots\ldots\ldots.\hat{y}_{2)}
}_{h_1+h_2-2r+1-\rho-\omega}K
\Bigg],
\nonu
\eea
which is equivalent to
(\ref{casethreeone}) and we observe that this is generalized to
Appendix (\ref{IdOddOdd4F3-1}). 
Note that by substituting the structure constants appearing in Appendix
$A$ into the ones in (\ref{casethreeone}), the second $\phi$ in $o_F$
and $o_B$ has different sign leading to the one generalized hypergeometric
function (not two).

Finally, the following relation holds
\bea
&&
\big\{ 
  \underbrace{
\hat{y}_{(1}\ldots\,\hat{y}_{1}
}_{h_1-\frac{1}{2}+\rho} 
\underbrace{
\hat{y}_{2}\ldots\,\hat{y}_{2)}
}_{h_1-\frac{1}{2}-\rho},\,
 \underbrace{
\hat{y}_{(1}\ldots\,\hat{y}_{1}
}_{h_2-\frac{1}{2}+\omega} 
\underbrace{
\hat{y}_{2}\ldots\,\hat{y}_{2)}
}_{h_2-\frac{1}{2}-\omega}K
\big\}
=
\nonu \\
&& -2\,\mi\, \sum_{r=1}^{[\frac{h_1+h_2}{2}]}\,(-1)^r\,
\frac{1}{(2r-1)!}\,
N_{2r}^{h_1+\frac{1}{2}, h_2+\frac{1}{2}}(\rho,\omega) \,
\Bigg[
\nonu\\
&&
\times
\Bigg({}
_4F_3\left[
\begin{array}{c|}
\frac{1}{2}  , -\frac{1}{2}  , \frac{-2r+1}{2}  ,\frac{-2r+2}{2}\\
\frac{1}{2}-h_1 ,  \frac{1}{2}-h_2 ,  h_1+h_2-2r+\frac{3}{2}
\end{array}   1\right]
\Bigg)
\,  \underbrace{
\hat{y}_{(1}\ldots\ldots\ldots.\hat{y}_{1}
}_{h_1+h_2-2r+\rho+\omega}\,
\underbrace{
\hat{y}_{2}\ldots\ldots\ldots.\hat{y}_{2)}
}_{h_1+h_2-2r-\rho-\omega}K
\nonumber\\
&&
+(2h_1+2h_2-4r+1)
\,
\Bigg({}
_4F_3\left[
\begin{array}{c|}
\frac{1}{2}  , -\frac{1}{2}  , \frac{-2r}{2}  ,\frac{-2r+1}{2}\\
\frac{1}{2}-h_1 ,  \frac{1}{2}-h_2 ,  h_1+h_2-2r+\frac{1}{2}
\end{array}   1\right]
\label{IdOddOdd4F3Odake}
\\
&& -\,_4F_3\left[
\begin{array}{c|}
\frac{1}{2}  , -\frac{1}{2}  , \frac{-2r+1}{2}  ,\frac{-2r+2}{2}\\
\frac{1}{2}-h_1 ,  \frac{1}{2}-h_2 ,  h_1+h_2-2r+\frac{3}{2}
\end{array}   1\right]
\Bigg)
\underbrace{
\hat{y}_{(1}\ldots\ldots\ldots.\hat{y}_{1}
}_{h_1+h_2-2r+\rho+\omega} \,
\underbrace{
\hat{y}_{2}\ldots\ldots\ldots.\hat{y}_{2)}
}_{h_1+h_2-2r-\rho-\omega}
\Bigg],
\nonu
\eea
which is equivalent to
(\ref{IdOddOddOdake}) and in Appendix (\ref{IdOddOdd4F3}) its
generalization is given. 


\section{$\mu$ -dependence in the (anti)commutators of the
  oscillators up to a total spin $11$ ($h_1+h_2 \leq 11$)}


\subsection{ The case-two with one bosonic and one fermionic oscillators}

\subsubsection{No operator $K$ dependence in the oscillators}

We continue to analyze the procedure in section $3$.
As done in (\ref{first}),
from (\ref{IdEvenEvenOdake}), we introduce the $\mu$ dependent
structure constant as follows:
\bea
&&\big[ 
  \underbrace{
\hat{y}_{(1}\ldots\,\hat{y}_{1}
}_{h_1-1+m} 
\underbrace{
\hat{y}_{2}\ldots\,\hat{y}_{2)}
}_{h_1-1-m},\,
 \underbrace{
\hat{y}_{(1}\ldots\,\hat{y}_{1}
}_{h_2-\frac{1}{2}+\rho} 
\underbrace{
\hat{y}_{2}\ldots\,\hat{y}_{2)}
}_{h_2-\frac{1}{2}-\rho}
\big]
=
\nonu \\
&& \sum_{r=1}^{[\frac{h_1+h_2-1}{2}]}\!\!
(-1)^r 
\Bigg[
 N^{h_1\,  h_2+\frac{1}{2}}_{2r}(m,\rho)\frac{(-\mi)}{(2r-1)!}\,
a_{2r}^{h_1, h_2+\frac{1}{2}}(\nu)
\,  \underbrace{
\hat{y}_{(1}\ldots\ldots\ldots\ldots\hat{y}_{1}
}_{h_1+h_2-2r-\frac{1}{2}+m+\rho} \,
\underbrace{
\hat{y}_{2}\ldots\ldots\ldots\ldots\hat{y}_{2)}
}_{h_1+h_2-2r-\frac{1}{2}-m-\rho} \Bigg]
\nonumber\\
&&
+
\sum_{r=1}^{[\frac{h_1+h_2}{2}]}\!\!
(-1)^r 
\Bigg[ N^{h_1\,  h_2+\frac{1}{2}}_{2r-1}(m,\rho)\frac{2(h_1-r)}{(2r-1)!}\,
b_{2r-1}^{h_1, h_2+\frac{1}{2}}(\nu)
\,  \underbrace{
\hat{y}_{(1}\ldots\ldots\ldots\ldots\hat{y}_{1}
}_{h_1+h_2-2r+\frac{1}{2}+m+\rho} \,
\underbrace{
\hat{y}_{2}\ldots\ldots\ldots\ldots\hat{y}_{2)}
}_{h_1+h_2-2r+\frac{1}{2}-m-\rho}K
\Bigg].
\nonu\\
\label{casetwoapp}
\eea

By starting with the spins $(h_1,h_2)=(2,1)$ and varying
the spins,
we can compute the commutators for several cases.
Then it turns out that
the structure constants associated with the oscillators (and without the
operator $K$ in Appendix (\ref{casetwoapp}))
can be determined as follows:
\bea
&&
a^{2,\frac{3}{2}}_{2}=2\,, \qquad
a^{2,\frac{5}{2}}_{2}=2\,, \qquad
a^{2,\frac{7}{2}}_{2}=2\,, \qquad
a^{2,\frac{9}{2}}_{2}=2\,, \qquad
a^{2,\frac{11}{2}}_{2}=2\,, \qquad
a^{2,\frac{13}{2}}_{2}=2\,, 
\nonu\\
&&
a^{2,\frac{15}{2}}_{2}=2\,, \qquad
a^{2,\frac{17}{2}}_{2}=2\,, \qquad
a^{3,\frac{3}{2}}_{2}=2\,, \qquad
a^{3,\frac{5}{2}}_{2}=2\,, 
a^{3,\frac{5}{2}}_{4}=-\frac{2}{9} (-3+\nu) (3+\nu)\,, 
\nonu\\
&&
a^{3,\frac{7}{2}}_{2}=2\,, \qquad
a^{3,\frac{7}{2}}_{4}=-\frac{2}{25} (-5+\nu) (5+\nu)\,, \qquad
a^{3,\frac{9}{2}}_{2}=2\,, 
\nonu\\
&&
a^{3,\frac{9}{2}}_{4}=-\frac{2}{49} (-7+\nu) (7+\nu)\,, \qquad
a^{3,\frac{11}{2}}_{2}=2\,, \qquad
a^{3,\frac{11}{2}}_{4}=-\frac{2}{81} (-9+\nu) (9+\nu)\,, 
\nonu\\
&&
a^{3,\frac{13}{2}}_{2}=2\,,  \qquad
a^{3,\frac{13}{2}}_{4}=-\frac{2}{121} (-11+\nu) (11+\nu)\,, \qquad
a^{3,\frac{15}{2}}_{2}=2\,, 
\nonu\\
&&
a^{3,\frac{15}{2}}_{4}=-\frac{2}{169} (-13+\nu) (13+\nu)\,, \qquad
a^{4,\frac{3}{2}}_{2}=2\,, \qquad
a^{4,\frac{5}{2}}_{2}=2\,, 
\nonu\\
&&
a^{4,\frac{5}{2}}_{4}=-\frac{2}{25} (-5+\nu) (5+\nu)\,, \qquad
a^{4,\frac{7}{2}}_{2}=2\,,  \qquad
a^{4,\frac{7}{2}}_{4}=-\frac{2}{175} (-175+3 \nu^2)\,, 
\nonu\\
&&
a^{4,\frac{7}{2}}_{6}=\frac{2}{225} (-5+\nu) (-3+\nu) (3+\nu) (5+\nu)\,, 
\nonu\\
&&
a^{4,\frac{9}{2}}_{2}=2\,,  
a^{4,\frac{9}{2}}_{4}=-\frac{2}{105} (-105+\nu^2)\,, 
a^{4,\frac{9}{2}}_{6}=\frac{2}{1225} (-7+\nu) (-5+\nu) (5+\nu) (7+\nu)\,, 
\nonu\\
&&
a^{4,\frac{11}{2}}_{2}=2\,,\qquad
a^{4,\frac{11}{2}}_{4}=-\frac{2}{165} (-165+\nu^2)\,, 
\nonu\\
&&
a^{4,\frac{11}{2}}_{6}=\frac{2}{3969} (-9+\nu) (-7+\nu) (7+\nu) (9+\nu)\,, \qquad
a^{4,\frac{13}{2}}_{2}=2\,, 
\nonu \\
&&
a^{4,\frac{13}{2}}_{4}=-\frac{2}{715} (-715+3 \nu^2)\,, \qquad
a^{4,\frac{13}{2}}_{6}=\frac{2}{9801} (-11+\nu) (-9+\nu) (9+\nu) (11+\nu)\,, 
\nonu\\
&&
a^{5,\frac{3}{2}}_{2}=2\,, \qquad
a^{5,\frac{5}{2}}_{2}=2\,, \qquad
a^{5,\frac{5}{2}}_{4}=-\frac{2}{49} (-7+\nu) (7+\nu)\,, 
\nonu\\
&&
a^{5,\frac{7}{2}}_{2}=2\,, 
a^{5,\frac{7}{2}}_{4}=-\frac{2}{105} (-105+\nu^2)\,, 
a^{5,\frac{7}{2}}_{6}=\frac{2}{1225} (-7+\nu) (-5+\nu) (5+\nu) (7+\nu)\,, 
\nonu\\
&&
a^{5,\frac{9}{2}}_{2}=2\,, \qquad
a^{5,\frac{9}{2}}_{4}=-\frac{2}{539} (-539+3 \nu^2)\,, 
a^{5,\frac{9}{2}}_{6}=\frac{2}{5145} (-7+\nu) (7+\nu) (-105+\nu^2)\,, 
\nonu\\
&&
a^{5,\frac{9}{2}}_{8}=-\frac{2}{11025} (-7+\nu) (-5+\nu) (-3+\nu) (3+\nu) (5+\nu) (7+\nu)\,, \qquad
a^{5,\frac{11}{2}}_{2}=2\,, 
\nonu\\
&&
a^{5,\frac{11}{2}}_{4}=-\frac{2}{273} (-273+\nu^2)\,, \qquad
a^{5,\frac{11}{2}}_{6}=\frac{2}{43659} (-9+\nu) (9+\nu) (-539+3 \nu^2)\,, 
\nonu\\
&&
a^{5,\frac{11}{2}}_{8}=-\frac{2}{99225} (-9+\nu) (-7+\nu) (-5+\nu) (5+\nu) (7+\nu) (9+\nu)\,, \qquad
a^{6,\frac{3}{2}}_{2}=2\,, 
\nonu\\
&&
a^{6,\frac{5}{2}}_{2}=2\,, \qquad
a^{6,\frac{5}{2}}_{4}=-\frac{2}{81} (-9+\nu) (9+\nu)\,, \qquad
a^{6,\frac{7}{2}}_{2}=2\,, 
a^{6,\frac{7}{2}}_{4}=-\frac{2}{165} (-165+\nu^2)\,, 
\nonu\\
&&
a^{6,\frac{7}{2}}_{6}=\frac{2}{3969} (-9+\nu) (-7+\nu) (7+\nu) (9+\nu)\,, 
a^{6,\frac{9}{2}}_{2}=2\,, 
a^{6,\frac{9}{2}}_{4}=-\frac{2}{273} (-273+\nu^2)\,, 
\nonu\\
&&
a^{6,\frac{9}{2}}_{6}=\frac{2}{43659} (-9+\nu) (9+\nu) (-539+3 \nu^2)\,, 
\nonu\\
&&
a^{6,\frac{9}{2}}_{8}=-\frac{2}{99225} (-9+\nu) (-7+\nu) (-5+\nu) (5+\nu) (7+\nu) (9+\nu)\,, \qquad
a^{7,\frac{3}{2}}_{2}=2\,, 
\nonu\\
&&
a^{7,\frac{5}{2}}_{2}=2\,, \qquad
a^{7,\frac{5}{2}}_{4}=-\frac{2}{121} (-11+\nu) (11+\nu)\,, \qquad
a^{7,\frac{7}{2}}_{2}=2\,, 
\nonu\\
&&
a^{7,\frac{7}{2}}_{4}=-\frac{2}{715} (-715+3 \nu^2)\,, \qquad
a^{7,\frac{7}{2}}_{6}=\frac{2}{9801} (-11+\nu) (-9+\nu) (9+\nu) (11+\nu)\,, 
\nonu\\
&&
a^{8,\frac{3}{2}}_{2}=2\,, \qquad
a^{8,\frac{5}{2}}_{2}=2\,, \qquad
a^{8,\frac{5}{2}}_{4}=-\frac{2}{169} (-13+\nu) (13+\nu)\,, \qquad
a^{9,\frac{3}{2}}_{2}=2\,.
\label{aval}
\eea
Compared to (\ref{avalue}),
the behavior of this coefficient looks different
because there exist some factorized coefficients where each factor
has only linear $\nu$ dependent term. 
For example, for fixed $h_1=3$ and $2r=4$ and different values for $h_2$,
we observe that there are factors
$(-3+\nu)(3+\nu), \cdots, (-13+\nu)(13+\nu)$.
For  fixed $h_1=4$ and $2r=6$ and different values for $h_2$,
there are factors
$(-5+\nu)(-3+\nu)(3+\nu)(5+\nu), \cdots, (-11+\nu)(-9+\nu)(9+\nu)(11+\nu)$.
Furthermore, for  fixed $h_1=5$ and $2r=8$, we have
two different sextic terms.
Then the question is how we can write down the linear combination of
two generalized hypergeometric functions leading to the above factorized
forms. From the experience of (\ref{avalue}), we can imagine
that the odd terms in $\nu$ of each generalized hypergeometric function
should be removed after adding them.  
That is, the linear, the third order term and the fifth order term
in $\nu$  of each generalized hypergeometric function
for the above sextic terms are removed.
Therefore we should find the appropriate generalized hypergeometric
functions which satisfy all these requirements \footnote{We have
  $\phi_5^{3,3}(\mu,2)= \frac{1}{9}(9-2\nu-\nu^2)$ corresponding to
  the coefficient $a_4^{3,\frac{5}{2}}$ and its expression by changing
  $\nu$ into $-\nu$. We can check that  $\phi_5^{3,3}(\mu,2)$ looks
  similar to  $\phi_4^{3,3}(\mu,1)$ in the sense that
  the four upper elements are the same while the last lower element
  is different from each other.
  This implies that the $\nu$ dependence is the same
  (the number of terms in the generalized hypergemetric function
  is the same) but their
  overall coefficients of each term are different.
  Let us emphasize that
  there are some shifts in the $r$ and $h_2$ in (\ref{3struct}).
  In this case, we need to substract $\frac{1}{2}$ from $h_2=3$
  and $1$ from $r=5$ in the $\phi_5^{3,3}$ in order to
  obtain the indices of the coefficient $a_4^{3,\frac{5}{2}}$.
  Then we can add these two and obtain the
  above quadratic term or  the coefficient $a_4^{3,\frac{5}{2}}$.
  Moreover we have $
  \phi_7^{4,4}(\mu,2)= \frac{1}{225}(225-76\nu-34\nu^2+4\nu^3+\nu^4)$
 corresponding to
 the coefficient $a_6^{4,\frac{7}{2}}$ and by similar analysis
 this will lead to the above quartic term where the odd $\nu$ terms
 are gone.
 For the
 $\phi_9^{5,5}(\mu,2)= \frac{1}{11025}(11025-4542\nu-1891\nu^2
  + 372\nu^3+83\nu^4-6\nu^5-\nu^6)$  corresponding to
  the coefficient $a_8^{5,\frac{9}{2}}$, we can obtain the sextic
  term by adding it to $\phi_9^{5,5}(1-\mu,2)$.}.

Similarly, 
the structure constants associated with the oscillators (and with the
operator $K$ in Appendix (\ref{casetwoapp}))
can be determined as follows:
\bea
&&
b^{2, \frac {7} {2}} _ {3} = -\frac{12 \nu}{25}\, , \qquad
b^{2, \frac {9} {2}} _ {3} = -\frac{12 \nu}{49}\, , \qquad
b^{2, \frac {11} {2}} _ {3} = -\frac{4 \nu}{27}\, , \qquad
b^{2, \frac {13} {2}} _ {3} = -\frac{12 \nu}{121}\, , 
b^{2, \frac {15} {2}} _ {3} = -\frac{12 \nu}{169}\, , 
\nonu\\
&&
b^{2, \frac {17} {2}} _ {3} = -\frac{4 \nu}{75}\, , \qquad
b^{3, \frac {5} {2}} _ {3} = -\frac{4 \nu}{15}\, , \qquad
b^{3, \frac {7} {2}} _ {3} = -\frac{4 \nu}{35}\, , \qquad
b^{3, \frac {9} {2}} _ {3} = -\frac{4 \nu}{63}\, , 
\nonu\\
&&
b^{3, \frac {9} {2}} _ {5} = \frac{8}{735} \nu (-47 + 3 \nu^2)\, , \qquad
b^{3, \frac {11} {2}} _ {3} = -\frac{4 \nu}{99}\, , \qquad
b^{3, \frac {11} {2}} _ {5} = \frac{40}{ 3969} (-5 + \nu) \nu (5 + \nu)\, , 
\nonu\\
&&
b^{3, \frac {13} {2}} _ {3} = -\frac{4 \nu}{143}\, , \qquad
b^{3, \frac {13} {2}} _ {5} = \frac{40}{9801} \nu (-37 + \nu^2)\, , \qquad
b^{3, \frac {15} {2}} _ {3} = -\frac{4 \nu}{195}\, , 
\nonu\\
&&
b^{3, \frac {15} {2}} _ {5} = \frac{40}{61347} \nu (-155 + 3 \nu^2)\, , \qquad
b^{4, \frac {5} {2}} _ {3} = -\frac{4 \nu}{35}\, , \qquad
b^{4, \frac {7} {2}} _ {3} = -\frac{4 \nu}{75}\, , 
\nonu\\
&&
b^{4, \frac {7} {2}} _ {5} = \frac{ 8}{525} (-5 + \nu) \nu (5 + \nu)\, , \qquad
b^{4, \frac {9} {2}} _ {3} = -\frac{12 \nu}{385}\, , \qquad
b^{4, \frac {9} {2}} _ {5} = \frac{8}{ 2205} (-7 + \nu) \nu (7 + \nu)\, , 
\nonu\\
&&
b^{4, \frac {11} {2}} _ {3} = -\frac{4 \nu}{195}\, , \qquad
b^{4, \frac {11} {2}} _ {5} = \frac{8}{6237} (-9 + \nu) \nu (9 + \nu)\, , 
b^{4, \frac {11} {2}} _ {7} = - \frac{4}{  4725} (-5 + \nu)^2 \nu (5 + \nu)^2\, , 
\nonu\\
&&
b^{4, \frac {13} {2}} _ {3} = -\frac{4 \nu}{275}\, , \qquad
b^{4, \frac {13} {2}} _ {5} = \frac{8}{14157} (-11 + \nu) \nu (11 + \nu)\, , 
\nonu\\
&&
b^{4, \frac {13} {2}} _ {7} = -\frac{4}{343035} \nu (22751 - 1070 \nu^2 +   15 \nu^4)\, , \qquad
b^{5, \frac {5} {2}} _ {3} = -\frac{4 \nu}{63}\, , \qquad
b^{5, \frac {7} {2}} _ {3} = -\frac{12 \nu}{385}\, , 
\nonu\\
&&
b^{5, \frac {7} {2}} _ {5} = \frac{8}{2205} (-7 + \nu) \nu (7 + \nu))\, , \qquad
b^{5, \frac {9} {2}} _ {3} = - \frac{12 \nu}{637}\, , \qquad
b^{5, \frac {9} {2}} _ {5} = \frac{8}{24255} \nu (-287 + 3 \nu^2)\, , 
\nonu\\
&&
b^{5, \frac {9} {2}} _ {7} = -\frac{4}{11025} (-7 + \nu) (-5 + \nu) \nu (5 + \nu) (7 + \nu)\, , \qquad
b^{5, \frac {11} {2}} _ {3} = -\frac{4 \nu}{315}\, ,
\nonu\\
&&
b^{5, \frac {11} {2}} _ {5} =\frac{8}{63063} \nu (-467 + 3 \nu^2)\, , \qquad
b^{5, \frac {11} {2}} _ {7} = -\frac{4}{72765} (-9 + \nu) (-7 + \nu) \nu (7 + \nu) (9 + \nu)\, , 
\nonu\\
&&
b^{6, \frac {5} {2}} _ {3} = -\frac{4 \nu}{99}\, , \qquad
b^{6, \frac {7} {2}} _ {3} = -\frac{4 \nu}{195}\, , \qquad
b^{6, \frac {7} {2}} _ {5} = \frac{8}{6237} (-9 + \nu) \nu (9 + \nu)\, , \qquad
b^{6, \frac {9} {2}} _ {3} = -\frac{4 \nu}{315}\, , 
\nonu\\
&&
b^{6, \frac {9} {2}} _ {5} =\frac{8}{63063} \nu (-467 + 3 \nu^2)\, , \qquad
b^{6, \frac {9} {2}} _ {7} = -\frac{4}{72765} (-9 + \nu) (-7 + \nu) \nu (7 + \nu) (9 + \nu)\, ,
\nonu\\
&&
b^{7, \frac {5} {2}} _ {3} = -\frac{4 \nu}{143}\, , \qquad
b^{7, \frac {7} {2}} _ {3} = -\frac{4 \nu}{275}\, , \qquad
b^{7, \frac {7} {2}} _ {5} =\frac{8}{14157} (-11 + \nu) \nu (11 + \nu)\, , 
\nonu\\
&&
b^{8, \frac {5} {2}} _ {3} = -\frac{4 \nu}{195}\, .
\label{bvalue1}
\eea
In particular, we observe that some of the coefficients in
Appendix (\ref{bvalue1})
are equal to the ones in (\ref{bvalue}).
Let us focus on the cubic term in $\nu$.
For example, the coefficient $b^{4, \frac {7} {2}} _ {5}$ in
Appendix (\ref{bvalue1})
is the same as $b^{4, 4} _ {6}$ in (\ref{bvalue}).
Then we can try to determine the explicit form for the
generalized hypergeometric function by following the
procedure around (\ref{bvalue}).
The corresponding quartic terms  $(\nu \mp 3)(\nu \pm 5)
(\nu \mp 5)(\nu \pm 7)$ for the above cubic term can arise in the
generalized hypergeometric functions.
Moreover, 
the sextic  terms  $(\nu \pm 5)(\nu \mp 7)(\nu \pm 7)
(\nu \mp 9)(\nu \pm 9)(\nu \mp 11)$
for the above quintic term in Appendix (\ref{bvalue1}) can arise in the
generalized hypergeometric functions
\footnote{
\label{example}
  We have
 $\phi_4^{3,3}(\mu,1)= -\frac{1}{15}(-3+\nu)(5+\nu)$,
   $\phi_6^{4,4}(\mu,1)= \frac{1}{525}(-5+\nu)(-3+\nu)(5+\nu)(7+\nu)$,
  and  $
   \phi_8^{5,5}(\mu,1)= -\frac{1}{33075}(-7+\nu)(-5+\nu)(-3+\nu)
   (5+\nu)(7+\nu)(9+\nu)$
   from the analysis of (\ref{avalue}). In this case,
   by subtracting the corresponding generalized hypergeometric functions
   replaced by $\nu \rightarrow -\nu$ respectively
   from the above, we obtain
   the coefficients $b_3^{3,\frac{5}{2}}$, $b_5^{4,\frac{7}{2}}$ and
   $b_7^{5,\frac{9}{2}}$.}.


Then we can rewrite these structure constants in terms of
generalized hypergeometric functions.
It turns out, from Appendices (\ref{aval}) and (\ref{bvalue1}),  that
\bea
a^{h_1,h_ 2+\frac{1}{2}} _ {2r}(\mu) 
& = & {}_4F_3\left[
\begin{array}{c|}
\frac{1}{2} + \mu \ ,  \frac{1}{2} - \mu  \ , \frac{2-2r}{ 2} \ , \frac{1-2r}{2}\\
\frac{3}{2}-h_1 \ ,  \frac{1}{2}-h_2\ , \frac{1}{2}+ h_1+h_2-2r
\end{array}  \ 1\right]
\nonu \\
& + & {}_4F_3\left[
\begin{array}{c|}
\frac{3}{2} - \mu \ ,  -\frac{1}{2} +\mu  \ , \frac{2-2r}{ 2} \ , \frac{1-2r}{2}\\
\frac{3}{2}-h_1 \ ,  \frac{1}{2}-h_2\ , \frac{1}{2}+ h_1+h_2-2r
\end{array}  \ 1\right]\,,
\nonu\\
b^{h_1,h_ 2+\frac{1}{2}} _ {2r-1}(\mu) 
& = & {}_4F_3\left[
\begin{array}{c|}
\frac{1}{2} + \mu \ ,  \frac{1}{2} - \mu  \ , \frac{2-2r}{ 2} \ , \frac{1-2r}{2}\\
\frac{3}{2}-h_1 \ ,  \frac{1}{2}-h_2\ , \frac{3}{2}+ h_1+h_2-2r
\end{array}  \ 1\right]
\nonu \\
& - & {}_4F_3\left[
\begin{array}{c|}
\frac{3}{2} - \mu \ ,  -\frac{1}{2} +\mu  \ , \frac{2-2r}{ 2} \ , \frac{1-2r}{2}\\
\frac{3}{2}-h_1 \ ,  \frac{1}{2}-h_2\ , \frac{3}{2}+ h_1+h_2-2r
\end{array}  \ 1\right]\,.
\label{secondabconstphi}
\eea
The structure constant $b^{h_1,h_ 2+\frac{1}{2}} _ {2r-1}(\mu) $
contains the generalized
hypergeometric function of ``saalschutzian'' in the sense that
the sum of upper elements plus $1$ is equal to the sum of
lower elements. On the other hand,
the structure constant $a^{h_1,h_ 2+\frac{1}{2}} _ {2r}(\mu)$
is not ``saalschutzian''.
In order to compare with (\ref{abconstphi}),
let us look at the first relation of Appendix (\ref{secondabconstphi}).
By replacing $h_2 \rightarrow h_2-1$ and $2r \rightarrow 2r-1$
in the generalized hypergeometric functions,
we observe that this becomes the first relation of (\ref{abconstphi})
where  the third and fourth upper elements of
generalized hypergeometric functions are shifted by $\frac{1}{2}$.
For the second relation of Appendix (\ref{secondabconstphi}),
by replacing $h_2 \rightarrow h_2-1$,
this  becomes the second relation of (\ref{abconstphi}).
We will see that the above two structure constants in
Appendix (\ref{secondabconstphi}) are denoted by
the second one in (\ref{3struct}).

\subsubsection{Operator $K$ dependence in the oscillators }

We can analyze the other case corresponding to (\ref{IdEvenOddOdake})
as follows.
By writing the $\nu$ dependent structure constants, we obtain 
\bea
&&\big[ 
  \underbrace{
\hat{y}_{(1}\ldots\,\hat{y}_{1}
}_{h_1-1+m} 
\underbrace{
\hat{y}_{2}\ldots\,\hat{y}_{2)}
}_{h_1-1-m} K,\,
 \underbrace{
\hat{y}_{(1}\ldots\,\hat{y}_{1}
}_{h_2-\frac{1}{2}+\rho} 
\underbrace{
\hat{y}_{2}\ldots\,\hat{y}_{2)}
}_{h_2-\frac{1}{2}-\rho}
\big]
=
\nonu \\
&& 
\sum_{r=1}^{[\frac{h_1+h_2}{2}]}
\,(-1)^r\,
\Bigg[
N^{h_1\,  h_2+\frac{1}{2}}_{2r-1}(m,\rho)\frac{1}{(2r-2)!}
\,a_{2r-1}^{h_1, h_2+\frac{1}{2}}(\nu)
\,  \underbrace{
\hat{y}_{(1}\ldots\ldots\ldots\ldots\hat{y}_{1}
}_{h_1+h_2-2r+\frac{1}{2}+m+\rho} \,
\underbrace{
\hat{y}_{2}\ldots\ldots\ldots\ldots\hat{y}_{2)}
}_{h_1+h_2-2r+\frac{1}{2}-m-\rho} K\Bigg]
\nonumber\\
&&
-\sum_{r=1}^{[\frac{h_1+h_2-1}{2}]}
\!\!(-1)^r \Bigg[
\mi\,N^{h_1\,  h_2+\frac{1}{2}}_{2r}(m,\rho)\, \frac{(h_1-r-\frac{1}{2})}{r(2r-1)!}\, 
\,b_{2r}^{h_1, h_2+\frac{1}{2}}(\nu)
  \underbrace{
\hat{y}_{(1}\ldots\ldots\ldots\ldots\hat{y}_{1}
}_{h_1+h_2-2r-\frac{1}{2}+m+\rho} 
\underbrace{
\hat{y}_{2}\ldots\ldots\ldots\ldots\hat{y}_{2)}
}_{h_1+h_2-2r-\frac{1}{2}-m-\rho}
\Bigg].
\nonu\\
\label{extraextra1}
\eea
As done before, the results for the coefficient
$a_{2r-1}^{h_1, h_2+\frac{1}{2}}(\nu)$ are, from
Appendix (\ref{extraextra1}), given by
\bea
&&
a^{1,\frac{5}{2}}_{1}=2\,,\qquad
a^{1,\frac{7}{2}}_{1}=2\,,\qquad
a^{1,\frac{9}{2}}_{1}=2\,,\qquad
a^{1,\frac{11}{2}}_{1}=2\,,\qquad
a^{1,\frac{13}{2}}_{1}=2\,,\qquad
a^{1,\frac{15}{2}}_{1}=2\,,
\nonu\\
&&
a^{1,\frac{17}{2}}_{1}=2\,,\qquad
a^{1,\frac{19}{2}}_{1}=2\,,\qquad
a^{2,\frac{3}{2}}_{1}=2\,,\qquad
a^{2,\frac{5}{2}}_{1}=2\,,\qquad
a^{2,\frac{7}{2}}_{1}=2\,,
\nonu\\
&&
a^{2,\frac{7}{2}}_{3}=-\frac{2}{25} (-5+\nu) (5+\nu)\,,\qquad
a^{2,\frac{9}{2}}_{1}=2\,,\qquad
a^{2,\frac{9}{2}}_{3}=-\frac{2}{49} (-7+\nu) (7+\nu)\,,
\nonu\\
&&
a^{2,\frac{11}{2}}_{1}=2\,,\qquad
a^{2,\frac{11}{2}}_{3}=-\frac{2}{81} (-9+\nu) (9+\nu)\,,\qquad
a^{2,\frac{13}{2}}_{1}=2\,,
\nonu\\
&&
a^{2,\frac{13}{2}}_{3}=-\frac{2}{121} (-11+\nu) (11+\nu)\,,\qquad
a^{2,\frac{15}{2}}_{1}=2\,,
a^{2,\frac{15}{2}}_{3}=-\frac{2}{169} (-13+\nu) (13+\nu)\,,
\nonu\\
&&
a^{2,\frac{17}{2}}_{1}=2\,,\qquad
a^{2,\frac{17}{2}}_{3}=-\frac{2}{225} (-15+\nu) (15+\nu)\,,\qquad
a^{3,\frac{3}{2}}_{1}=2\,,\qquad
a^{3,\frac{5}{2}}_{1}=2\,,
\nonu\\
&&
a^{3,\frac{5}{2}}_{3}=-\frac{2}{45} (-45+\nu^2)\,,\qquad
a^{3,\frac{7}{2}}_{1}=2\,,\qquad
a^{3,\frac{7}{2}}_{3}=-\frac{2}{105} (-105+\nu^2)\,,\qquad
a^{3,\frac{9}{2}}_{1}=2\,,
\nonu\\
&&
a^{3,\frac{9}{2}}_{3}=-\frac{2}{189} (-189+\nu^2)\,,\qquad
a^{3,\frac{9}{2}}_{5}=\frac{2}{1225} (-7+\nu) (-5+\nu) (5+\nu) (7+\nu)\,,
\nonu\\
&&
a^{3,\frac{11}{2}}_{1}=2\,,\qquad
a^{3,\frac{11}{2}}_{3}=-\frac{2}{297} (-297+\nu^2)\,,
\nonu\\
&&
a^{3,\frac{11}{2}}_{5}=\frac{2}{3969} (-9+\nu) (-7+\nu) (7+\nu) (9+\nu)\,,\qquad
a^{3,\frac{13}{2}}_{1}=2\,,
\nonu\\
&&
a^{3,\frac{13}{2}}_{3}=-\frac{2}{429} (-429+\nu^2)\,,\qquad
a^{3,\frac{13}{2}}_{5}=\frac{2}{9801} (-11+\nu) (-9+\nu) (9+\nu) (11+\nu)\,,
\nonu\\
&&
a^{3,\frac{15}{2}}_{1}=2\,,\qquad
a^{3,\frac{15}{2}}_{3}=-\frac{2}{585} (-585+\nu^2)\,,
\nonu\\
&&
a^{3,\frac{15}{2}}_{5}=\frac{2}{20449} (-13+\nu) (-11+\nu) (11+\nu) (13+\nu)\,,\qquad
a^{4,\frac{3}{2}}_{1}=2\,,\qquad
a^{4,\frac{5}{2}}_{1}=2\,,
\nonu\\
&&
a^{4,\frac{5}{2}}_{3}=-\frac{2}{105} (-105+\nu^2)\,,\qquad
a^{4,\frac{7}{2}}_{1}=2\,,\qquad
a^{4,\frac{7}{2}}_{3}=-\frac{2}{225} (-15+\nu) (15+\nu)\,,
\nonu\\
&&
a^{4,\frac{7}{2}}_{5}=\frac{2}{2625} (-5+\nu) (5+\nu) (-105+\nu^2)\,,\qquad
a^{4,\frac{9}{2}}_{1}=2\,,
a^{4,\frac{9}{2}}_{3}=-\frac{2}{385} (-385+\nu^2)\,,
\nonu\\
&&
a^{4,\frac{9}{2}}_{5}=\frac{2 }{11025}(-15+\nu) (-7+\nu) (7+\nu) (15+\nu)\,,\qquad
a^{4,\frac{11}{2}}_{1}=2\,,
\nonu\\
&&
a^{4,\frac{11}{2}}_{3}=-\frac{2}{585} (-585+\nu^2)\,,\qquad
a^{4,\frac{11}{2}}_{5}=\frac{2}{31185} (-9+\nu) (9+\nu) (-385+\nu^2)\,,
\nonu\\
&&
a^{4,\frac{11}{2}}_{7}=-\frac{2}{99225} (-9+\nu) (-7+\nu) (-5+\nu) (5+\nu) (7+\nu) (9+\nu)\,,\qquad
a^{4,\frac{13}{2}}_{1}=2\,,
\nonu\\
&&
a^{4,\frac{13}{2}}_{3}=-\frac{2}{825} (-825+\nu^2)\,,\qquad
a^{4,\frac{13}{2}}_{5}=\frac{2}{70785} (-11+\nu) (11+\nu) (-585+\nu^2)\,,
\nonu\\
&&
a^{4,\frac{13}{2}}_{7}=-\frac{2}{480249} (-11+\nu) (-9+\nu) (-7+\nu) (7+\nu) (9+\nu) (11+\nu)\,,\qquad
a^{5,\frac{3}{2}}_{1}=2\,,
\nonu\\
&&
a^{5,\frac{5}{2}}_{1}=2\,,\qquad
a^{5,\frac{5}{2}}_{3}=-\frac{2}{189} (-189+\nu^2)\,,\qquad
a^{5,\frac{7}{2}}_{1}=2\,,
a^{5,\frac{7}{2}}_{3}=-\frac{2}{385} (-385+\nu^2)\,,
\nonu \\
%
&&
a^{5,\frac{7}{2}}_{5}=\frac{2}{11025} (-15+\nu) (-7+\nu) (7+\nu) (15+\nu)\,,\qquad
a^{5,\frac{9}{2}}_{1}=2\,,
\nonu\\
&&
a^{5,\frac{9}{2}}_{3}=-\frac{2}{637} (-637+\nu^2)\,,\qquad
a^{5,\frac{9}{2}}_{5}=\frac{2}{40425} (40425-554 \nu^2+\nu^4)\,,
\nonu\\
&&
a^{5,\frac{9}{2}}_{7}=-\frac{2}{231525} (-7+\nu) (-5+\nu) (5+\nu) (7+\nu) (-189+\nu^2)\,,\qquad
a^{5,\frac{11}{2}}_{1}=2\,,
\nonu\\
&&
a^{5,\frac{11}{2}}_{3}=-\frac{2}{945} (-945+\nu^2)\,,\qquad
a^{5,\frac{11}{2}}_{5}=\frac{2}{105105} (105105-914 \nu^2+\nu^4)\,,
\nonu\\
&&
a^{5,\frac{11}{2}}_{7}=-\frac{2}{1528065} (-9+\nu) (-7+\nu) (7+\nu) (9+\nu) (-385+\nu^2)\,,\qquad
a^{6,\frac{3}{2}}_{1}=2\,,
\nonu\\
&&
a^{6,\frac{5}{2}}_{1}=2\,,\qquad
a^{6,\frac{5}{2}}_{3}=-\frac{2}{297} (-297+\nu^2)\,,\qquad
a^{6,\frac{7}{2}}_{1}=2\,,\qquad
a^{6,\frac{7}{2}}_{3}=-\frac{2}{585} (-585+\nu^2)\,,
\nonu\\
&&
a^{6,\frac{7}{2}}_{5}=\frac{2}{31185} (-9+\nu) (9+\nu) (-385+\nu^2)\,,\qquad
a^{6,\frac{9}{2}}_{1}=2\,,
a^{6,\frac{9}{2}}_{3}=-\frac{2}{945} (-945+\nu^2)\,,
\nonu\\
&&
a^{6,\frac{9}{2}}_{5}=\frac{2}{105105} (105105-914 \nu^2+\nu^4)\,,
\nonu\\
&&
a^{6,\frac{9}{2}}_{7}=-\frac{2}{1528065} (-9+\nu) (-7+\nu) (7+\nu) (9+\nu) (-385+\nu^2)\,,
a^{7,\frac{3}{2}}_{1}=2\,,
a^{7,\frac{5}{2}}_{1}=2\,,
\nonu\\
&&
a^{7,\frac{5}{2}}_{3}=-\frac{2}{429} (-429+\nu^2)\,,\qquad
a^{7,\frac{7}{2}}_{1}=2\,,\qquad
a^{7,\frac{7}{2}}_{3}=-\frac{2}{825} (-825+\nu^2)\,,
\nonu\\
&&
a^{7,\frac{7}{2}}_{5}=\frac{2}{70785} (-11+\nu) (11+\nu) (-585+\nu^2)\,,\qquad
a^{8,\frac{3}{2}}_{1}=2\,,\qquad
a^{8,\frac{5}{2}}_{1}=2\,,
\nonu\\
&&
a^{8,\frac{5}{2}}_{3}=-\frac{2}{585} (-585+\nu^2)\,,\qquad
a^{9,\frac{3}{2}}_{1}=2\,.
\label{ava1}
\eea
We can easily see that some of the coefficients in Appendix (\ref{ava1})
overlap with the ones in Appendix (\ref{aval}).
For example, the coefficient
$a^{2,\frac{7}{2}}_{3}$ of Appendix (\ref{ava1}) is exactly the same as
the coefficient $a^{3,\frac{7}{2}}_{3}$ of Appendix (\ref{aval}).
We can do similar analysis and will arrive at the explicit $\nu$
dependence in the generalized hypergeometric functions we are
considering.

Similarly, the coefficient $b_{2r}^{h_1, h_2+\frac{1}{2}}(\nu)$
can be summarized by
\bea
&&
b^{2,\frac{3}{2}}_{2}=-\frac{4 \nu}{3}\,,\qquad
b^{2,\frac{5}{2}}_{2}=-\frac{4 \nu}{15}\,,\qquad
b^{2,\frac{7}{2}}_{2}=-\frac{4 \nu}{35}\,,\qquad
b^{2,\frac{9}{2}}_{2}=-\frac{4 \nu}{63}\,,\qquad
b^{2,\frac{11}{2}}_{2}=-\frac{4 \nu}{99}\,,
\nonu\\
&&
b^{2,\frac{13}{2}}_{2}=-\frac{4 \nu}{143}\,,\qquad
b^{2,\frac{15}{2}}_{2}=-\frac{4 \nu}{195}\,,\qquad
b^{2,\frac{17}{2}}_{2}=-\frac{4 \nu}{255}\,,\qquad
b^{3,\frac{3}{2}}_{2}=-\frac{4 \nu}{15}\,,
b^{3,\frac{5}{2}}_{2}=-\frac{4 \nu}{63}\,,
\nonu\\
&&
b^{3,\frac{5}{2}}_{4}=\frac{8}{45} (-3+\nu) \nu (3+\nu)\,,\qquad
b^{3,\frac{7}{2}}_{2}=-\frac{4 \nu}{135}\,,\qquad
b^{3,\frac{7}{2}}_{4}=\frac{8}{525} (-5+\nu) \nu (5+\nu)\,,
\nonu\\
&&
b^{3,\frac{9}{2}}_{2}=-\frac{4 \nu}{231}\,,\qquad
b^{3,\frac{9}{2}}_{4}=\frac{8}{2205} (-7+\nu) \nu (7+\nu)\,,\qquad
b^{3,\frac{11}{2}}_{2}=-\frac{4 \nu}{351}\,,
\nonu\\
&&
b^{3,\frac{11}{2}}_{4}=\frac{8}{6237} (-9+\nu) \nu (9+\nu)\,,\qquad
b^{3,\frac{13}{2}}_{2}=-\frac{4 \nu}{495}\,,
b^{3,\frac{13}{2}}_{4}=\frac{8}{14157} (-11+\nu) \nu (11+\nu)\,,
\nonu\\
&&
b^{3,\frac{15}{2}}_{2}=-\frac{4 \nu}{663}\,,\qquad
b^{3,\frac{15}{2}}_{4}=\frac{8}{27885} (-13+\nu) \nu (13+\nu)\,,\qquad
b^{4,\frac{3}{2}}_{2}=-\frac{4 \nu}{35}\,,
\nonu\\
&&
b^{4,\frac{5}{2}}_{2}=-\frac{4 \nu}{135}\,,\qquad
b^{4,\frac{5}{2}}_{4}=\frac{8}{525} (-5+\nu) \nu (5+\nu)\,,\qquad
b^{4,\frac{7}{2}}_{2}=-\frac{4 \nu}{275}\,,
\nonu\\
&&
b^{4,\frac{7}{2}}_{4}=\frac{8}{4725} \nu (-85+\nu^2)\,,\qquad
b^{4,\frac{7}{2}}_{6}=-\frac{4}{525} (-5+\nu) (-3+\nu) \nu (3+\nu) (5+\nu)\,,
\nonu\\
&&
b^{4,\frac{9}{2}}_{2}=-\frac{4 \nu}{455}\,,\qquad
b^{4,\frac{9}{2}}_{4}=\frac{8}{17325} (-13+\nu) \nu (13+\nu)\,,
\nonu\\
&&
b^{4,\frac{9}{2}}_{6}=-\frac{4}{11025} (-7+\nu) (-5+\nu) \nu (5+\nu) (7+\nu)\,,\qquad
b^{4,\frac{11}{2}}_{2}=-\frac{4 \nu}{675}\,,
\nonu\\
&&
b^{4,\frac{11}{2}}_{4}=\frac{8}{45045} \nu (-277+\nu^2)\,,\qquad
b^{4,\frac{11}{2}}_{6}=-\frac{4}{72765} (-9+\nu) (-7+\nu) \nu (7+\nu) (9+\nu)\,,
\nonu\\
&&
b^{4,\frac{13}{2}}_{2}=-\frac{4 \nu}{935}\,,\qquad
b^{4,\frac{13}{2}}_{4}=\frac{8}{96525} \nu (-409+\nu^2)\,,
\nonu\\
&&
b^{4,\frac{13}{2}}_{6}=-\frac{4}{297297} (-11+\nu) (-9+\nu) \nu (9+\nu) (11+\nu)\,,
b^{5,\frac{3}{2}}_{2}=-\frac{4 \nu}{63}\,,\qquad
b^{5,\frac{5}{2}}_{2}=-\frac{4 \nu}{231}\,,
\nonu\\
&&
b^{5,\frac{5}{2}}_{4}=\frac{8}{2205} (-7+\nu) \nu (7+\nu)\,,\qquad
b^{5,\frac{7}{2}}_{2}=-\frac{4 \nu}{455}\,,
b^{5,\frac{7}{2}}_{4}=\frac{8}{17325} (-13+\nu) \nu (13+\nu)\,,
\nonu\\
&&
b^{5,\frac{7}{2}}_{6}=-\frac{4 }{11025}(-7+\nu) (-5+\nu) \nu (5+\nu) (7+\nu)\,,\qquad
b^{5,\frac{9}{2}}_{2}=-\frac{4 \nu}{735}\,,
\nonu\\
&&
b^{5,\frac{9}{2}}_{4}=\frac{24}{175175} \nu (-329+\nu^2)\,,
b^{5,\frac{9}{2}}_{6}=-\frac{4}{169785} (-13+\nu) (-7+\nu) \nu (7+\nu) (13+\nu)\,,
\nonu\\
&&
b^{5,\frac{9}{2}}_{8}=\frac{16}{99225} (-7+\nu) (-5+\nu) (-3+\nu) \nu (3+\nu) (5+\nu) (7+\nu)\,,
b^{5,\frac{11}{2}}_{2}=-\frac{4 \nu}{1071}\,,
\nonu\\
&&
b^{5,\frac{11}{2}}_{4}=\frac{8}{143325} (-23+\nu) \nu (23+\nu)\,,
b^{5,\frac{11}{2}}_{6}=-\frac{4}{945945} (-9+\nu) \nu (9+\nu) (-329+\nu^2)\,,
\nonu \\
%
&&
b^{5,\frac{11}{2}}_{8}=\frac{16}{3274425} (-9+\nu) (-7+\nu) (-5+\nu) \nu (5+\nu) (7+\nu) (9+\nu)\,,\qquad
b^{6,\frac{3}{2}}_{2}=-\frac{4 \nu}{99}\,,
\nonu\\
&&
b^{6,\frac{5}{2}}_{2}=-\frac{4 \nu}{351}\,,\qquad
b^{6,\frac{5}{2}}_{4}=\frac{8}{6237} (-9+\nu) \nu (9+\nu)\,,\qquad
b^{6,\frac{7}{2}}_{2}=-\frac{4 \nu}{675}\,,
\nonu\\
&&
b^{6,\frac{7}{2}}_{4}=\frac{8}{45045} \nu (-277+\nu^2)\,,\qquad
b^{6,\frac{7}{2}}_{6}=-\frac{4}{72765} (-9+\nu) (-7+\nu) \nu (7+\nu) (9+\nu)\,,
\nonu\\
&&
b^{6,\frac{9}{2}}_{2}=-\frac{4 \nu}{1071}\,,\qquad
b^{6,\frac{9}{2}}_{4}=\frac{8}{143325} (-23+\nu) \nu (23+\nu)\,,
\nonu\\
&&
b^{6,\frac{9}{2}}_{6}=-\frac{4}{945945} (-9+\nu) \nu (9+\nu) (-329+\nu^2)\,,
\nonu\\
&&
b^{6,\frac{9}{2}}_{8}=\frac{16}{3274425} (-9+\nu) (-7+\nu) (-5+\nu) \nu (5+\nu) (7+\nu) (9+\nu)\,,\qquad
b^{7,\frac{3}{2}}_{2}=-\frac{4 \nu}{143}\,,
\nonu\\
&&
b^{7,\frac{5}{2}}_{2}=-\frac{4 \nu}{495}\,,\qquad
b^{7,\frac{5}{2}}_{4}=\frac{8}{14157} (-11+\nu) \nu (11+\nu)\,,\qquad
b^{7,\frac{7}{2}}_{2}=-\frac{4 \nu}{935}\,,
\nonu\\
&&
b^{7,\frac{7}{2}}_{4}=\frac{8}{96525} \nu (-409+\nu^2)\,,\qquad
b^{7,\frac{7}{2}}_{6}=-\frac{4}{297297} (-11+\nu) (-9+\nu) \nu (9+\nu) (11+\nu)\,,
\nonu\\
&&
b^{8,\frac{3}{2}}_{2}=-\frac{4 \nu}{195}\,,\qquad
b^{8,\frac{5}{2}}_{2}=-\frac{4 \nu}{663}\,,\qquad
b^{8,\frac{5}{2}}_{4}=\frac{8}{27885} (-13+\nu) \nu (13+\nu)\,,
\nonu\\
&&
b^{9,\frac{3}{2}}_{2}=-\frac{4 \nu}{255}\,.
\label{bva1}
\eea
Some of the coefficients in Appendix (\ref{bva1})
overlap with the ones in (\ref{bvalue}).
For example, the coefficient
$b^{3,\frac{7}{2}}_{4}$ of Appendix (\ref{bva1}) is exactly the same as
the coefficient $b^{4,4}_{6}$ of  (\ref{bvalue}).
We can do similar analysis and the explicit $\nu$
dependence in the generalized hypergeometric functions can be
determined.

From the results of Appendices (\ref{ava1}) and (\ref{bva1}), the general
expressions are given by
\bea
a^{h_1,h_ 2+\frac{1}{2}} _ {2r-1}(\mu) 
& = & {}_4F_3\left[
\begin{array}{c|}
\frac{1}{2} + \mu \ ,  \frac{1}{2} - \mu  \ , \frac{3-2r}{ 2} \ , \frac{2-2r}{2}\\
\frac{3}{2}-h_1 \ ,  \frac{1}{2}-h_2\ , \frac{3}{2}+ h_1+h_2-2r
\end{array}  \ 1\right]
\nonu \\
& + & {}_4F_3\left[
\begin{array}{c|}
\frac{3}{2} - \mu \ ,  -\frac{1}{2} +\mu  \ , \frac{3-2r}{ 2} \ , \frac{2-2r}{2}\\
\frac{3}{2}-h_1 \ ,  \frac{1}{2}-h_2\ , \frac{3}{2}+ h_1+h_2-2r
\end{array}  \ 1\right]\,,
\nonu\\
b^{h_1,h_ 2+\frac{1}{2}} _ {2r}(\mu) 
&=& {}_4F_3\left[
\begin{array}{c|}
\frac{1}{2} + \mu \ ,  \frac{1}{2} - \mu  \ , \frac{1-2r}{ 2} \ , \frac{-2r}{2}\\
\frac{3}{2}-h_1 \ ,  \frac{1}{2}-h_2\ , \frac{1}{2}+ h_1+h_2-2r
\end{array}  \ 1\right]
\nonu \\
& - & {}_4F_3\left[
\begin{array}{c|}
\frac{3}{2} - \mu \ ,  -\frac{1}{2} +\mu  \ , \frac{1-2r}{ 2} \ , \frac{-2r}{2}\\
\frac{3}{2}-h_1 \ ,  \frac{1}{2}-h_2\ , \frac{1}{2}+ h_1+h_2-2r
\end{array}  \ 1\right]\,.
\label{fullab}
\eea
We can check that by replacing $2r$ by $(2r+1)$ for the coefficient
$a$ and $2r$ by $(2r-1)$ for the coefficient $b$,
the above expression Appendix (\ref{fullab}) becomes the ones in
Appendix (\ref{secondabconstphi}) and we can see this result in
(\ref{IdEvenOdd-1}) and (\ref{IdEvenOdd}).


\subsection{ The case-three with two fermionic oscillators}

\subsubsection{No operator $K$ dependence in the oscillators}

From (\ref{casethreeone}), we introduce
$\mu$ dependent structure constants as follows:
\bea
&&
\big\{ 
  \underbrace{
\hat{y}_{(1}\ldots\,\hat{y}_{1}
}_{h_1-\frac{1}{2}+\rho} 
\underbrace{
\hat{y}_{2}\ldots\,\hat{y}_{2)}
}_{h_1-\frac{1}{2}-\rho},\,
 \underbrace{
\hat{y}_{(1}\ldots\,\hat{y}_{1}
}_{h_2-\frac{1}{2}+\omega} 
\underbrace{
\hat{y}_{2}\ldots\,\hat{y}_{2)}
}_{h_2-\frac{1}{2}-\omega}
\big\}
=
\nonu \\
&& \sum_{r=1}^{[\frac{h_1+h_2+1}{2}]}\,(-1)^r\,
 N^{h_1+\frac{1}{2}\,  h_2+\frac{1}{2}}_{2r-1}(\rho,\omega)\,
\frac{(-1)}{(2r-2)!}\,
\Bigg[
a_{2r-1}^{h_1+\frac{1}{2}, h_2+\frac{1}{2}}(\nu)
\,  \underbrace{
\hat{y}_{(1}\ldots\ldots\ldots.\hat{y}_{1}
}_{h_1+h_2-2r+1+\rho+\omega} \,
\underbrace{
\hat{y}_{2}\ldots\ldots\ldots.\hat{y}_{2)}
}_{h_1+h_2-2r+1-\rho-\omega}
\nonumber\\
&&
-\frac{2(h_1+h_2-r+1)}{(2r-1)}\,
b_{2r-1}^{h_1+\frac{1}{2}, h_2+\frac{1}{2}}(\nu)
\,  \underbrace{
\hat{y}_{(1}\ldots\ldots\ldots.\hat{y}_{1}
}_{h_1+h_2-2r+1+\rho+\omega} \,
\underbrace{
\hat{y}_{2}\ldots\ldots\ldots.\hat{y}_{2)}
}_{h_1+h_2-2r+1-\rho-\omega}K
\Bigg].
\label{casethreeapp}
\eea
We can compute the anticommutator for several values
of $(h_1,h_2)$.
We can determine the structure constants having
the operator $K$ independent oscillators in Appendix
(\ref{casethreeapp})
as follows:
\bea
&&
a^{\frac{3}{2},\frac{3}{2}}_{1}=2\,,\qquad
a^{\frac{3}{2},\frac{5}{2}}_{1}=2\,,\qquad
a^{\frac{3}{2},\frac{7}{2}}_{1}=2\,,\qquad
a^{\frac{3}{2},\frac{9}{2}}_{1}=2\,,\qquad
a^{\frac{3}{2},\frac{11}{2}}_{1}=2\,,\qquad
a^{\frac{3}{2},\frac{13}{2}}_{1}=2\,,
\nonu\\
&&
a^{\frac{3}{2},\frac{15}{2}}_{1}=2\,,\qquad
a^{\frac{3}{2},\frac{17}{2}}_{1}=2\,,\qquad
a^{\frac{3}{2},\frac{19}{2}}_{1}=2\,,\qquad
a^{\frac{5}{2},\frac{3}{2}}_{1}=2\,,
\nonu\\
&&
a^{\frac{5}{2},\frac{5}{2}}_{1}=2\,,\qquad
a^{\frac{5}{2},\frac{5}{2}}_{3}=-\frac{2}{45} (-45+\nu^2)\,,\qquad
a^{\frac{5}{2},\frac{7}{2}}_{1}=2\,,\qquad
a^{\frac{5}{2},\frac{7}{2}}_{3}=-\frac{2}{105} (-105+\nu^2)\,,
\nonu\\
&&
a^{\frac{5}{2},\frac{9}{2}}_{1}=2\,,\qquad
a^{\frac{5}{2},\frac{9}{2}}_{3}=-\frac{2}{189} (-189+\nu^2)\,,\qquad
a^{\frac{5}{2},\frac{11}{2}}_{1}=2\,,
a^{\frac{5}{2},\frac{11}{2}}_{3}=-\frac{2}{297} (-297+\nu^2)\,,
\nonu\\
&&
a^{\frac{5}{2},\frac{13}{2}}_{1}=2\,,\qquad
a^{\frac{5}{2},\frac{13}{2}}_{3}=-\frac{2}{429} (-429+\nu^2)\,,\qquad
a^{\frac{5}{2},\frac{15}{2}}_{1}=2\,,
a^{\frac{5}{2},\frac{15}{2}}_{3}=-\frac{2}{585} (-585+\nu^2)\,,
\nonu\\
&&
a^{\frac{5}{2},\frac{17}{2}}_{1}=2\,,\qquad
a^{\frac{5}{2},\frac{17}{2}}_{3}=-\frac{2}{765} (-765+\nu^2)\,,\qquad
a^{\frac{7}{2},\frac{3}{2}}_{1}=2\,,\qquad
a^{\frac{7}{2},\frac{5}{2}}_{1}=2\,,
\nonu\\
&&
a^{\frac{7}{2},\frac{5}{2}}_{3}=-\frac{2}{105} (-105+\nu^2)\,,\qquad
a^{\frac{7}{2},\frac{7}{2}}_{1}=2\,,\qquad
a^{\frac{7}{2},\frac{7}{2}}_{3}=-\frac{2}{225} (-15+\nu) (15+\nu)\,,
\nonu\\
&&
a^{\frac{7}{2},\frac{7}{2}}_{5}=\frac{2}{2625} (-5+\nu) (5+\nu) (-105+\nu^2)\,,\qquad
a^{\frac{7}{2},\frac{9}{2}}_{1}=2\,,
a^{\frac{7}{2},\frac{9}{2}}_{3}=-\frac{2}{385} (-385+\nu^2)\,,
\nonu\\
&&
a^{\frac{7}{2},\frac{9}{2}}_{5}=\frac{2}{11025} (-15+\nu) (-7+\nu) (7+\nu) (15+\nu)\,,\qquad
a^{\frac{7}{2},\frac{11}{2}}_{1}=2\,,
\nonu\\
&&
a^{\frac{7}{2},\frac{11}{2}}_{3}=-\frac{2}{585} (-585+\nu^2)\,,\qquad
a^{\frac{7}{2},\frac{11}{2}}_{5}=\frac{2}{31185} (-9+\nu) (9+\nu) (-385+\nu^2)\,,
\nonu\\
&&
a^{\frac{7}{2},\frac{13}{2}}_{1}=2\,,\qquad
a^{\frac{7}{2},\frac{13}{2}}_{3}=-\frac{2}{825} (-825+\nu^2)\,,
\nonu\\
&&
a^{\frac{7}{2},\frac{13}{2}}_{5}=\frac{2}{70785} (-11+\nu) (11+\nu) (-585+\nu^2)\,,\qquad
a^{\frac{7}{2},\frac{15}{2}}_{1}=2\,,
\nonu\\
&&
a^{\frac{7}{2},\frac{15}{2}}_{3}=-\frac{2}{1105} (-1105+\nu^2)\,,\qquad
a^{\frac{7}{2},\frac{15}{2}}_{5}=\frac{2}{139425} (-13+\nu) (13+\nu) (-825+\nu^2)\,,
\nonu\\
&&
a^{\frac{9}{2},\frac{3}{2}}_{1}=2\,,\qquad
a^{\frac{9}{2},\frac{5}{2}}_{1}=2\,,\qquad
a^{\frac{9}{2},\frac{5}{2}}_{3}=-\frac{2}{189} (-189+\nu^2)\,,\qquad
a^{\frac{9}{2},\frac{7}{2}}_{1}=2\,,
\nonu\\
&&
a^{\frac{9}{2},\frac{7}{2}}_{3}=-\frac{2}{385} (-385+\nu^2)\,,\qquad
a^{\frac{9}{2},\frac{7}{2}}_{5}=\frac{2}{11025} (-15+\nu) (-7+\nu) (7+\nu) (15+\nu)\,,
\nonu\\
&&
a^{\frac{9}{2},\frac{9}{2}}_{1}=2\,,\qquad
a^{\frac{9}{2},\frac{9}{2}}_{3}=-\frac{2}{637} (-637+\nu^2)\,,\qquad
a^{\frac{9}{2},\frac{9}{2}}_{5}=\frac{2}{40425} (40425-554 \nu^2+\nu^4)\,,
\nonu\\
&&
a^{\frac{9}{2},\frac{9}{2}}_{7}=-\frac{2}{231525} (-7+\nu) (-5+\nu) (5+\nu) (7+\nu) (-189+\nu^2)\,,\qquad
a^{\frac{9}{2},\frac{11}{2}}_{1}=2\,,
\nonu\\
&&
a^{\frac{9}{2},\frac{11}{2}}_{3}=-\frac{2}{945} (-945+\nu^2)\,,\qquad
a^{\frac{9}{2},\frac{11}{2}}_{5}=\frac{2}{105105} (105105-914 \nu^2+\nu^4)\,,
\nonu\\
&&
a^{\frac{9}{2},\frac{11}{2}}_{7}=-\frac{2}{1528065} (-9+\nu) (-7+\nu) (7+\nu) (9+\nu) (-385+\nu^2)\,,\qquad
a^{\frac{9}{2},\frac{13}{2}}_{1}=2\,,
\nonu\\
&&
a^{\frac{9}{2},\frac{13}{2}}_{3}=-\frac{2}{1309} (-1309+\nu^2)\,,\qquad
a^{\frac{9}{2},\frac{13}{2}}_{5}=\frac{2}{225225} (225225-1354 \nu^2+\nu^4)\,,
\nonu\\
&&
a^{\frac{9}{2},\frac{13}{2}}_{7}=-\frac{2}{6243237} (-11+\nu) (-9+\nu) (9+\nu) (11+\nu) (-637+\nu^2)\,,\qquad
a^{\frac{11}{2},\frac{3}{2}}_{1}=2\,,
\nonu\\
&&
a^{\frac{11}{2},\frac{5}{2}}_{1}=2\,,\qquad
a^{\frac{11}{2},\frac{5}{2}}_{3}=-\frac{2}{297} (-297+\nu^2)\,,\qquad
a^{\frac{11}{2},\frac{7}{2}}_{1}=2\,,
a^{\frac{11}{2},\frac{7}{2}}_{3}=-\frac{2}{585} (-585+\nu^2)\,,
\nonu \\
%
&&
a^{\frac{11}{2},\frac{7}{2}}_{5}=\frac{2}{31185} (-9+\nu) (9+\nu) (-385+\nu^2)\,,\qquad
a^{\frac{11}{2},\frac{9}{2}}_{1}=2\,,
\nonu\\
&&
a^{\frac{11}{2},\frac{9}{2}}_{3}=-\frac{2}{945} (-945+\nu^2)\,,\qquad
a^{\frac{11}{2},\frac{9}{2}}_{5}=\frac{2}{105105} (105105-914 \nu^2+\nu^4)\,,
\nonu\\
&&
a^{\frac{11}{2},\frac{9}{2}}_{7}=-\frac{2}{1528065} (-9+\nu) (-7+\nu) (7+\nu) (9+\nu) (-385+\nu^2)\,,\qquad
a^{\frac{11}{2},\frac{11}{2}}_{1}=2\,,
\nonu\\
&&
a^{\frac{11}{2},\frac{11}{2}}_{3}=-\frac{2}{1377} (-1377+\nu^2)\,,\qquad
a^{\frac{11}{2},\frac{11}{2}}_{5}=\frac{2}{257985} (257985-1474 \nu^2+\nu^4)\,,
\nonu\\
&&
a^{\frac{11}{2},\frac{11}{2}}_{7}=-\frac{2}{8513505} (-9+\nu) (9+\nu) (105105-914 \nu^2+\nu^4)\,,
\nonu\\
&&
a^{\frac{11}{2},\frac{11}{2}}_{9}=\frac{2}{29469825} (-9+\nu) (-7+\nu) (-5+\nu) (5+\nu) (7+\nu) (9+\nu) (-297+\nu^2)\,,
\nonu\\
&&
a^{\frac{13}{2},\frac{3}{2}}_{1}=2\,,\qquad
a^{\frac{13}{2},\frac{5}{2}}_{1}=2\,,\qquad
a^{\frac{13}{2},\frac{5}{2}}_{3}=-\frac{2}{429} (-429+\nu^2)\,,\qquad
a^{\frac{13}{2},\frac{7}{2}}_{1}=2\,,
\nonu\\
&&
a^{\frac{13}{2},\frac{7}{2}}_{3}=-\frac{2}{825} (-825+\nu^2)\,,\qquad
a^{\frac{13}{2},\frac{7}{2}}_{5}=\frac{2}{70785} (-11+\nu) (11+\nu) (-585+\nu^2)\,,
\nonu\\
&&
a^{\frac{13}{2},\frac{9}{2}}_{1}=2\,,\qquad
a^{\frac{13}{2},\frac{9}{2}}_{3}=-\frac{2}{1309} (-1309+\nu^2)\,,
\nonu\\
&&
a^{\frac{13}{2},\frac{9}{2}}_{5}=\frac{2}{225225} (225225-1354 \nu^2+\nu^4)\,,
\nonu\\
&&
a^{\frac{13}{2},\frac{9}{2}}_{7}=-\frac{2}{6243237} (-11+\nu) (-9+\nu) (9+\nu) (11+\nu) (-637+\nu^2)\,,\qquad
a^{\frac{15}{2},\frac{3}{2}}_{1}=2\,,
\nonu\\
&&
a^{\frac{15}{2},\frac{5}{2}}_{1}=2\,,\qquad
a^{\frac{15}{2},\frac{5}{2}}_{3}=-\frac{2}{585} (-585+\nu^2)\,,\qquad
a^{\frac{15}{2},\frac{7}{2}}_{1}=2\,,
\nonu\\
&&
a^{\frac{15}{2},\frac{7}{2}}_{3}=-\frac{2}{1105} (-1105+\nu^2)\,,\qquad
a^{\frac{15}{2},\frac{7}{2}}_{5}=\frac{2}{139425} (-13+\nu) (13+\nu) (-825+\nu^2)\,,
\nonu\\
&&
a^{\frac{17}{2},\frac{3}{2}}_{1}=2\,,\qquad
a^{\frac{17}{2},\frac{5}{2}}_{1}=2\,,\qquad
a^{\frac{17}{2},\frac{5}{2}}_{3}=-\frac{2}{765} (-765+\nu^2)\,,\qquad
a^{\frac{19}{2},\frac{3}{2}}_{1}=2\,.
\label{Aval}
\eea
The first nontrivial $\nu$ dependence
appears in $a^{\frac{5}{2},\frac{5}{2}}_{3}$.
From the previous experience, we expect that there is a linear
$\nu$ dependence in the generalized hypergeometric functions.
By adding the two of them, we obtain the above coefficient which has
the constant term as well as the quadratic term in $\nu$.
The first nontrivial quartic term in $\nu$ dependence
appears in $a^{\frac{7}{2},\frac{7}{2}}_{5}$.
There will be  a linear term
and the third order term in $\nu$ dependence in
the generalized hypergeometric functions.
But it is nontrivial to obtain their explicit forms.
We observe that there are sextic term in  the coefficient
$a^{\frac{9}{2},\frac{9}{2}}_{7}$
and octic term in  the coefficient $a^{\frac{11}{2},\frac{11}{2}}_{9}$.
We should determine the odd power of $\nu$ terms in the
generalized hypergeometric functions before we add them
\footnote{As found in the footnote \ref{example}, we obtain,
  for the second argument $2$ of generalized hypergeometric function,
  $\phi_4^{3,3}(\mu,2)=\frac{1}{45}(45-2\nu-\nu^2)$,
  $\phi_6^{4,4}(\mu,2)=\frac{1}{2625}(2625-268\nu-130\nu^2+4\nu^3+
  \nu^4)$ and $\phi_8^{5,5}(\mu,2)=\frac{1}{231525}(231525-32622\nu-
  15211\nu^2+1092\nu^3+263\nu^4-6\nu^5-\nu^6)$.
For example,  we can check that  $\phi_4^{3,3}(\mu,2)$ looks
  similar to  $\phi_4^{3,3}(\mu,1)$ in the sense that
  the three lower elements are the same while
  the one upper element
  is different from each other.
  Because the fourth upper element in the former is the same as
  the third upper element in the latter, the number of terms in the
  generalized hypergeometric functions is the same with different
  coefficients.
  There are odd $\nu$ terms in the above expressions.
  By taking
  $\nu \rightarrow -\nu$, we obtain the corresponding
generalized hypergeometric functions 
and as done before, we obtain the coefficients $a_3^{\frac{5}{2},
  \frac{5}{2}}$, $a_5^{\frac{7}{2},
  \frac{7}{2}}$ and $a_7^{\frac{9}{2},
  \frac{9}{2}}$ respectively.}.

The other structure constants 
can be obtained as follows:
\bea
&&
b^{\frac{5}{2},\frac{5}{2}}_{3}=-\frac{4 \nu}{15}\,,\qquad
b^{\frac{5}{2},\frac{7}{2}}_{3}=-\frac{4 \nu}{35}\,,\qquad
b^{\frac{5}{2},\frac{9}{2}}_{3}=-\frac{4 \nu}{63}\,,\qquad
b^{\frac{5}{2},\frac{11}{2}}_{3}=-\frac{4 \nu}{99}\,,
b^{\frac{5}{2},\frac{13}{2}}_{3}=-\frac{4 \nu}{143}\,,
\nonu\\
&&
b^{\frac{5}{2},\frac{15}{2}}_{3}=-\frac{4 \nu}{195}\,,\qquad
b^{\frac{5}{2},\frac{17}{2}}_{3}=-\frac{4 \nu}{255}\,,\qquad
b^{\frac{7}{2},\frac{5}{2}}_{3}=-\frac{4 \nu}{35}\,,\qquad
b^{\frac{7}{2},\frac{7}{2}}_{3}=-\frac{4 \nu}{75}\,,
\nonu\\
&&
b^{\frac{7}{2},\frac{7}{2}}_{5}=\frac{8}{525}(-5+\nu) \nu (5+\nu)\,,\qquad
b^{\frac{7}{2},\frac{9}{2}}_{3}=-\frac{12\nu}{385}\,,\qquad
b^{\frac{7}{2},\frac{9}{2}}_{5}=\frac{8}{2205} (-7+\nu) \nu (7+\nu)\,,
\nonu\\
&&
b^{\frac{7}{2},\frac{11}{2}}_{3}=-\frac{4 \nu}{195}\,,\qquad
b^{\frac{7}{2},\frac{11}{2}}_{5}=\frac{8}{6237} (-9+\nu) \nu (9+\nu)\,,\qquad
b^{\frac{7}{2},\frac{13}{2}}_{3}=-\frac{4\nu}{275}\,,
\nonu\\
&&
b^{\frac{7}{2},\frac{13}{2}}_{5}=\frac{8}{14157} (-11+\nu) \nu (11+\nu)\,,\qquad
b^{\frac{7}{2},\frac{15}{2}}_{3}=-\frac{12\nu}{1105}\,,
\nonu\\
&&
b^{\frac{7}{2},\frac{15}{2}}_{5}=\frac{8}{27885} (-13+\nu) \nu (13+\nu)\,,\qquad
b^{\frac{9}{2},\frac{5}{2}}_{3}=-\frac{4 \nu}{63}\,,\qquad
b^{\frac{9}{2},\frac{7}{2}}_{3}=-\frac{12\nu}{385}\,,
\nonu\\
&&
b^{\frac{9}{2},\frac{7}{2}}_{5}=\frac{8}{2205} (-7+\nu) \nu (7+\nu)\,,\qquad
b^{\frac{9}{2},\frac{9}{2}}_{3}=-\frac{12\nu}{637}\,,\qquad
b^{\frac{9}{2},\frac{9}{2}}_{5}=\frac{8}{24255} \nu (-287+3 \nu^2)\,,
\nonu\\
&&
b^{\frac{9}{2},\frac{9}{2}}_{7}=-\frac{4}{11025} (-7+\nu) (-5+\nu) \nu (5+\nu) (7+\nu)\,,\qquad
b^{\frac{9}{2},\frac{11}{2}}_{3}=-\frac{4\nu}{315}\,,
\nonu\\
&&
b^{\frac{9}{2},\frac{11}{2}}_{5}=\frac{8}{63063} \nu (-467+3 \nu^2)\,,\qquad
b^{\frac{9}{2},\frac{11}{2}}_{7}=-\frac{4}{72765} (-9+\nu) (-7+\nu) \nu (7+\nu) (9+\nu)\,,
\nonu\\
&&
b^{\frac{9}{2},\frac{13}{2}}_{3}=-\frac{12\nu}{1309}\,,\qquad
b^{\frac{9}{2},\frac{13}{2}}_{5}=\frac{8}{45045} \nu (-229+\nu^2)\,,
\nonu\\
&&
b^{\frac{9}{2},\frac{13}{2}}_{7}=-\frac{4}{297297} (-11+\nu) (-9+\nu) \nu (9+\nu) (11+\nu)\,,\qquad
b^{\frac{11}{2},\frac{5}{2}}_{3}=-\frac{4 \nu}{99}\,,
\nonu\\
&&
b^{\frac{11}{2},\frac{7}{2}}_{3}=-\frac{4 \nu}{195}\,,\qquad
b^{\frac{11}{2},\frac{7}{2}}_{5}=\frac{8}{6237} (-9+\nu) \nu (9+\nu)\,,\qquad
b^{\frac{11}{2},\frac{9}{2}}_{3}=-\frac{4\nu}{315}\,,
\nonu\\
&&
b^{\frac{11}{2},\frac{9}{2}}_{5}=\frac{8}{63063} \nu (-467+3 \nu^2)\,,\qquad
b^{\frac{11}{2},\frac{9}{2}}_{7}=-\frac{4}{72765} (-9+\nu) (-7+\nu) \nu (7+\nu) (9+\nu)\,,
\nonu\\
&&
b^{\frac{11}{2},\frac{11}{2}}_{3}=-\frac{4\nu}{459}\,,\qquad
b^{\frac{11}{2},\frac{11}{2}}_{5}=\frac{8 }{51597}\nu (-249+\nu^2)\,,
\nonu\\
&&
b^{\frac{11}{2},\frac{11}{2}}_{7}=-\frac{4}{1216215} (-9+\nu) \nu (9+\nu) (-467+3 \nu^2)\,,
\nonu\\
&&
b^{\frac{11}{2},\frac{11}{2}}_{9}=\frac{16}{3274425} (-9+\nu) (-7+\nu) (-5+\nu) \nu (5+\nu) (7+\nu) (9+\nu)\,,
b^{\frac{13}{2},\frac{5}{2}}_{3}=-\frac{4 \nu}{143}\,,
\nonu\\
&&
b^{\frac{13}{2},\frac{7}{2}}_{3}=-\frac{4\nu}{275}\,,\qquad
b^{\frac{13}{2},\frac{7}{2}}_{5}=\frac{8}{14157} (-11+\nu) \nu (11+\nu)\,,
b^{\frac{13}{2},\frac{9}{2}}_{3}=-\frac{12\nu}{1309}\,,
\nonu\\
&&
b^{\frac{13}{2},\frac{9}{2}}_{5}=\frac{8}{45045} \nu (-229+\nu^2)\,,
b^{\frac{13}{2},\frac{9}{2}}_{7}=-\frac{4}{297297} (-11+\nu) (-9+\nu) \nu (9+\nu) (11+\nu)\,,
\nonu\\
&&
b^{\frac{15}{2},\frac{5}{2}}_{3}=-\frac{4 \nu}{195}\,,\qquad
b^{\frac{15}{2},\frac{7}{2}}_{3}=-\frac{12\nu}{1105}\,,\qquad
b^{\frac{15}{2},\frac{7}{2}}_{5}=\frac{8}{27885} (-13+\nu) \nu (13+\nu)\,,\qquad
\nonu\\
&&
b^{\frac{17}{2},\frac{5}{2}}_{3}=-\frac{4 \nu}{255}\,.
\label{Bval}
\eea
It is easy to observe that some of the coefficients in
Appendix (\ref{Bval})
are equal to the ones in (\ref{bvalue}).
Let us focus on the cubic term in $\nu$:
the coefficient $b^{\frac{7}{2}, \frac {7} {2}} _ {5}$
in Appendix (\ref{Bval})
is the same as $b^{4, 4} _ {6}$ in (\ref{bvalue}).
Then we can try to determine the explicit form for the
generalized hypergeometric function by following the
procedure around (\ref{bvalue}).
The corresponding quartic terms  $(\nu \mp 3)(\nu \pm 5)
(\nu \mp 5)(\nu \pm 7)$ for the above cubic term can arise in the
generalized hypergeometric functions.
Moreover, 
the sextic  terms $(\nu \pm 3)(\nu \mp 5)(\nu \pm 5)
(\nu \mp 7)(\nu \pm 7)(\nu \mp 9)$
for the above quintic term in Appendix (\ref{Bval}) can arise in the
generalized hypergeometric functions.
Similarly,
the octic  terms $(\nu \pm 3)(\nu \mp 5)(\nu \pm 5)
(\nu \mp 7)(\nu \pm 7)(\nu \mp 9)(\nu \pm 9)(\nu \mp 11)$
 for the above heptic term in Appendix (\ref{Bval}) can arise.
 We need to obtain the explicit forms, by trying to vary the various
 elements in order to match with the odd powers of $\nu$ in
 Appendix (\ref{Bval}), for the
 generalized hypergeometric functions.

 Then we have the following relations from Appendices
 (\ref{Aval}) and (\ref{Bval})
\bea
a^{h_1+\frac{1}{2},h_ 2+\frac{1}{2}} _ {2r-1}(\mu)
& = &
{}_4F_3\left[
\begin{array}{c|}
\frac{1}{2} + \mu \ ,  \frac{1}{2} - \mu  \ , \frac{3-2r}{ 2} \ , \frac{2-2r}{2}\\
\frac{1}{2}-h_1 \ ,  \frac{1}{2}-h_2\ , \frac{5}{2}+ h_1+h_2-2r
\end{array}  \ 1\right]
\nonu \\
& + & {}_4F_3\left[
\begin{array}{c|}
\frac{3}{2} - \mu \ ,  -\frac{1}{2} +\mu  \ , \frac{3-2r}{ 2} \ , \frac{2-2r}{2}\\
\frac{1}{2}-h_1 \ ,  \frac{1}{2}-h_2\ , \frac{5}{2}+ h_1+h_2-2r
\end{array}  \ 1\right]\,,
\nonu\\
b^{h_1+\frac{1}{2},h_ 2+\frac{1}{2}} _ {2r-1}(\mu) 
 & = &
{}_4F_3\left[
\begin{array}{c|}
\frac{1}{2} + \mu \ ,  \frac{1}{2} - \mu  \ , \frac{2-2r}{ 2} \ , \frac{1-2r}{2}\\
\frac{1}{2}-h_1 \ ,  \frac{1}{2}-h_2\ , \frac{5}{2}+ h_1+h_2-2r
\end{array}  \ 1\right]
\nonu \\
& - &
{}_4F_3\left[
\begin{array}{c|}
\frac{3}{2} - \mu \ ,  -\frac{1}{2} +\mu  \ , \frac{2-2r}{ 2} \ , \frac{1-2r}{2}\\
\frac{1}{2}-h_1 \ ,  \frac{1}{2}-h_2\ , \frac{5}{2}+ h_1+h_2-2r
\end{array}  \ 1\right]\,.
\label{appenrela}
\eea
The structure constant $a^{h_1+\frac{1}{2},h_ 2+\frac{1}{2}} _ {2r-1}$
is not ``saalschutzian'' type while
the structure constant $b^{h_1+\frac{1}{2},h_ 2+\frac{1}{2}} _ {2r-1} $
is ``saalschutzian'' type.

We realize that after collecting $(h_1+1)$ and $(h_2+1)$ in the right hand side
of the first relation of Appendix (\ref{appenrela}),
we obtain the first relation
of (\ref{abconstphi}) by a shift $\frac{1}{2}$ in the upper
third and fourth elements of generalized hypergeometric function.
Similarly,
 after collecting $(h_1+1)$ and $(h_2+1)$ in the right hand side
of the second relation of Appendix (\ref{appenrela}),
we obtain the second relation
of (\ref{abconstphi}).
This observation is encoded in (\ref{spf}).
We will see that the above two structure constants in Appendix
(\ref{appenrela}) are denoted by
the third one in (\ref{3struct}).

\subsubsection{Operator $K$ dependence in the oscillators}

We can analyze the other case corresponding to (\ref{IdOddOdd})
as follows.
By introducing the $\nu$ dependent structure constants,
we have 
\bea
&&
\big\{ 
  \underbrace{
\hat{y}_{(1}\ldots\,\hat{y}_{1}
}_{h_1-\frac{1}{2}+\rho} 
\underbrace{
\hat{y}_{2}\ldots\,\hat{y}_{2)}
}_{h_1-\frac{1}{2}-\rho},\,
 \underbrace{
\hat{y}_{(1}\ldots\,\hat{y}_{1}
}_{h_2-\frac{1}{2}+\omega} 
\underbrace{
\hat{y}_{2}\ldots\,\hat{y}_{2)}
}_{h_2-\frac{1}{2}-\omega} K
\big\}
=
 \sum_{r=1}^{[\frac{h_1+h_2}{2}]}\,\mi\,(-1)^r\,
 N^{h_1+\frac{1}{2}\,  h_2+\frac{1}{2}}_{2r}(\rho,\omega)\,
 \nonu \\
 && \times  \Bigg[
-\frac{1}{(2r-1)!}\,
a_{2r}^{h_1+\frac{1}{2}, h_2+\frac{1}{2}}(\nu)
\,  \underbrace{
\hat{y}_{(1}\ldots\ldots\ldots.\hat{y}_{1}
}_{h_1+h_2-2r+\rho+\omega}\,
\underbrace{
\hat{y}_{2}\ldots\ldots\ldots.\hat{y}_{2)}
}_{h_1+h_2-2r-\rho-\omega} K
\nonumber\\
&&
+\frac{2(h_1+h_2-r)+1}{(2r)!}
\,b_{2r}^{h_1+\frac{1}{2}, h_2+\frac{1}{2}}(\nu)
\,  \underbrace{
\hat{y}_{(1}\ldots\ldots\ldots.\hat{y}_{1}
}_{h_1+h_2-2r+\rho+\omega} \,
\underbrace{
\hat{y}_{2}\ldots\ldots\ldots.\hat{y}_{2)}
}_{h_1+h_2-2r-\rho-\omega}
\Bigg].
\label{threeext}
\eea
It turns out that from the analysis of Appendix (\ref{threeext}) 
\bea
&&
a^{\frac{3}{2},\frac{3}{2}}_{2}=2\,,\qquad
a^{\frac{3}{2},\frac{5}{2}}_{2}=2\,,\qquad
a^{\frac{3}{2},\frac{7}{2}}_{2}=2\,,\qquad
a^{\frac{3}{2},\frac{9}{2}}_{2}=2\,,\qquad
a^{\frac{3}{2},\frac{11}{2}}_{2}=2\,,\qquad
a^{\frac{3}{2},\frac{13}{2}}_{2}=2\,,
\nonu\\
&&
a^{\frac{3}{2},\frac{15}{2}}_{2}=2\,,\qquad
a^{\frac{3}{2},\frac{17}{2}}_{2}=2\,,\qquad
a^{\frac{3}{2},\frac{19}{2}}_{2}=2\,,\qquad
a^{\frac{5}{2},\frac{3}{2}}_{2}=2\,,\qquad
a^{\frac{5}{2},\frac{5}{2}}_{2}=2\,,
\nonu\\
&&
a^{\frac{5}{2},\frac{5}{2}}_{4}=-\frac{2}{9} (-3+\nu) (3+\nu)\,,\qquad
a^{\frac{5}{2},\frac{7}{2}}_{2}=2\,,\qquad
a^{\frac{5}{2},\frac{7}{2}}_{4}=-\frac{2}{25} (-5+\nu) (5+\nu)\,,
\nonu\\
&&
a^{\frac{5}{2},\frac{9}{2}}_{2}=2\,,\qquad
a^{\frac{5}{2},\frac{9}{2}}_{4}=-\frac{2}{49} (-7+\nu) (7+\nu)\,,\qquad
a^{\frac{5}{2},\frac{11}{2}}_{2}=2\,,
\nonu\\
&&
a^{\frac{5}{2},\frac{11}{2}}_{4}=-\frac{2}{81} (-9+\nu) (9+\nu)\,,\qquad
a^{\frac{5}{2},\frac{13}{2}}_{2}=2\,,\qquad
a^{\frac{5}{2},\frac{13}{2}}_{4}=-\frac{2}{121} (-11+\nu) (11+\nu)\,,
\nonu\\
&&
a^{\frac{5}{2},\frac{15}{2}}_{2}=2\,,\qquad
a^{\frac{5}{2},\frac{15}{2}}_{4}=-\frac{2}{169} (-13+\nu) (13+\nu)\,,\qquad
a^{\frac{5}{2},\frac{17}{2}}_{2}=2\,,
\nonu\\
&&
a^{\frac{5}{2},\frac{17}{2}}_{4}=-\frac{2}{225} (-15+\nu) (15+\nu)\,,\qquad
a^{\frac{7}{2},\frac{3}{2}}_{2}=2\,,\qquad
a^{\frac{7}{2},\frac{5}{2}}_{2}=2\,,
\nonu\\
&&
a^{\frac{7}{2},\frac{5}{2}}_{4}=-\frac{2}{25} (-5+\nu) (5+\nu)\,,\qquad
a^{\frac{7}{2},\frac{7}{2}}_{2}=2\,,\qquad
a^{\frac{7}{2},\frac{7}{2}}_{4}=-\frac{2}{175} (-175+3 \nu^2)\,,
\nonu\\
&&
a^{\frac{7}{2},\frac{7}{2}}_{6}=\frac{2}{225} (-5+\nu) (-3+\nu) (3+\nu) (5+\nu)\,,\qquad
a^{\frac{7}{2},\frac{9}{2}}_{2}=2\,,
\nonu\\
&&
a^{\frac{7}{2},\frac{9}{2}}_{4}=-\frac{2}{105} (-105+\nu^2)\,,\qquad
a^{\frac{7}{2},\frac{9}{2}}_{6}=\frac{2}{1225} (-7+\nu) (-5+\nu) (5+\nu) (7+\nu)\,,
\nonu\\
&&
a^{\frac{7}{2},\frac{11}{2}}_{2}=2\,,\qquad
a^{\frac{7}{2},\frac{11}{2}}_{4}=-\frac{2}{165} (-165+\nu^2)\,,
\nonu\\
&&
a^{\frac{7}{2},\frac{11}{2}}_{6}=\frac{2}{3969} (-9+\nu) (-7+\nu) (7+\nu) (9+\nu)\,,\qquad
a^{\frac{7}{2},\frac{13}{2}}_{2}=2\,,
\nonu\\
&&
a^{\frac{7}{2},\frac{13}{2}}_{4}=-\frac{2}{715} (-715+3 \nu^2)\,,\qquad
a^{\frac{7}{2},\frac{13}{2}}_{6}=\frac{2}{9801} (-11+\nu) (-9+\nu) (9+\nu) (11+\nu)\,,
\nonu\\
&&
a^{\frac{7}{2},\frac{15}{2}}_{2}=2\,,\qquad
a^{\frac{7}{2},\frac{15}{2}}_{4}=-\frac{2}{325} (-325+\nu^2)\,,
\nonu\\
&&
a^{\frac{7}{2},\frac{15}{2}}_{6}=\frac{2}{20449} (-13+\nu) (-11+\nu) (11+\nu) (13+\nu)\,,\qquad
a^{\frac{9}{2},\frac{3}{2}}_{2}=2\,,\qquad
a^{\frac{9}{2},\frac{5}{2}}_{2}=2\,,
\nonu\\
&&
a^{\frac{9}{2},\frac{5}{2}}_{4}=-\frac{2}{49} (-7+\nu) (7+\nu)\,,\qquad
a^{\frac{9}{2},\frac{7}{2}}_{2}=2\,,\qquad
a^{\frac{9}{2},\frac{7}{2}}_{4}=-\frac{2}{105} (-105+\nu^2)\,,
\nonu\\
&&
a^{\frac{9}{2},\frac{7}{2}}_{6}=\frac{2}{1225} (-7+\nu) (-5+\nu) (5+\nu) (7+\nu)\,,\qquad
a^{\frac{9}{2},\frac{9}{2}}_{2}=2\,,
\nonu\\
&&
a^{\frac{9}{2},\frac{9}{2}}_{4}=-\frac{2}{539} (-539+3 \nu^2)\,,\qquad
a^{\frac{9}{2},\frac{9}{2}}_{6}=\frac{2}{5145} (-7+\nu) (7+\nu) (-105+\nu^2)\,,
\nonu\\
&&
a^{\frac{9}{2},\frac{9}{2}}_{8}=-\frac{2}{11025} (-7+\nu) (-5+\nu) (-3+\nu) (3+\nu) (5+\nu) (7+\nu)\,,\qquad
a^{\frac{9}{2},\frac{11}{2}}_{2}=2\,,
\nonu\\
&&
a^{\frac{9}{2},\frac{11}{2}}_{4}=-\frac{2}{273} (-273+\nu^2)\,,\qquad
a^{\frac{9}{2},\frac{11}{2}}_{6}=\frac{2}{43659} (-9+\nu) (9+\nu) (-539+3 \nu^2)\,,
\nonu \\
%
&&
a^{\frac{9}{2},\frac{11}{2}}_{8}=-\frac{2}{99225} (-9+\nu) (-7+\nu) (-5+\nu) (5+\nu) (7+\nu) (9+\nu)\,,\qquad
a^{\frac{9}{2},\frac{13}{2}}_{2}=2\,,
\nonu\\
&&
a^{\frac{9}{2},\frac{13}{2}}_{4}=-\frac{2}{385} (-385+\nu^2)\,,\qquad
a^{\frac{9}{2},\frac{13}{2}}_{6}=\frac{2}{33033} (-11+\nu) (11+\nu) (-273+\nu^2)\,,
\nonu\\
&&
a^{\frac{9}{2},\frac{13}{2}}_{8}=-\frac{2}{480249} (-11+\nu) (-9+\nu) (-7+\nu) (7+\nu) (9+\nu) (11+\nu)\,,\qquad
a^{\frac{11}{2},\frac{3}{2}}_{2}=2\,,
\nonu\\
&&
a^{\frac{11}{2},\frac{5}{2}}_{2}=2\,,\qquad
a^{\frac{11}{2},\frac{5}{2}}_{4}=-\frac{2}{81} (-9+\nu) (9+\nu)\,,\qquad
a^{\frac{11}{2},\frac{7}{2}}_{2}=2\,,
\nonu\\
&&
a^{\frac{11}{2},\frac{7}{2}}_{4}=-\frac{2}{165} (-165+\nu^2)\,,\qquad
a^{\frac{11}{2},\frac{7}{2}}_{6}=\frac{2}{3969} (-9+\nu) (-7+\nu) (7+\nu) (9+\nu)\,,
\nonu\\
&&
a^{\frac{11}{2},\frac{9}{2}}_{2}=2\,,\qquad
a^{\frac{11}{2},\frac{9}{2}}_{4}=-\frac{2}{273} (-273+\nu^2)\,,
\nonu\\
&&
a^{\frac{11}{2},\frac{9}{2}}_{6}=\frac{2}{43659} (-9+\nu) (9+\nu) (-539+3 \nu^2)\,,
\nonu\\
&&
a^{\frac{11}{2},\frac{9}{2}}_{8}=-\frac{2}{99225} (-9+\nu) (-7+\nu) (-5+\nu) (5+\nu) (7+\nu) (9+\nu)\,,\qquad
a^{\frac{11}{2},\frac{11}{2}}_{2}=2\,,
\nonu\\
&&
a^{\frac{11}{2},\frac{11}{2}}_{4}=-\frac{2}{405} (-405+\nu^2)\,,\qquad
a^{\frac{11}{2},\frac{11}{2}}_{6}=\frac{2}{567567} (567567-6430 \nu^2+15 \nu^4)\,,
\nonu\\
&&
a^{\frac{11}{2},\frac{11}{2}}_{8}=-\frac{2}{654885} (-9+\nu) (-7+\nu) (7+\nu) (9+\nu) (-165+\nu^2)\,,
\nonu\\
&&
a^{\frac{11}{2},\frac{11}{2}}_{10}=\frac{2}{893025} (-9+\nu) (-7+\nu) (-5+\nu) (-3+\nu) (3+\nu) (5+\nu) (7+\nu) (9+\nu)\,,
\nonu\\
&&
a^{\frac{13}{2},\frac{3}{2}}_{2}=2\,,\qquad
a^{\frac{13}{2},\frac{5}{2}}_{2}=2\,,\qquad
a^{\frac{13}{2},\frac{5}{2}}_{4}=-\frac{2}{121} (-11+\nu) (11+\nu)\,,\qquad
a^{\frac{13}{2},\frac{7}{2}}_{2}=2\,,
\nonu\\
&&
a^{\frac{13}{2},\frac{7}{2}}_{4}=-\frac{2}{715} (-715+3 \nu^2)\,,\qquad
a^{\frac{13}{2},\frac{7}{2}}_{6}=\frac{2}{9801} (-11+\nu) (-9+\nu) (9+\nu) (11+\nu)\,,
\nonu\\
&&
a^{\frac{13}{2},\frac{9}{2}}_{2}=2\,,\qquad
a^{\frac{13}{2},\frac{9}{2}}_{4}=-\frac{2}{385} (-385+\nu^2)\,,
\nonu\\
&&
a^{\frac{13}{2},\frac{9}{2}}_{6}=\frac{2}{33033} (-11+\nu) (11+\nu) (-273+\nu^2)\,,
\nonu\\
&&
a^{\frac{13}{2},\frac{9}{2}}_{8}=-\frac{2}{480249} (-11+\nu) (-9+\nu) (-7+\nu) (7+\nu) (9+\nu) (11+\nu)\,,\qquad
a^{\frac{15}{2},\frac{3}{2}}_{2}=2\,,
\nonu\\
&&
a^{\frac{15}{2},\frac{5}{2}}_{2}=2\,,\qquad
a^{\frac{15}{2},\frac{5}{2}}_{4}=-\frac{2}{169} (-13+\nu) (13+\nu)\,,\qquad
a^{\frac{15}{2},\frac{7}{2}}_{2}=2\,,
\nonu\\
&&
a^{\frac{15}{2},\frac{7}{2}}_{4}=-\frac{2}{325} (-325+\nu^2)\,,\qquad
a^{\frac{15}{2},\frac{7}{2}}_{6}=\frac{2}{20449} (-13+\nu) (-11+\nu) (11+\nu) (13+\nu)\,,
\nonu\\
&&
a^{\frac{17}{2},\frac{3}{2}}_{2}=2\,,\qquad
a^{\frac{17}{2},\frac{5}{2}}_{2}=2\,,\qquad
a^{\frac{17}{2},\frac{5}{2}}_{4}=-\frac{2}{225} (-15+\nu) (15+\nu)\,,\,
a^{\frac{19}{2},\frac{3}{2}}_{2}=2\,.
\label{Ava1}
\eea
Some of the coefficients in Appendix (\ref{Ava1})
overlap with the ones in Appendix (\ref{aval}).
For example, the coefficient
$a^{\frac{5}{2},\frac{7}{2}}_{4}$ of Appendix (\ref{Ava1}) is exactly the same as
the coefficient $a^{3,\frac{7}{2}}_{4}$ of Appendix (\ref{aval}).
Note that there is a shift in $h_1$.
The explicit $\nu$
dependence in the generalized hypergeometric functions can be
determined similarly.

Furthermore we have
\bea
&&
b^{\frac{3}{2},\frac{3}{2}}_{2}=-\frac{4 \nu}{3}\,,\qquad
b^{\frac{3}{2},\frac{5}{2}}_{2}=-\frac{4 \nu}{15}\,,\qquad
b^{\frac{3}{2},\frac{7}{2}}_{2}=-\frac{4 \nu}{35}\,,\qquad
b^{\frac{3}{2},\frac{9}{2}}_{2}=-\frac{4 \nu}{63}\,,\qquad
b^{\frac{3}{2},\frac{11}{2}}_{2}=-\frac{4 \nu}{99}\,,
\nonu\\
&&
b^{\frac{3}{2},\frac{13}{2}}_{2}=-\frac{4 \nu}{143}\,,\qquad
b^{\frac{3}{2},\frac{15}{2}}_{2}=-\frac{4 \nu}{195}\,,\qquad
b^{\frac{3}{2},\frac{17}{2}}_{2}=-\frac{4 \nu}{255}\,,\qquad
b^{\frac{3}{2},\frac{19}{2}}_{2}=-\frac{4 \nu}{323}\,,
\nonu\\
&&
b^{\frac{5}{2},\frac{3}{2}}_{2}=-\frac{4 \nu}{15}\,,\qquad
b^{\frac{5}{2},\frac{5}{2}}_{2}=-\frac{4 \nu}{63}\,,\qquad
b^{\frac{5}{2},\frac{5}{2}}_{4}=\frac{8}{45} (-3+\nu) \nu (3+\nu)\,,\qquad
b^{\frac{5}{2},\frac{7}{2}}_{2}=-\frac{4 \nu}{135}\,,
\nonu\\
&&
b^{\frac{5}{2},\frac{7}{2}}_{4}=\frac{8}{525} (-5+\nu) \nu (5+\nu)\,,\qquad
b^{\frac{5}{2},\frac{9}{2}}_{2}=-\frac{4 \nu}{231}\,,\qquad
b^{\frac{5}{2},\frac{9}{2}}_{4}=\frac{8}{2205} (-7+\nu) \nu (7+\nu)\,,
\nonu\\
&&
b^{\frac{5}{2},\frac{11}{2}}_{2}=-\frac{4 \nu}{351}\,,\qquad
b^{\frac{5}{2},\frac{11}{2}}_{4}=\frac{8}{6237} (-9+\nu) \nu (9+\nu)\,,\qquad
b^{\frac{5}{2},\frac{13}{2}}_{2}=-\frac{4 \nu}{495}\,,
\nonu\\
&&
b^{\frac{5}{2},\frac{13}{2}}_{4}=\frac{8}{14157} (-11+\nu) \nu (11+\nu)\,,\qquad
b^{\frac{5}{2},\frac{15}{2}}_{2}=-\frac{4 \nu}{663}\,,
\nonu\\
&&
b^{\frac{5}{2},\frac{15}{2}}_{4}=\frac{8}{27885} (-13+\nu) \nu (13+\nu)\,,\qquad
b^{\frac{5}{2},\frac{17}{2}}_{2}=-\frac{4 \nu}{855}\,,
\nonu\\
&&
b^{\frac{5}{2},\frac{17}{2}}_{4}=\frac{8}{49725} (-15+\nu) \nu (15+\nu)\,,\qquad
b^{\frac{7}{2},\frac{3}{2}}_{2}=-\frac{4 \nu}{35}\,,\qquad
b^{\frac{7}{2},\frac{5}{2}}_{2}=-\frac{4 \nu}{135}\,,
\nonu\\
&&
b^{\frac{7}{2},\frac{5}{2}}_{4}=\frac{8}{525} (-5+\nu) \nu (5+\nu)\,,\qquad
b^{\frac{7}{2},\frac{7}{2}}_{2}=-\frac{4 \nu}{275}\,,\qquad
b^{\frac{7}{2},\frac{7}{2}}_{4}=\frac{8}{4725} \nu (-85+\nu^2)\,,
\nonu\\
&&
b^{\frac{7}{2},\frac{7}{2}}_{6}=-\frac{4}{525} (-5+\nu) (-3+\nu) \nu (3+\nu) (5+\nu)\,,\qquad
b^{\frac{7}{2},\frac{9}{2}}_{2}=-\frac{4 \nu}{455}\,,
\nonu\\
&&
b^{\frac{7}{2},\frac{9}{2}}_{4}=\frac{8}{17325} (-13+\nu) \nu (13+\nu)\,,
b^{\frac{7}{2},\frac{9}{2}}_{6}=-\frac{4}{11025} (-7+\nu) (-5+\nu) \nu (5+\nu) (7+\nu)\,,
\nonu\\
&&
b^{\frac{7}{2},\frac{11}{2}}_{2}=-\frac{4 \nu}{675}\,,\qquad
b^{\frac{7}{2},\frac{11}{2}}_{4}=\frac{8}{45045} \nu (-277+\nu^2)\,,
\nonu\\
&&
b^{\frac{7}{2},\frac{11}{2}}_{6}=-\frac{4}{72765} (-9+\nu) (-7+\nu) \nu (7+\nu) (9+\nu)\,,\qquad
b^{\frac{7}{2},\frac{13}{2}}_{2}=-\frac{4 \nu}{935}\,,
\nonu\\
&&
b^{\frac{7}{2},\frac{13}{2}}_{4}=\frac{8}{96525} \nu (-409+\nu^2)\,,
\nonu\\
&&
b^{\frac{7}{2},\frac{13}{2}}_{6}=-\frac{4}{297297} (-11+\nu) (-9+\nu) \nu (9+\nu) (11+\nu)\,,\qquad
b^{\frac{7}{2},\frac{15}{2}}_{2}=-\frac{4 \nu}{1235}\,,
\nonu\\
&&
b^{\frac{7}{2},\frac{15}{2}}_{4}=\frac{8}{182325} \nu (-565+\nu^2)\,,
\nonu\\
&&
b^{\frac{7}{2},\frac{15}{2}}_{6}=-\frac{4}{920205} (-13+\nu) (-11+\nu) \nu (11+\nu) (13+\nu)\,,\qquad
b^{\frac{9}{2},\frac{3}{2}}_{2}=-\frac{4 \nu}{63}\,,
\nonu\\
&&
b^{\frac{9}{2},\frac{5}{2}}_{2}=-\frac{4 \nu}{231}\,,\qquad
b^{\frac{9}{2},\frac{5}{2}}_{4}=\frac{8}{2205} (-7+\nu) \nu (7+\nu)\,,\qquad
b^{\frac{9}{2},\frac{7}{2}}_{2}=-\frac{4 \nu}{455}\,,
\nonu\\
&&
b^{\frac{9}{2},\frac{7}{2}}_{4}=\frac{8}{17325} (-13+\nu) \nu (13+\nu)\,,
b^{\frac{9}{2},\frac{7}{2}}_{6}=-\frac{4}{11025} (-7+\nu) (-5+\nu) \nu (5+\nu) (7+\nu)\,,
\nonu\\
&&
b^{\frac{9}{2},\frac{9}{2}}_{2}=-\frac{4 \nu}{735}\,,\qquad
b^{\frac{9}{2},\frac{9}{2}}_{4}=\frac{24}{175175} \nu (-329+\nu^2)\,,
\nonu \\
%
&&
b^{\frac{9}{2},\frac{9}{2}}_{6}=-\frac{4}{169785} (-13+\nu) (-7+\nu) \nu (7+\nu) (13+\nu)\,,
\nonu\\
&&
b^{\frac{9}{2},\frac{9}{2}}_{8}=\frac{16}{99225} (-7+\nu) (-5+\nu) (-3+\nu) \nu (3+\nu) (5+\nu) (7+\nu)\,,\qquad
b^{\frac{9}{2},\frac{11}{2}}_{2}=-\frac{4 \nu}{1071}\,,
\nonu\\
&&
b^{\frac{9}{2},\frac{11}{2}}_{4}=\frac{8}{143325} (-23+\nu) \nu (23+\nu)\,,
\nonu\\
&&
b^{\frac{9}{2},\frac{11}{2}}_{6}=-\frac{4}{945945} (-9+\nu) \nu (9+\nu) (-329+\nu^2)\,,
\nonu\\
&&
b^{\frac{9}{2},\frac{11}{2}}_{8}=\frac{16}{3274425} (-9+\nu) (-7+\nu) (-5+\nu) \nu (5+\nu) (7+\nu) (9+\nu)\,,
b^{\frac{9}{2},\frac{13}{2}}_{2}=-\frac{4 \nu}{1463}\,,
\nonu\\
&&
b^{\frac{9}{2},\frac{13}{2}}_{4}=\frac{8}{294525} \nu (-769+\nu^2)\,,
\nonu\\
&&
b^{\frac{9}{2},\frac{13}{2}}_{6}=-\frac{4}{3468465} (-23+\nu) (-11+\nu) \nu (11+\nu) (23+\nu)\,,
\nonu\\
&&
b^{\frac{9}{2},\frac{13}{2}}_{8}=\frac{16}{31216185} (-11+\nu) (-9+\nu) (-7+\nu) \nu (7+\nu) (9+\nu) (11+\nu)\,,
b^{\frac{11}{2},\frac{3}{2}}_{2}=-\frac{4 \nu}{99}\,,
\nonu\\
&&
b^{\frac{11}{2},\frac{5}{2}}_{2}=-\frac{4 \nu}{351}\,,\qquad
b^{\frac{11}{2},\frac{5}{2}}_{4}=\frac{8}{6237} (-9+\nu) \nu (9+\nu)\,,\qquad
b^{\frac{11}{2},\frac{7}{2}}_{2}=-\frac{4 \nu}{675}\,,
\nonu\\
&&
b^{\frac{11}{2},\frac{7}{2}}_{4}=\frac{8}{45045} \nu (-277+\nu^2)\,,\qquad
b^{\frac{11}{2},\frac{7}{2}}_{6}=-\frac{4}{72765} (-9+\nu) (-7+\nu) \nu (7+\nu) (9+\nu)\,,
\nonu\\
&&
b^{\frac{11}{2},\frac{9}{2}}_{2}=-\frac{4 \nu}{1071}\,,\qquad
b^{\frac{11}{2},\frac{9}{2}}_{4}=\frac{8}{143325} (-23+\nu) \nu (23+\nu)\,,
\nonu\\
&&
b^{\frac{11}{2},\frac{9}{2}}_{6}=-\frac{4}{945945} (-9+\nu) \nu (9+\nu) (-329+\nu^2)\,,
\nonu\\
&&
b^{\frac{11}{2},\frac{9}{2}}_{8}=\frac{16}{3274425} (-9+\nu) (-7+\nu) (-5+\nu) \nu (5+\nu) (7+\nu) (9+\nu)\,,
b^{\frac{11}{2},\frac{11}{2}}_{2}=-\frac{4 \nu}{1539}\,,
\nonu\\
&&
b^{\frac{11}{2},\frac{11}{2}}_{4}=\frac{8}{337365} \nu (-837+\nu^2)\,,\qquad
b^{\frac{11}{2},\frac{11}{2}}_{6}=-\frac{4}{4729725} \nu (-641+\nu^2) (-129+\nu^2)\,,
\nonu\\
&&
b^{\frac{11}{2},\frac{11}{2}}_{8}=\frac{16}{76621545} (-9+\nu) (-7+\nu) \nu (7+\nu) (9+\nu) (-277+\nu^2)\,,
\nonu\\
&&
b^{\frac{11}{2},\frac{11}{2}}_{10}=-\frac{4}{1964655} (-9+\nu) (-7+\nu) (-5+\nu) (-3+\nu) \nu (3+\nu) (5+\nu) (7+\nu) (9+\nu)\,,
\nonu\\
&&
b^{\frac{13}{2},\frac{3}{2}}_{2}=-\frac{4 \nu}{143}\,,\qquad
b^{\frac{13}{2},\frac{5}{2}}_{2}=-\frac{4 \nu}{495}\,,\qquad
b^{\frac{13}{2},\frac{5}{2}}_{4}=\frac{8}{14157} (-11+\nu) \nu (11+\nu)\,,
\nonu\\
&&
b^{\frac{13}{2},\frac{7}{2}}_{2}=-\frac{4 \nu}{935}\,,\qquad
b^{\frac{13}{2},\frac{7}{2}}_{4}=\frac{8}{96525} \nu (-409+\nu^2)\,,
\nonu\\
&&
b^{\frac{13}{2},\frac{7}{2}}_{6}=-\frac{4}{297297} (-11+\nu) (-9+\nu) \nu (9+\nu) (11+\nu)\,,\qquad
b^{\frac{13}{2},\frac{9}{2}}_{2}=-\frac{4 \nu}{1463}\,,
\nonu\\
&&
b^{\frac{13}{2},\frac{9}{2}}_{4}=\frac{8}{294525} \nu (-769+\nu^2)\,,
\nonu \\
%
%
&&
b^{\frac{13}{2},\frac{9}{2}}_{6}=-\frac{4}{3468465} (-23+\nu) (-11+\nu) \nu (11+\nu) (23+\nu)\,,
\nonu\\
&&
b^{\frac{13}{2},\frac{9}{2}}_{8}=\frac{16}{31216185} (-11+\nu) (-9+\nu) (-7+\nu) \nu (7+\nu) (9+\nu) (11+\nu)\,,
\nonu\\
&&
b^{\frac{15}{2},\frac{3}{2}}_{2}=-\frac{4 \nu}{195}\,,\qquad
b^{\frac{15}{2},\frac{5}{2}}_{2}=-\frac{4 \nu}{663}\,,\qquad
b^{\frac{15}{2},\frac{5}{2}}_{4}=\frac{8}{27885} (-13+\nu) \nu (13+\nu)\,,
\nonu\\
&&
b^{\frac{15}{2},\frac{7}{2}}_{2}=-\frac{4 \nu}{1235}\,,\qquad
b^{\frac{15}{2},\frac{7}{2}}_{4}=\frac{8}{182325} \nu (-565+\nu^2)\,,
\nonu\\
&&
b^{\frac{15}{2},\frac{7}{2}}_{6}=-\frac{4}{920205} (-13+\nu) (-11+\nu) \nu (11+\nu) (13+\nu)\,,
b^{\frac{17}{2},\frac{3}{2}}_{2}=-\frac{4 \nu}{255}\,,
\nonu\\
&&
b^{\frac{17}{2},\frac{5}{2}}_{2}=-\frac{4 \nu}{855}\,,\qquad
b^{\frac{17}{2},\frac{5}{2}}_{4}=\frac{8}{49725} (-15+\nu) \nu (15+\nu)\,,\qquad
b^{\frac{19}{2},\frac{3}{2}}_{2}=-\frac{4 \nu}{323}\,.
\label{Bva1}
\eea
Some of the coefficients in Appendix (\ref{Bva1})
overlap with the ones in  (\ref{bvalue}).
For example, the coefficient
$b^{\frac{5}{2},\frac{7}{2}}_{4}$ of Appendix (\ref{Bva1}) is exactly the same as
the coefficient $b^{4,4}_{6}$ of  (\ref{bvalue}).
The explicit $\nu$
dependence in the generalized hypergeometric functions can be
determined similarly.

From Appendices (\ref{Ava1}) and (\ref{Bva1}),
we obtain
\bea
a^{h_1+\frac{1}{2},h_ 2+\frac{1}{2}} _ {2r}(\mu)
& = &
{}_4F_3\left[
\begin{array}{c|}
\frac{1}{2} + \mu \ ,  \frac{1}{2} - \mu  \ , \frac{2-2r}{ 2} \ , \frac{1-2r}{2}\\
\frac{1}{2}-h_1 \ ,  \frac{1}{2}-h_2\ , \frac{3}{2}+ h_1+h_2-2r
\end{array}  \ 1\right]
\nonu \\
& + & {}_4F_3\left[
\begin{array}{c|}
\frac{3}{2} - \mu \ ,  -\frac{1}{2} +\mu  \ , \frac{2-2r}{ 2} \ , \frac{1-2r}{2}\\
\frac{1}{2}-h_1 \ ,  \frac{1}{2}-h_2\ , \frac{3}{2}+ h_1+h_2-2r
\end{array}  \ 1\right]\,,
\nonu\\
b^{h_1+\frac{1}{2},h_ 2+\frac{1}{2}} _ {2r}(\mu) 
& = &
{}_4F_3\left[
\begin{array}{c|}
\frac{1}{2} + \mu \ ,  \frac{1}{2} - \mu  \ , \frac{1-2r}{ 2} \ , \frac{-2r}{2}\\
\frac{1}{2}-h_1 \ ,  \frac{1}{2}-h_2\ , \frac{3}{2}+ h_1+h_2-2r
\end{array}  \ 1\right]
\nonu \\
& - &
{}_4F_3\left[
\begin{array}{c|}
\frac{3}{2} - \mu \ ,  -\frac{1}{2} +\mu  \ , \frac{1-2r}{ 2} \ , \frac{-2r}{2}\\
\frac{1}{2}-h_1 \ ,  \frac{1}{2}-h_2\ , \frac{3}{2}+ h_1+h_2-2r
\end{array}  \ 1\right]\,.
\label{threeextra}
\eea
It is easy to see that if we replace $2r$ by $(2r-1)$, then the above
Appendix
(\ref{threeextra}) becomes the ones in Appendix (\ref{appenrela}). 
This can be read from (\ref{IdOddOdd-1}) and (\ref{IdOddOdd}).

\section{
The (anti)commutators between the oscillators with nonzero
$\mu$  in terms of
generalized hypergeometric functions
}

$\bullet$ The case-one with two bosonic oscillators

We summarize the result (\ref{first}) with (\ref{abconstphi})
as follows:
\bea
&&\big[ 
  \underbrace{
\hat{y}_{(1}\ldots\,\hat{y}_{1}
}_{h_1-1+m} 
\underbrace{
\hat{y}_{2}\ldots\,\hat{y}_{2)}
}_{h_1-1-m},\,
 \underbrace{
\hat{y}_{(1}\ldots\,\hat{y}_{1}
}_{h_2-1+n} 
\underbrace{
\hat{y}_{2}\ldots\,\hat{y}_{2)}
}_{h_2-1-n}
\big]
=
\nonu \\
&& -\mi \, \sum_{r=1}^{[\frac{h_1+h_2-1}{2}]} \,(-1)^r \,\frac{1}{(2r-1)!}\,N_{2r}^{h_1, h_2}(m,n) 
\Bigg[
\,
\Bigg( {}
_4F_3\left[
\begin{array}{c|}
\frac{1}{2}+\mu  , \frac{1}{2}-\mu  , \frac{-2r+1}{2}  ,\frac{-2r+2}{2}\\
\frac{3}{2}-h_1 ,  \frac{3}{2}-h_2 ,  h_1+h_2-2r+\frac{1}{2}
\end{array}   1\right]
\nonu \\
&& +\,_4F_3\left[
\begin{array}{c|}
\frac{3}{2}-\mu  , -\frac{1}{2}+\mu  , \frac{-2r+1}{2}  ,\frac{-2r+2}{2}\\
\frac{3}{2}-h_1 ,  \frac{3}{2}-h_2 ,  h_1+h_2-2r+\frac{1}{2}
\end{array}   1\right]
\Bigg)
\, \underbrace{
\hat{y}_{(1}\ldots\ldots\ldots\ldots\hat{y}_{1}
}_{h_1+h_2-2r-1+m+n} \,
\underbrace{
\hat{y}_{2}\ldots\ldots\ldots\ldots\hat{y}_{2)}
}_{h_1+h_2-2r-1-m-n}
\nonumber\\
&&
-
\Bigg( {}
_4F_3\left[
\begin{array}{c|}
\frac{1}{2}+\mu  , \frac{1}{2}-\mu  , \frac{-2r+1}{2}  ,\frac{-2r+2}{2}\\
\frac{3}{2}-h_1 ,  \frac{3}{2}-h_2 ,  h_1+h_2-2r+\frac{1}{2}
\end{array}   1\right]
\nonu \\
&& -\,_4F_3\left[
\begin{array}{c|}
\frac{3}{2}-\mu  , -\frac{1}{2}+\mu  , \frac{-2r+1}{2}  ,\frac{-2r+2}{2}\\
\frac{3}{2}-h_1 ,  \frac{3}{2}-h_2 ,  h_1+h_2-2r+\frac{1}{2}
\end{array}   1\right]
\Bigg)\,
\nonu \\
&& \times \underbrace{
\hat{y}_{(1}\ldots\ldots\ldots\ldots\hat{y}_{1}
}_{h_1+h_2-2r-1+m+n} \,
\underbrace{
\hat{y}_{2}\ldots\ldots\ldots\ldots\hat{y}_{2)}
}_{h_1+h_2-2r-1-m-n}K
\Bigg]
\,,
\label{IdEvenEven4F3}
\eea
which can be led to
(\ref{IdEvenEven}) and moreover reduces to Appendix
(\ref{IdEvenEven4F3Odake})
for vanishing $\mu$.


$\bullet$ The case-two with one bosonic and one fermionic oscillators

From the result of Appendix (\ref{secondabconstphi}),
we can present Appendix (\ref{casetwoapp})
as follows:
\bea
&&\big[ 
  \underbrace{
\hat{y}_{(1}\ldots\,\hat{y}_{1}
}_{h_1-1+m} 
\underbrace{
\hat{y}_{2}\ldots\,\hat{y}_{2)}
}_{h_1-1-m},\,
 \underbrace{
\hat{y}_{(1}\ldots\,\hat{y}_{1}
}_{h_2-\frac{1}{2}+\rho} 
\underbrace{
\hat{y}_{2}\ldots\,\hat{y}_{2)}
}_{h_2-\frac{1}{2}-\rho}
\big]
=
\nonu \\
&& \sum_{r=1}^{[\frac{h_1+h_2-1}{2}]}\Bigg[
\,
\frac{-\mi\,(-1)^r }{(2r-1)!}\,
N_{2r}^{h_1, h_2+\frac{1}{2}}(m,\rho) 
\,
\Bigg( {}
_4F_3\left[
\begin{array}{c|}
\frac{1}{2}+\mu  , \frac{1}{2}-\mu  , \frac{-2r+1}{2}  ,\frac{-2r+2}{2}\\
\frac{3}{2}-h_1 ,  \frac{1}{2}-h_2 ,  h_1+h_2-2r+\frac{1}{2}
\end{array}   1\right]
\nonu \\
&& +\,_4F_3\left[
\begin{array}{c|}
\frac{3}{2}-\mu  , -\frac{1}{2}+\mu  , \frac{-2r+1}{2}  ,\frac{-2r+2}{2}\\
\frac{3}{2}-h_1 ,  \frac{1}{2}-h_2 ,  h_1+h_2-2r+\frac{1}{2}
\end{array}   1\right]
\Bigg)
\,
\underbrace{
\hat{y}_{(1}\ldots\ldots\ldots\ldots\hat{y}_{1}
}_{h_1+h_2-2r-\frac{1}{2}+m+\rho} \,
\underbrace{
\hat{y}_{2}\ldots\ldots\ldots\ldots\hat{y}_{2)}
}_{h_1+h_2-2r-\frac{1}{2}-m-\rho} \Bigg]
\nonumber\\
&&
+
 \sum_{r=1}^{[\frac{h_1+h_2}{2}]}\Bigg[
\,
\frac{(-1)^r\,2(h_1-r)}{(2r-1)!}\,
N_{2r-1}^{h_1, h_2+\frac{1}{2}}(m,\rho) 
\,
\Bigg({}
_4F_3\left[
\begin{array}{c|}
\frac{1}{2}+\mu  , \frac{1}{2}-\mu  , \frac{-2r+1}{2}  ,\frac{-2r+2}{2}\\
\frac{3}{2}-h_1 ,  \frac{1}{2}-h_2 ,  h_1+h_2-2r+\frac{3}{2}
\end{array}   1\right]
\nonu \\
&& -\,_4F_3\left[
\begin{array}{c|}
\frac{3}{2}-\mu  , -\frac{1}{2}+\mu  , \frac{-2r+1}{2}  ,\frac{-2r+2}{2}\\
\frac{3}{2}-h_1 ,  \frac{1}{2}-h_2 ,  h_1+h_2-2r+\frac{3}{2}
\end{array}   1\right]
\Bigg)
\,
\nonu \\
&& \times \underbrace{
\hat{y}_{(1}\ldots\ldots\ldots\ldots\hat{y}_{1}
}_{h_1+h_2-2r+\frac{1}{2}+m+\rho} \,
\underbrace{
\hat{y}_{2}\ldots\ldots\ldots\ldots\hat{y}_{2)}
}_{h_1+h_2-2r+\frac{1}{2}-m-\rho}K
\Bigg].
\label{IdEvenOdd4F3-1}
\eea
This can be rewritten as (\ref{IdEvenOdd-1}).

In order to reproduce Appendix (\ref{case2other}) from
Appendix (\ref{IdEvenOdd4F3-1})
by taking $\mu \rightarrow 0 $, we consider
the following expression
\bea
&&
\frac{1}{(2r)!}\,
\Bigg( (2r+1)\,
_4F_3\left[
\begin{array}{c|}
\frac{1}{2} \ , \frac{1}{2} \ , \frac{-2r}{2} \ ,\frac{-2r+1}{2}\\
\frac{3}{2}-h_1\ ,  \frac{1}{2}-h_2\ ,  h_1+h_2-2r+\frac{1}{2}
\end{array}  \ 1\right]
\nonu\\
&&
\hspace{1cm}
+(2r-1)\,
_4F_3\left[
\begin{array}{c|}
\frac{3}{2} \ , -\frac{1}{2} \ , \frac{-2r}{2} \ ,\frac{-2r+1}{2}\\
\frac{3}{2}-h_1\ ,  \frac{1}{2}-h_2\ ,  h_1+h_2-2r+\frac{1}{2}
\end{array}  \ 1\right]
\Bigg),
\label{combi1}
\eea
which
contains
\bea
\frac{(2r+1)}{(2r)!}
(\tfrac{1}{2})_n(\tfrac{1}{2})_n(-r)_n(\tfrac{-2r+1}{2})_n
\,+
\frac{(2r-1)}{(2r)!}
(\tfrac{3}{2})_n(\tfrac{-1}{2})_n(-r)_n(\tfrac{-2r+1}{2})_n\,.
\label{term1}
\eea
Here we take $(n+1)$-th term of generalized hypergeometric function.
On the other hand,
the following combination  
\bea
&&
\frac{1}{(2r-1)!}\,
\Bigg( {}
_4F_3\left[
\begin{array}{c|}
\frac{1}{2} \ , \frac{1}{2} \ , \frac{-2r+1}{2} \ ,\frac{-2r+2}{2}\\
\frac{3}{2}-h_1\ ,  \frac{1}{2}-h_2\ ,  h_1+h_2-2r+\frac{1}{2}
\end{array}  \ 1\right]
\nonu \\
&& +\,_4F_3\left[
\begin{array}{c|}
\frac{3}{2} \ , -\frac{1}{2} \ , \frac{-2r+1}{2} \ ,\frac{-2r+2}{2}\\
\frac{3}{2}-h_1\ ,  \frac{1}{2}-h_2\ ,  h_1+h_2-2r+\frac{1}{2}
\end{array}  \ 1\right]
\Bigg)
\label{combi2}
\eea
contains 
\bea
\frac{1}{(2r-1)!}
(\tfrac{1}{2})_n(\tfrac{1}{2})_n(\tfrac{-2r+1}{2})_n(1-r)_n
\,+
\frac{1}{(2r-1)!}
(\tfrac{3}{2})_n(\tfrac{-1}{2})_n(\tfrac{-2r+1}{2})_n(1-r)_n\,
\label{term2}
\eea
%
in the $(n+1)$-th term.
Note that the lower three elements in the generalized hypergeometric
functions in Appendix (\ref{combi1}) and Appendix
(\ref{combi2}) are common.  

Then the relations Appendix (\ref{term1}) and Appendix
(\ref{term2})
are equivalent to each other
\bea
\Big( (2r+1)(\tfrac{1}{2})_n(\tfrac{1}{2})_n  +(2r-1)(\tfrac{-1}{2})_n (\tfrac{3}{2})_n\Big) (-r)_n
=2r\Big((\tfrac{1}{2})_n(\tfrac{1}{2})_n  +(\tfrac{-1}{2})_n (\tfrac{3}{2})_n \Big)(1-r)_n\,.
\label{iden}
\eea
It is easy to check this identity Appendix (\ref{iden}) from 
\bea
(\tfrac{1}{2})_n& = & -2(n-\tfrac{1}{2}) (\tfrac{-1}{2})_n\,,
\qquad
(\tfrac{3}{2})_n=2(n+\tfrac{1}{2}) (\tfrac{1}{2})_n\,,
\nonu \\
(\tfrac{1}{2})_n(\tfrac{1}{2})_n & = &-
\frac{n-\frac{1}{2}}{n+\frac{1}{2}} \,(\tfrac{-1}{2})_n(\tfrac{3}{2})_n\,,
\qquad
\frac{(1-r)_n}{(-r)_n}=1-\frac{n}{r}\,.
\label{iden1}
\eea
Therefore,
 Appendix (\ref{case2other}) can be obtained from 
Appendix (\ref{IdEvenOdd4F3-1}) if we take $\mu \rightarrow 0$.

Similarly, although we have not presented here for the complete
expressions, the next one can be written as 
\bea
&&\big[ 
  \underbrace{
\hat{y}_{(1}\ldots\,\hat{y}_{1}
}_{h_1-1+m} 
\underbrace{
\hat{y}_{2}\ldots\,\hat{y}_{2)}
}_{h_1-1-m} K ,\,
 \underbrace{
\hat{y}_{(1}\ldots\,\hat{y}_{1}
}_{h_2-\frac{1}{2}+\rho} 
\underbrace{
\hat{y}_{2}\ldots\,\hat{y}_{2)}
}_{h_2-\frac{1}{2}-\rho}
\big]
=
\nonu \\
&& \sum_{r=1}^{[\frac{h_1+h_2}{2}]}\,
\Bigg[
\frac{(-1)^r }{(2r-2)!}\,
N_{2r-1}^{h_1, h_2+\frac{1}{2}}(m,\rho) 
\,
\Bigg({}
_4F_3\left[
\begin{array}{c|}
\frac{1}{2}+\mu  , \frac{1}{2}-\mu  , \frac{-2r+2}{2}  ,\frac{-2r+3}{2}\\
\frac{3}{2}-h_1 ,  \frac{1}{2}-h_2 ,  h_1+h_2-2r+\frac{3}{2}
\end{array}   1\right]
 \nonu\\
&& +\,_4F_3\left[
\begin{array}{c|}
\frac{3}{2}-\mu  , -\frac{1}{2}+\mu  , \frac{-2r+2}{2}  ,\frac{-2r+3}{2}\\
\frac{3}{2}-h_1 ,  \frac{1}{2}-h_2 ,  h_1+h_2-2r+\frac{3}{2}
\end{array}   1\right]
\Bigg) \,
\label{IdEvenOdd4F3}
 \nonu\\
&& \times \underbrace{
\hat{y}_{(1}\ldots\ldots\ldots\ldots\hat{y}_{1}
}_{h_1+h_2-2r+\frac{1}{2}+m+\rho} \,
\underbrace{
\hat{y}_{2}\ldots\ldots\ldots\ldots\hat{y}_{2)}
}_{h_1+h_2-2r+\frac{1}{2}-m-\rho}K \Bigg]
\\
&&
- \mi \sum_{r=1}^{[\frac{h_1+h_2-1}{2}]}\,\Bigg[
\frac{(-1)^r(h_1-r-\frac{1}{2})}{r (2r-1)!}
\,
N_{2r}^{h_1, h_2+\frac{1}{2}}(m,\rho) 
\,
\Bigg( {}
_4F_3\left[
\begin{array}{c|}
\frac{1}{2}+\mu  , \frac{1}{2}-\mu  , \frac{-2r}{2}  ,\frac{-2r+1}{2}\\
\frac{3}{2}-h_1 ,  \frac{1}{2}-h_2 ,  h_1+h_2-2r+\frac{1}{2}
\end{array}   1\right]
\nonu\\
&& -\,_4F_3\left[
\begin{array}{c|}
\frac{3}{2}-\mu  , -\frac{1}{2}+\mu  , \frac{-2r}{2}  ,\frac{-2r+1}{2}\\
\frac{3}{2}-h_1 ,  \frac{1}{2}-h_2 ,  h_1+h_2-2r+\frac{1}{2}
\end{array}   1\right]
\Bigg)\,
\underbrace{
\hat{y}_{(1}\ldots\ldots\ldots\ldots\hat{y}_{1}
}_{h_1+h_2-2r-\frac{1}{2}+m+\rho} \,
\underbrace{
\hat{y}_{2}\ldots\ldots\ldots\ldots\hat{y}_{2)}
}_{h_1+h_2-2r-\frac{1}{2}-m-\rho}
\Bigg],
\nonu
\eea
which can be rewritten as  (\ref{IdEvenOdd}) and this leads to
Appendix (\ref{IdEvenOdd4F3Odake}) when we take $\mu \rightarrow 0$.


$\bullet$ The case-three with two fermionic oscillators

By combining the previous result Appendix (\ref{casethreeapp}) with
Appendix (\ref{appenrela}),
we obtain
\bea
&&
\big\{ 
  \underbrace{
\hat{y}_{(1}\ldots\,\hat{y}_{1}
}_{h_1-\frac{1}{2}+\rho} 
\underbrace{
\hat{y}_{2}\ldots\,\hat{y}_{2)}
}_{h_1-\frac{1}{2}-\rho},\,
 \underbrace{
\hat{y}_{(1}\ldots\,\hat{y}_{1}
}_{h_2-\frac{1}{2}+\omega} 
\underbrace{
\hat{y}_{2}\ldots\,\hat{y}_{2)}
}_{h_2-\frac{1}{2}-\omega}
\big\}
\nonumber\\
&&=
 \sum_{r=1}^{[\frac{h_1+h_2+1}{2}]}\,(-1)^r\,
N_{2r-1}^{h_1+\frac{1}{2}, h_2+\frac{1}{2}}(\rho,\omega) \,
\Bigg[
-\frac{1}{(2r-2)!}
\nonu\\
&&
\times
\Bigg({}
_4F_3\left[
\begin{array}{c|}
\frac{1}{2}+\mu  , \frac{1}{2}-\mu  , \frac{-2r+2}{2}  ,\frac{-2r+3}{2}\\
\frac{1}{2}-h_1 ,  \frac{1}{2}-h_2 ,  h_1+h_2-2r+\frac{5}{2}
\end{array}   1\right]
\nonu \\
&& +\,_4F_3\left[
\begin{array}{c|}
\frac{3}{2}-\mu  , -\frac{1}{2}+\mu  , \frac{-2r+2}{2}  ,\frac{-2r+3}{2}\\
\frac{1}{2}-h_1 ,  \frac{1}{2}-h_2 ,  h_1+h_2-2r+\frac{5}{2}
\end{array}   1\right]
\Bigg)
  \,
\underbrace{
\hat{y}_{(1}\ldots\ldots\ldots.\hat{y}_{1}
}_{h_1+h_2-2r+1+\rho+\omega} \,
\underbrace{
\hat{y}_{2}\ldots\ldots\ldots.\hat{y}_{2)}
}_{h_1+h_2-2r+1-\rho-\omega}
\nonumber\\
&&
+\frac{2(h_1\!+\!h_2\!-\!r\!+\!1)}{(2r-1)!}
\,
\Bigg({}
_4F_3\left[
\begin{array}{c|}
\frac{1}{2}+\mu  , \frac{1}{2}-\mu  , \frac{-2r+1}{2}  ,\frac{-2r+2}{2}\\
\frac{1}{2}-h_1 ,  \frac{1}{2}-h_2 ,  h_1+h_2-2r+\frac{5}{2}
\end{array}   1\right]
\nonu \\
&& -\,_4F_3\left[
\begin{array}{c|}
\frac{3}{2}-\mu  , -\frac{1}{2}+\mu  , \frac{-2r+1}{2}  ,\frac{-2r+2}{2}\\
\frac{1}{2}-h_1\ ,  \frac{1}{2}-h_2\ ,  h_1+h_2-2r+\frac{5}{2}
\end{array}   1\right]
\Bigg)
\nonu\\
&&
\times \, \underbrace{
\hat{y}_{(1}\ldots\ldots\ldots.\hat{y}_{1}
}_{h_1+h_2-2r+1+\rho+\omega} \,
\underbrace{
\hat{y}_{2}\ldots\ldots\ldots.\hat{y}_{2)}
}_{h_1+h_2-2r+1-\rho-\omega}K
\Bigg],
\label{IdOddOdd4F3-1}
\eea
which can be expressed as
(\ref{IdOddOdd-1}).
By taking $\mu \rightarrow 0$, we obtain the previous Appendix
(\ref{equivIdOddOdd4F3-1}).

There is a relation 
\bea
&&_4F_3\left[
\begin{array}{c|}
\frac{1}{2} \ , -\frac{1}{2} \ , \frac{-2r+2}{2} \ ,\frac{-2r+3}{2}\\
\frac{1}{2}-h_1\ ,  \frac{1}{2}-h_2\ ,  h_1+h_2-2r+\frac{5}{2}
\end{array}  \ 1\right]
\nonu\\
&&=
\frac{1}{2}
\Bigg({}
_4F_3\left[
\begin{array}{c|}
\frac{1}{2} \ , \frac{1}{2} \ , \frac{-2r+2}{2} \ ,\frac{-2r+3}{2}\\
\frac{1}{2}-h_1\ ,  \frac{1}{2}-h_2\ ,  h_1+h_2-2r+\frac{5}{2}
\end{array}  \ 1\right]
\nonu \\
&& +\,_4F_3\left[
\begin{array}{c|}
\frac{3}{2} \ , -\frac{1}{2} \ , \frac{-2r+2}{2} \ ,\frac{-2r+3}{2}\\
\frac{1}{2}-h_1\ ,  \frac{1}{2}-h_2\ ,  h_1+h_2-2r+\frac{5}{2}
\end{array}  \ 1\right]
\Bigg).
\label{Iden}
\eea
This Appendix (\ref{Iden}) can be checked, similar to
Appendix (\ref{iden1}),
by realizing that 
\bea
 2 \,(\tfrac{-1}{2})_n(\tfrac{1}{2})_n
&=&(\tfrac{1}{2})_n(\tfrac{1}{2})_n +(\tfrac{-1}{2})_n(\tfrac{3}{2})_n\,
\nonu\\
&=&
-2(n-\tfrac{1}{2})(\tfrac{-1}{2})_n(\tfrac{1}{2})_n +2(n+\tfrac{1}{2})(\tfrac{-1}{2})_n(\tfrac{1}{2})_n.
\label{iden2}
\eea
Then we observe that when we take ${\mu} \rightarrow 0$,
the relation Appendix (\ref{IdOddOdd4F3-1}) becomes the relation
Appendix (\ref{equivIdOddOdd4F3-1}) with
Appendix (\ref{iden2}).

Finally, we obtain
\bea
&&
\big\{ 
  \underbrace{
\hat{y}_{(1}\ldots\,\hat{y}_{1}
}_{h_1-\frac{1}{2}+\rho} 
\underbrace{
\hat{y}_{2}\ldots\,\hat{y}_{2)}
}_{h_1-\frac{1}{2}-\rho},\,
 \underbrace{
\hat{y}_{(1}\ldots\,\hat{y}_{1}
}_{h_2-\frac{1}{2}+\omega} 
\underbrace{
\hat{y}_{2}\ldots\,\hat{y}_{2)}
}_{h_2-\frac{1}{2}-\omega}K
\big\}
\nonumber\\
&&=
\mi\, \sum_{r=1}^{[\frac{h_1+h_2}{2}]}\,(-1)^r\,
N_{2r}^{h_1+\frac{1}{2}, h_2+\frac{1}{2}}(\rho,\omega) \,
\Bigg[
-\frac{1}{(2r-1)!}
\nonu\\
&&
\times
\Bigg({}
_4F_3\left[
\begin{array}{c|}
\frac{1}{2}+\mu  , \frac{1}{2}-\mu  , \frac{-2r+1}{2}  ,\frac{-2r+2}{2}\\
\frac{1}{2}-h_1 ,  \frac{1}{2}-h_2 ,  h_1+h_2-2r+\frac{3}{2}
\end{array}  \ 1\right]
\nonu \\
&& +\,_4F_3\left[
\begin{array}{c|}
\frac{3}{2}-\mu  , -\frac{1}{2}+\mu  , \frac{-2r+1}{2}  ,\frac{-2r+2}{2}\\
\frac{1}{2}-h_1 ,  \frac{1}{2}-h_2 ,  h_1+h_2-2r+\frac{3}{2}
\end{array}  \ 1\right]
\Bigg)
 \, \underbrace{
\hat{y}_{(1}\ldots\ldots\ldots.\hat{y}_{1}
}_{h_1+h_2-2r+\rho+\omega}\,
\underbrace{
\hat{y}_{2}\ldots\ldots\ldots.\hat{y}_{2)}
}_{h_1+h_2-2r-\rho-\omega}K
\nonumber\\
&&
+\frac{2(h_1+h_2-r)+1}{(2r)!}
\,
\Bigg({}
_4F_3\left[
\begin{array}{c|}
\frac{1}{2}+\mu  , \frac{1}{2}-\mu  , \frac{-2r}{2}  ,\frac{-2r+1}{2}\\
\frac{1}{2}-h_1\ ,  \frac{1}{2}-h_2\ ,  h_1+h_2-2r+\frac{3}{2}
\end{array}   1\right]
\label{IdOddOdd4F3}
\\
&& -\,_4F_3\left[
\begin{array}{c|}
\frac{3}{2}-\mu  , -\frac{1}{2}+\mu  , \frac{-2r}{2}  ,\frac{-2r+1}{2}\\
\frac{1}{2}-h_1 ,  \frac{1}{2}-h_2 ,  h_1+h_2-2r+\frac{3}{2}
\end{array}   1\right]
\Bigg)
\,  \underbrace{
\hat{y}_{(1}\ldots\ldots\ldots.\hat{y}_{1}
}_{h_1+h_2-2r+\rho+\omega} \,
\underbrace{
\hat{y}_{2}\ldots\ldots\ldots.\hat{y}_{2)}
}_{h_1+h_2-2r-\rho-\omega}
\Bigg],
\nonu
\eea
which is equal to
(\ref{IdOddOdd}) by using the relations (\ref{3struct}).
When we take ${\mu} \rightarrow 0$,
the relation Appendix (\ref{IdOddOdd4F3}) becomes the relation
Appendix (\ref{IdOddOdd4F3Odake}).

\section{The remaining (anti)commutators between the ${\cal N}=4$ higher spin generators}

By following the procedure in section $4$,
the remaining ten (anti)commutators
are described as follows
\footnote{For convenience, we present some
  values of structure constants here. The structure constants
  $\mathrm{BB}^{h_1,h_2}_{r,\,+}$ contain $(\nu^2-15329555)$, $(\nu^8-
  18824252 \nu^6+80986952555566 \nu^4-112329283433337991596 \nu^2+
  46136714396921922707472825)$ and $(-197+\nu)(-195+\nu) \cdots
  (-3+\nu)(3+\nu)(5+\nu) \cdots
  (197+\nu)(-199+\nu^2)$ for $r=3, 10$ and $r=199$
  together with $h_1=h_2=100$ respectively. 
 The structure constants
 $\mathrm{BB}^{h_1,h_2}_{r,\,-}$ contain
 $\nu$, $\nu(\nu^6-14118203 \nu^4+40493570399155 \nu^2-
 28082401845512944761)$, and $(-197+\nu)(-195+\nu) \cdots (-3+\nu)\nu(3+\nu)(5+\nu) \cdots
 (197+\nu)$  for above values.
 The structure constants
 $\mathrm{BF}^{h_1,h_2+\frac{1}{2}}_{r,\,+}$ contain
 $(15368364-\nu^2)$, $(5 \nu^8-94369560 \nu^6+407046639208272 \nu^4-
 565989214693985201408 \nu^2+233034068993325158419660800)$ and
 $(-198+\nu)(-196+\nu) \cdots (-2+\nu)(2+\nu)(4+\nu) \cdots
 (198+\nu)$  for above values of $r, h_1, h_2$.
 The structure constants
 $\mathrm{BF}^{h_1,h_2+\frac{1}{2}}_{r,\,-}$ contain
 $\nu$, $\nu  (\nu^8-55732440 \nu^6+394459539349008 \nu^4-
 756192834594836501248 \nu^2+394300540856824919467204608)$ and
 $(-198+\nu)(-196+\nu) \cdots (-4+\nu)\nu(4+\nu)(6+\nu) \cdots
  (198+\nu)$  for $r=3, 10$ and $r=198$
 with $h_1=h_2=100$ respectively.
 Finally,
the structure constants
$\mathrm{FF}^{h_1+\frac{1}{2},h_2+\frac{1}{2}}_{r,\,+}$ contain
$(15485580-\nu^2)$ and $(\nu^8-19021880 \nu^6+82687450448656 \nu^4-
115867186219647371520 \nu^2+48074201136264436025131008)$
 for $r=3$ and $10$
 with $h_1=h_2=100$ respectively.
 The structure constants
$\mathrm{FF}^{h_1+\frac{1}{2},h_2+\frac{1}{2}}_{r,\,-}$ contain
 $\nu$ and $\nu  (\nu^8-56171720 \nu^6+400685503988368 \nu^4-
 774118850666523886080 \nu^2+406779762639433803784851456)$
 for above values. In this case there is no simple factorized form compared to other cases.
The even and odd powers of $\nu$ are observed alternatively.}.

\subsection{ The four (anti)commutators with common
$\Phi^{(s_1),i}_{\frac{1}{2},\,\rho}$}

The four kinds of
(anti)commutators are given in this subsection. 
By using the relations
(\ref{IdOddOdd-1}) and (\ref{IdOddOdd}),
we obtain 
\bea
&&
\big\{\Phi^{(s_1),i}_{\frac{1}{2},\,\rho},\,\Phi^{(s_2),j}_{\frac{1}{2},\,\omega}\big\}=
-\delta^{ij}
\sum_{r=1}^{[\frac{s_1+s_2}{2}]}\,\mi (-1)^r\,(2s_1-1)(2s_2-1)
\Bigg[
\nonumber\\
&&
\frac{2(s_1+s_2-r+1)}{(2r-1)(2s_1+2s_2-4r+3)}\,\mathrm{FF}_{2r-1,\,-}^{s_1+\frac{1}{2}, s_2+\frac{1}{2}}(\rho,\omega;\mu)
\Phi^{(s_1+s_2-2r+2)}_{0,\,\rho+\omega}
\nonumber\\
&&
+
\bigg(
\frac{1}{2(s_1+s_2-2r)-1}\,\mathrm{FF}_{2r-1,\,+}^{s_1+\frac{1}{2}, s_2+\frac{1}{2}}(\rho,\omega;\mu)
\nonumber\\
&&
+\frac{2\nu(s_1+s_2+1-r)}{(2r-1)(2s_1+2s_2-4r+3)}\,\mathrm{FF}_{2r-1,\,-}^{s_1+\frac{1}{2}, s_2+\frac{1}{2}}(\rho,\omega;\mu)
\bigg)
\tilde{\Phi}^{(s_1+s_2-2r)}_{2,\,\rho+\omega}
\Bigg]
\nonumber\\
&&
-\sum_{r=1}^{[\frac{s_1+s_2}{2}]}\mi (-1)^r \frac{(2s_1-1)(2s_2-1)}{2(s_1+s_2-2r)-1}
\Bigg[
\mathrm{FF}_{2r,\,+}^{s_1+\frac{1}{2}, s_2+\frac{1}{2}}(\rho,\omega;\mu)\,
\Phi^{(s_1+s_2-2r),ij}_{1,\,\rho+\omega}
\nonumber\\
&&
+
\frac{2(s_1+s_2-r)+1}{2r}\,\mathrm{FF}_{2r,\,-}^{s_1+\frac{1}{2},s_2+\frac{1}{2}}(\rho,\omega;\mu)\,
\Phi^{(s_1+s_2-2r),ij}_{1,\,\rho+\omega}
\Bigg].
\label{halfhalf}
\end{eqnarray}
It is obvious that for the indices $i=j$, the square of Pauli matrix
is proportional to the $2 \times 2$ identity matrix and this leads to
the above (bosonic) ${\cal N}=4$ higher spin generators having
$\delta^{ij}$
tensor.
For the indices $i \neq j$,
the similar analysis to (\ref{nontrivial})
provides the above ${\cal N}=4$ higher spin generators with $SO(4)$ adjoint
indices. There is a linear $\nu$ dependence in the coefficient.

With (\ref{IdEvenOdd-1}) and (\ref{IdEvenOdd}), 
we determine the following commutator 
\begin{eqnarray}
&&
\big[\Phi^{(s_1),i}_{\frac{1}{2},\,\rho},\,\Phi^{(s_2),jk}_{1,\,m}\big]=
\Bigg(-\delta^{ij}\,
\mi \, (2s_1-1)(2s_2-1)\Bigg[
\nonumber\\
&&
\sum_{r=1}^{[\frac{s_1+s_2}{2}]}(-1)^r\,\,\frac{2(s_2-r)+1}{2r(2s_1+2s_2-4r-1)}\,\mathrm{BF}_{2r,\,-}^{s_2+1, s_1+\frac{1}{2}}(m,\rho;\mu)
\tilde{\Phi}^{(s_1+s_2-2r),k}_{\frac{3}{2},\,\rho+m}
\nonumber\\
&&
+\sum_{r=1}^{[\frac{s_1+s_2+1}{2}]}
(-1)^r\,\frac{1}{2(s_1+s_2-2r)+3}\,\mathrm{BF}_{2r-1,\,+}^{s_2+1, s_1+\frac{1}{2}}(m,\rho;\mu)
\Phi^{(s_1+s_2-2r+2),k}_{\frac{1}{2},\,\rho+m}
\Bigg]
-(j \longleftrightarrow k ) \Bigg)
\nonumber\\
&&
+\epsilon^{ijkl}\,
\mi \,(2s_1-1)(2s_2-1)\Bigg[
\sum_{r=1}^{[\frac{s_1+s_2}{2}]}\,(-1)^r\,
\frac{1}{2(s_1+s_2-2r)-1}\,\mathrm{BF}_{2r,\,+}^{s_2+1, s_1+\frac{1}{2}}(m,\rho;\mu)
\tilde{\Phi}^{(s_1+s_2-2r),l}_{\frac{3}{2},\,\rho+m}
\nonumber\\
&&+
\sum_{r=1}^{[\frac{s_1+s_2+1}{2}]}\,(-1)^r\,
\frac{2(s_2-r+1)}{(2r-1)(2s_1+2s_2-4r+3)}\,\mathrm{BF}_{2r-1,\,-}^{s_2+1, s_1+\frac{1}{2}}(m,\rho;\mu)
\Phi^{(s_1+s_2+2-2r),l}_{\frac{1}{2},\,\rho+m}
\Bigg].
\label{half1}
\end{eqnarray}
We should use the simple relations between the Pauli matrices
in order to deal with the $2 \times 2$ matrices in the tensor product.

We obtain the anticommutator with
(\ref{IdOddOdd-1}) and (\ref{IdOddOdd})
\begin{eqnarray}
&&
\big\{\Phi^{(s_1),i}_{\frac{1}{2},\,\rho},\,\Phi^{(s_2),j}_{\frac{3}{2},\omega}\big\}=
\delta^{ij}
\,\mi  \,(2s_1-1)(2s_2-1)
\Bigg[
\nonumber\\
&&
\sum_{r=1}^{[\frac{s_1+s_2+1}{2}]}\,(-1)^r
\frac{1}{2(s_1+s_2-2r)+3}\,\mathrm{FF}_{2r,\,+}^{s_1+\frac{1}{2}, s_2+\frac{3}{2}}(\rho,\omega;\mu)
\,
\Phi^{(s_1+s_2-2r+2)}_{0,\,\rho+\omega}
\nonumber\\
&&
+\sum_{r=1}^{[\frac{s_1+s_2}{2}]}\,(-1)^r
\frac{1}{2(s_1+s_2-2r)-1}
\bigg(
\frac{\nu}{(2s_1+2s_2-4r+3)}\,\mathrm{FF}_{2r,\,+}^{s_1+\frac{1}{2}, s_2+\frac{3}{2}}(\rho,\omega;\mu)
\nonumber\\
&&
+\frac{2(s_1+s_2-r)+3}{2r}\,\mathrm{FF}_{2r,\,-}^{s_1+\frac{1}{2}, s_2+\frac{3}{2}}(\rho,\omega;\mu)
\bigg)
\tilde{\Phi}^{(s_1+s_2-2r)}_{2,\,\rho+\omega}\Bigg]
\nonumber\\
&&
-\sum_{r=1}^{[\frac{s_1+s_2+1}{2}]}\,\frac{\mi \,(-1)^r\,(2s_1-1)(2s_2-1)}{2(s_1+s_2-2r)+3}
\Bigg[
\nonumber\\
&&
\frac{2(s_1+s_2-r)+4}{(2r-1)}\,\mathrm{FF}_{2r-1,\,-}^{s_1+\frac{1}{2}, s_2+\frac{3}{2}}(\rho,\omega;\mu)\,
\Phi^{(s_1+s_2-2r+2),ij}_{1,\,\rho+\omega}
\nonumber\\
&&
+
\mathrm{FF}_{2r-1,\,+}^{s_1+\frac{1}{2}, s_2+\frac{3}{2}}(\rho,\omega;\mu)\,
\Phi^{(s_1+s_2-2r+2),ij}_{1,\,\rho+\omega}
\Bigg].
\label{half3half}
\eea
Again the linear $\nu$ dependence appears in the coefficient.

By applying the relation (\ref{IdEvenOdd-1}), the following
commutator can be determined
\begin{eqnarray}
&&
\big[\Phi^{(s_1),i}_{\frac{1}{2},\,\rho},\,\tilde{\Phi}^{(s_2)}_{2,\,m}\big]=
\mi \, (2s_1-1)(2s_2-1)
\Bigg[\,
\nonu\\
&&
\sum_{r=1}^{[\frac{s_1+s_2+1}{2}]} \,  (-1)^r \frac{1}{2(s_1+s_2-2r)+3}\,
\mathrm{BF}_{2r,\,+}^{s_2+2,  s_1+\frac{1}{2}}(m,\rho;\mu)
\,
\Phi^{(s_1+s_2-2r+2),i}_{\frac{1}{2},\,\rho+m}
\nonumber\\
&&
-\sum_{r=1}^{[\frac{s_1+s_2}{2}]} \,  (-1)^r \frac{1}{2(s_1+s_2-2r)+3}\,
 \frac{2(s_2-r+2)}{(2r-1)}\,\mathrm{BF}_{2r-1,\,-}^{s_2+2,s_1+\frac{1}{2}}(m,\rho;\mu)
\,
\tilde{\Phi}^{(s_1+s_2-2r+2),i}_{\frac{3}{2},\,\rho+m}\,
\Bigg].
\nonu\\
\label{half2}
\end{eqnarray}

\subsection{ The three (anti)commutators with common
$\Phi^{(s_1),ij}_{1,\,m}$}

The three kinds of
commutators are given in this subsection. 
By using (\ref{IdEvenEven}) and (\ref{IdAntiEvenEven}),
we obtain the commutator
\begin{eqnarray}
&&
\big[\Phi^{(s_1),ij}_{1,\,m},\,\Phi^{(s_2),kl}_{1,\,n}\big]=
-(\delta^{ik}\delta^{jl} -\delta^{il}\delta^{jk})
\mi \,(2s_1-1)(2s_2-1)
\Bigg[
\nonumber\\
&&\sum_{r=1}^{[\frac{s_1+s_2+1}{2}]}\,(-1)^r
\frac{1}{2(s_1+s_2-2r)+3}
\mathrm{BB}_{2r,\,-}^{s_1+1,s_2+1}(m,n;\mu)
\,
\Phi^{(s_1+s_2-2r+2)}_{0,\,m+n}
\nonumber\\
&&+
\sum_{r=1}^{[\frac{s_1+s_2}{2}]}\,(-1)^r
\frac{1}{2(s_1+s_2-2r)-1}
\nonumber\\
&&
\times 
\bigg(
\mathrm{BB}_{2r,\,+}^{s_1+1,s_2+1}(m,n;\mu)
+\frac{\nu}{2(s_1+s_2-2r)+3}\,\mathrm{BB}_{2r,\,-}^{s_1+1,s_2+1}(m,n;\mu)
\bigg)\,
\tilde{\Phi}^{(s_1+s_2-2r)}_{2,\,m+n}\Bigg]
\nonumber\\
&&
-\epsilon^{ijkl}
\, \mi \,(2s_1-1)(2s_2-1)
\Bigg[
\nonumber\\
&&\sum_{r=1}^{[\frac{s_1+s_2+1}{2}]}\,(-1)^r
\frac{1}{2(s_1+s_2-2r)+3}
\,
\mathrm{BB}_{2r,\,+}^{s_1+1,s_2+1}(m,n;\mu)
\,
\Phi^{(s_1+s_2-2r+2)}_{0,\,m+n}\,
\nonumber\\
&&+ \sum_{r=1}^{[\frac{s_1+s_2}{2}]}\,(-1)^r \frac{1}{2(s_1+s_2-2r)-1}
\nonumber\\
&&
\times 
\bigg(
\frac{\nu}{2(s_1+s_2-2r)+3}\mathrm{BB}_{2r,\,+}^{s_1+1,s_2+1}(m,n;\mu)
+\mathrm{BB}_{2r,\,-}^{s_1+1,s_2+1}(m,n;\mu)
\bigg)\,
\tilde{\Phi}^{(s_1+s_2-2r)}_{2,\,m+n}\Bigg]
\nonumber\\
&&
+\Bigg(\delta^{ik}
\sum_{r=1}^{[\frac{s_1+s_2+2}{2}]}\,\frac{\mi \,(-1)^r(2s_1-1)(2s_2-1)}{2(s_1+s_2-2r)+3}
\Bigg[
\mathrm{BB}_{2r-1,\,+}^{s_1+1,s_2+1}(m,n;\mu)
\,\Phi^{(s_1+s_2-2r+2),jl}_{1,\,m+n}
\nonumber\\
&&+
\mathrm{BB}_{2r-1,\,-}^{s_1+1,s_2+1}(m,n;\mu)
\,\Phi^{(s_1+s_2-2r+2),jl}_{1,\,m+n}
\Bigg]
\nonumber\\
&&
-(  k  \longleftrightarrow l   )
- (  i  \longleftrightarrow j   )
+(  i  \longleftrightarrow j,\,  k
  \longleftrightarrow l     ) \Bigg).
\label{Com11}
\end{eqnarray}
The linear $\nu$ dependence appears in the coefficient
coming from the replacement of the ${\cal N}=4$ higher spin generators
in (\ref{HSbasis}).

By using (\ref{IdEvenOdd-1}) and (\ref{IdEvenOdd}),
we determine the following commutator
\begin{eqnarray}
&&
\big[\Phi^{(s_1),ij}_{1,\,m},\,\tilde{\Phi}^{(s_2),k}_{\frac{3}{2},\,\rho}\big]=
\Bigg(-\delta^{ik}\,
\mi  (2s_1-1)(2s_2-1)
\Bigg[
\nonumber\\
&&
\sum_{r=1}^{[\frac{s_1+s_2+1}{2}]}\,(-1)^r\,\frac{1}{2(s_1+s_2-2r)+3}\,
\frac{2(s_1-r)+1}{2r}\,\mathrm{BF}_{2r,\,-}^{s_1+1,s_2+\frac{3}{2}}(m,\rho;\mu)
\,\Phi^{(s_1+s_2-2r+2),j}_{\frac{1}{2},\,m+\rho}
\nonumber\\
&&
-
\sum_{r=1}^{[\frac{s_1+s_2+2}{2}]}\,(-1)^r\,\frac{1}{2(s_1+s_2-2r)+3}\,
\mathrm{BF}_{2r-1,\,+}^{s_1+1,s_2+\frac{3}{2}}(m,\rho;\mu)
\,\tilde{\Phi}^{(s_1+s_2-2r+2),j}_{\frac{3}{2},\,m+\rho}
\Bigg]
-( i \longleftrightarrow j ) \Bigg)
\nonumber\\
&&
+\epsilon^{ijkl}\,\mi \,
(2s_1-1)(2s_2-1)
\Bigg[
\nonumber\\
&&
\sum_{r=1}^{[\frac{s_1+s_2+1}{2}]}\,(-1)^r \frac{1}{2(s_1+s_2-2r)+3}
\mathrm{BF}_{2r,\,+}^{s_1+1,s_2+\frac{3}{2}}(m,\rho;\mu)
\,\Phi^{(s_1+s_2-2r+2),l}_{\frac{1}{2},\,m+\rho}
\nonumber\\
&&
-\sum_{r=1}^{[\frac{s_1+s_2+2}{2}]}\,(-1)^r \frac{1}{2(s_1+s_2-2r)+3}
\frac{2(s_1-r+1)}{(2r-1)}\,\mathrm{BF}_{2r-1,\,-}^{s_1+1,s_2+\frac{3}{2}}(m,\rho;\mu)
\,
\tilde{\Phi}^{(s_1+s_2-2r+2),l}_{\frac{3}{2},\,m+\rho}
\Bigg].
\nonu\\
\label{13half}
\end{eqnarray}

From (\ref{IdEvenEven}), we obtain 
\begin{eqnarray}
&&
\big[\Phi^{(s_1),ij}_{1,\,m},\,\tilde{\Phi}^{(s_2)}_{2,\,n}\big]=
\nonumber\\
&&
-\sum_{r=1}^{[\frac{s_1+s_2+2}{2}]}\,
 \frac{\mi\,(-1)^r (2s_1-1)(2s_2-1)}{2(s_1+s_2-2r)+3}
\Bigg[
\mathrm{BB}_{2r,\,+}^{s_1+1,s_2+2}(m,n;\mu)\,
 \Phi^{(s_1+s_2+2-2r),ij}_{1,\,m+n}
\nonumber\\
&&
+\mathrm{BB}_{2r,\,-}^{s_1+1,s_2+2}(m,n;\mu)\,
\Phi^{(s_1+s_2+2-2r),ij}_{1,\,m+n}
\Bigg].
\label{12}
\end{eqnarray}

\subsection{ The two (anti)commutators with common
$\tilde{\Phi}^{(s_1),i}_{\frac{3}{2},\,\rho}$}

The two kinds of
(anti)commutators are given.
The following anticommutator can be obtained
\begin{eqnarray}
&&
  \big\{ \tilde{\Phi}^{(s_1),i}_{\frac{3}{2},\,\rho},\,
  \tilde{\Phi}^{(s_2),j}_{\frac{3}{2},\omega}\big\}=
-\delta^{ij}
\sum_{r=1}^{[\frac{s_1+s_2+2}{2}]}\,\mi \,(-1)^r(2s_1-1)(2s_2-1)
\Bigg[
\nonumber\\
&&
\frac{2(s_1+s_2-r+3)}{(2r-1)(2s_1+2s_2-4r+7)}\,\mathrm{FF}_{2r-1,\,-}^{s_1+\frac{3}{2},s_2+\frac{3}{2}}(\rho,\omega;\mu)
\,
\Phi^{(s_1+s_2-2r+4)}_{0,\,\rho+\omega}
\nonumber\\
&&
+\frac{1}{2(s_1+s_2-2r)+3}
\bigg(
\mathrm{FF}_{2r-1,\,+}^{s_1+\frac{3}{2}, s_2+\frac{3}{2}}(\rho,\omega;\mu)
\nonumber\\
&&
+\frac{2\nu(s_1+s_2+3-r)}{(2r-1)(2s_1+2s_2-4r+7)}\,\mathrm{FF}_{2r-1,\,-}^{s_1+\frac{3}{2},s_2+\frac{3}{2}}(\rho,\omega;\mu)
\bigg)
\tilde{\Phi}^{(s_1+s_2-2r+2)}_{2,\,\rho+\omega}
\Bigg]
\nonumber\\
&&
-\sum_{r=1}^{[\frac{s_1+s_2+2}{2}]}\,\frac{\mi \, (-1)^r(2s_1-1)(2s_2-1)}{2(s_1+s_2-2r)+3}
\Bigg[
\mathrm{FF}_{2r,\,+}^{s_1+\frac{3}{2},s_2+\frac{3}{2}}(\rho,\omega;\mu)\,
\Phi^{(s_1+s_2-2r+2),ij}_{1,\,\rho+\omega}
\nonumber\\
&&
+
\frac{2(s_1+s_2-r)+5}{2r}\,\mathrm{FF}_{2r,\,-}^{s_1+\frac{3}{2},s_2+\frac{3}{2})}(\rho,
\omega;\mu)\,
\tilde{\Phi}^{(s_1+s_2-2r+2),ij}_{1,\,\rho+\omega}
\Bigg],
\label{3half3half}
\end{eqnarray}
where (\ref{IdOddOdd-1}) and (\ref{IdOddOdd}) are used.

We have the next commutator 
\begin{eqnarray}
&&
\big[ \tilde{\Phi}^{(s_1),i}_{\frac{3}{2},\,\rho},\,\tilde{\Phi}^{(s_2)}_{2,\,m}\big]=
\sum_{r=1}^{[\frac{s_1+s_2+3}{2}]} \,\mi \,(-1)^r (2s_1-1)(2s_2-1)
\Bigg[
\nonumber\\
&&
\frac{2(s_2-r+2)}{(2r-1)(2s_1+2s_2-4r+7)}\,\mathrm{BF}_{2r-1,\,-}^{s_2+2,s_1+\frac{3}{2}}(m,\rho;\mu)
\,
\Phi^{(s_1+s_2-2r+4),i}_{\frac{1}{2},\,\rho+m}
\nonumber\\
&&
+\frac{1}{2(s_1+s_2-2r)+3}\mathrm{BF}_{2r,\,+}^{s_2+2,s_1+\frac{3}{2}}(m,\rho;\mu)
\,
\tilde{\Phi}^{(s_1+s_2-2r+2),i}_{\frac{3}{2},\,\rho+m}\,
\Bigg],
\label{3half2}
\end{eqnarray}
where (\ref{IdEvenOdd-1}) is used.

\subsection{ The final commutator}

With (\ref{IdEvenEven}), the final commutator is given by
\begin{eqnarray}
&&
\big[\tilde{\Phi}^{(s_1)}_{2,\,m},\,\tilde{\Phi}^{(s_2)}_{2,\,n}\big]=
-\sum_{r=1}^{[\frac{s_1+s_2+3}{2}]}
\,
\mi\,(-1)^r (2s_1-1)(2s_2-1)
\nonu\\
&&
\times
\Bigg[
\frac{1}{2(s_1+s_2-2r)+7}\,
\mathrm{BB}_{2r,\,-}^{s_1+2,s_2+2}(m,n;\mu)
\,
 \Phi^{(s_1+s_2-2r+4)}_{0,\,m+n}
 \nonu \\
 && +\frac{1}{2(s_1+s_2-2r)+3}
\bigg(
\mathrm{BB}_{2r,\,+}^{s_1+2,s_2+2}(m,n;\mu)
\nonu\\
&&
+\frac{\nu}{2(s_1+s_2-2r)+7}\,
\mathrm{BB}_{2r,\,-}^{s_1+2,s_2+2}(m,n;\mu)
 \bigg) 
 \tilde{\Phi}^{(s_1+s_2-2r+2)}_{2,\,m+n}
\Bigg].
\label{22comm}
\end{eqnarray}
The $\nu$ dependence appears in the coefficient.

The complete $15$ (anti)commutators are given by (\ref{00}),
(\ref{0half}),(\ref{01}),(\ref{03half}),(\ref{02}), Appendices
(\ref{halfhalf}), (\ref{half1}), (\ref{half3half}),
 (\ref{half2}),  (\ref{Com11}),  (\ref{13half}),
 (\ref{12}),  (\ref{3half3half}),
 (\ref{3half2}), and  (\ref{22comm}).

\section{The ${\cal N}=2$ higher spin algebra $shs[\lambda]$
}

In this Appendix, we describe the embedding of $shs[\mu]$
inside $shs_2[\mu]$ by taking the algebraically obvious one
where the $2 \times 2$ identity matrix is taken.
Although this is accidental (not physical
because the different higher spin multiplets are mixed),
the nontrivial relations between the two kinds of
structure constants below
satisfy in general. They do not depend on the
accidental or physical embeddings we take.
We refer to the paper \cite{EGR} for the physical embedding
of $shs[\mu]$
inside $shs_2[\mu]$ and see also the footnote \ref{refereepoint}
for the corresponding $2 \times 2$ matrices.

\subsection{Review }

The (anti)commutation relations of
the ${\cal N}=2$ higher spin algebra  $shs[\lambda]$
generated by bosonic generators $T^j$ and $U^j$ of integer spins
$(j+1)$ and
fermionic generators $\Psi^j$ and $\bar{\Psi}^j$
of half integer spins $(j+1)$ with $j =0, \frac{1}{2}, 1,
\frac{3}{2}, \cdots $
can be summarized by \cite{FL} (in the notation of \cite{CG})
\begin{align}
 [T^j_m,T^{j'}_{m'}] & =\sum_{j'',m''} f^{jj'j''}_{TTT}C^{jj'j''}_{mm'm''} T^{j''}_{m''} , \qquad
[U^j_m, U^{j'}_{m'}]  =\sum_{j'',m''} f_{UUU}^{jj'j''}C^{jj'j''}_{mm'm''} U^{j''}_{m''} ,
 \notag
\\
[T^j_m,\Psi^{j'}_{r'}] & =\sum_{j'',r''} f_{T\Psi\Psi}^{jj'j''}C^{jj'j''}_{mr'r''} \Psi^{j''}_{r''} , \qquad \quad \,
[T^j_m, \bar{\Psi}^{j'}_{r'}]  =\sum_{j'',r''} f_{T\bar{\Psi}\bar{\Psi}}^{jj'j''}C^{jj'j''}_{mr'r''} 
\bar{\Psi}^{j''}_{r''}\ ,
\notag\\ 
[U^j_m,\Psi^{j'}_{r'}] & =\sum_{j'',r''} f_{U\Psi\Psi}^{jj'j''}C^{jj'j''}_{mr'r''} \Psi^{j''}_{r''} , \qquad \quad \,
[U^j_m, \bar{\Psi}^{j'}_{r'}] =\sum_{j'',r''} f_{U\bar{\Psi}\bar{\Psi}}^{jj'j''}C^{jj'j''}_{mr'r''} 
\bar{\Psi}^{j''}_{r''} ,
\notag
\\ 
\{\Psi^j_r, \bar{\Psi}^{j'}_{r'}\} & =\sum_{j'',m''} C^{jj'j''}_{rr'm''} 
\left( f_{\Psi\bar{\Psi}T}^{jj'j''} T^{j''}_{m''}  + f_{\Psi\bar{\Psi}U}^{jj'j''} U^{j''}_{m''}\right).
\label{shsRacah}
\end{align}
Of course, the summation $\sum_{j'',m''}$ (or other similar ones)
above is only valid for the selection rule
of $3j$-symbol in Clebsch-Gordan coefficients Appendix (\ref{Clebschdf})
\footnote{One of them is given by $|j-j'| \leq  j'' \leq j+j'$.}.
The structure constants in Appendix (\ref{shsRacah}) are given by 
\begin{align}
f^{jj'j''}_{TTT} &= (1-\epsilon^{jj'j''}) F^{jj'j''}_{000}(\pm 1 \mp \mu)\ ,&
f^{jj'j''}_{UUU} &= (1-\epsilon^{jj'j''}) F^{jj'j''}_{000}(\pm \mu)\ ,
\nonu
\\
f^{jj'j''}_{T\Psi\Psi} &= -\epsilon^{jj'j''} f^{jj'j''}_{T\bar{\Psi}\bar{\Psi}}\ ,&
f^{jj'j''}_{T\bar{\Psi}\bar{\Psi}} &= +F^{jj'j''}_{0\, \frac{1}{2}\,\frac{1}{2}}(-\mu)\ ,
\nonu
\\
f^{jj'j''}_{U\Psi\Psi} &= -\epsilon^{jj'j''} f^{jj'j''}_{U\bar{\Psi}\bar{\Psi}}\ ,&
f^{jj'j''}_{U\bar{\Psi}\bar{\Psi}} &= -\epsilon^{jj'j''}F^{jj'j''}_{0\,-
  \frac{1}{2}\,-\frac{1}{2}}(1-\mu)
\ , \nonu \\
f^{jj'j''}_{\Psi\bar{\Psi}T}&=-\epsilon^{jj'j''}F^{jj'j''}_{\frac{1}{2}\,-
  \frac{1}{2}\,0}(1-\mu)\ ,&
f^{jj'j''}_{\Psi\bar{\Psi}U}&=-F^{jj'j''}_{-\frac{1}{2}\,\frac{1}{2}\,0}(-\mu)\ ,
\label{fs}
\end{align}
with $
\epsilon^{jj'j''}=(-1)^{j+j'-j''}$.
The symbols  $F^{jj'j''}_{l\,  l'\, l''}(\mu)$
in Appendix (\ref{fs}) are given by \cite{FL,CG}
\bea
F^{jj'j''}_{l  l' l''}(\mu) & = &
\sqrt{2j''+1}\Delta^{jj'j''}\sum_t \Bigg[
  (-1)^t \prod_{p=1}^{j+j'-j''-t}(\mu-j''+l''-p)\prod_{q=1}^{t}(\mu+j''+l''+q)
\nonu
\\
&\times &
\frac{\sqrt{ (j+l)!(j-l)!(j'+l')!(j'-l')!(j''+l'')!(j''-l'')!}}{t!(j+j'-j''-t)!(t+j''-j-l')!(t+j''-j'+l)!(j-l-t)!(j'+l'-t)!} \Bigg],
\nonu \\
\label{Fsym}
\eea
where we assume $l+l'=l''$ and 
\bea
\Delta^{jj'j''} \equiv \sqrt{\frac{(j+j'-j'')!(j+j''-j')!(j'+j''-j)!}{(
    j+j'+j''+1)!}}.
\label{Delta}
\eea
The dummy variable $t$ runs from the maximum value
among $(0,-j''+j+l',-j''+j'-l)$
to the minimum value among $(j+j'-j'',j-l,j'+l')$.
The $SL(2)$ Clebsch-Gordan coefficient (or $3j$ symbol)
appearing in Appendix (\ref{shsRacah}) is given by
\bea
&& C^{j j' j''}_{m m' m''}
=
(-1)^{j-j'+m''}
\sqrt{2j''+1}
\left(\!\!
\begin{array}{c c c}
j & j' & j''\\
m & m' &- m''
 \end{array}
\!\!
\right)
\nonu\\
&& =
\Delta^{j j' j''}
\sqrt{(2j''+1)(j+m)!(j-m)!(j'+m')!(j'-m')!(j''+m'')!(j''-m'')!}
\nonu\\
&& \times
\sum_{t} \Bigg[\frac{(-1)^t}{t!(j+j'-j''-t)!(j-m-t)!(j'+m'-t)!}
\nonu \\
&& \times  \frac{1}{(-j'+j''+m+t)!(-j+j''-m'+t)!} \Bigg],
\label{Clebschdf}
\eea
where 
the summation is over all nonnegative integers $t$ such that
the arguments in the factorials are nonnegative \cite{dlmf}
or the $SL(2)$ Clebsch-Gordan coefficient 
can be wrritten by using the hypergeometric function \cite{PRS}
and the expression Appendix (\ref{Delta}) is used.

Then the ${\cal N}=2$ higher spin algebra
is dscribed by Appendix (\ref{shsRacah}) with
Appendices (\ref{fs}), (\ref{Fsym})
and (\ref{Clebschdf}).

\subsection{How to obtain the
  ${\cal N}=2$ higher spin algebra $shs[\lambda]$ from
 ${\cal N}=4$ higher spin algebra $shs_2[\lambda]$}

According to the definition (\ref{capitalN}),
there are no negative factorials in $N_r^{h_1, h_2}(m,n) $
because of $h_1-1\pm m, \, h_2-1\pm n,\,  r-1-k  \geq 0$.
Then, we have \cite{PRS}
\bea
& & N_r^{h_1, h_2}(m,n)  =
\label{N}
\\
&&
\sum_{k=0}^{r-1}
\frac{(-1)^k (r-1)! (h_1-1+m)!  (h_1-1-m)! (h_2-1+n)! (h_2-1-n)! }
{k!(r-1-k)! (h_1+m-r+k)! (h_1-1-m-k)! (h_2-1+n-k)!(h_2-n-r+k)!}.
\nonu
\eea
On the other hand,
the $SL(2)$ Clebsch-Gordan coefficient  can be written as
\bea
&& C^{h_1-1, h_2-1, h_1+h_2-r-1}_{m, n, m+n} =
\nonu\\
&&
\Delta^{h_1-1,h_2-1,h_1+h_2-r-1}\, \sqrt{(2h_1-2)!(2h_2-2)!(2h_1+2h_2-2r-1)!}
\, \times 
\label{CG} \\
&&
\sum_{t}\frac{(-1)^t  \alpha^{h_1-1}_m \alpha^{h_2-1}_n
  \alpha^{h_1+h_2-r-1}_{m+n}}{t!(r-1-t)!(h_1-1-m-t)!
  (h_2-1+n-t)!(h_1+m-r+t)!(h_2-n-r+t)!}.
\nonu 
 \eea
 Here we introduce the scale factor \cite{CG}
 \bea
\alpha_m^j \equiv \sqrt{\frac{(j-m)!(j+m)!}{(2j)!}}.
\label{alpha}
\eea
 
From Appendices (\ref{N}) and (\ref{CG}), we arrive at
\bea
C^{h_1-1, h_2-1, h_1+h_2-r-1}_{m, n, m+n}
& = & 
\frac{\Delta^{h_1-1,h_2-1,h_1+h_2-r-1}\,\alpha^{h_1+h_2-r-1}_{m+n}}{(r-1)!\,\alpha^{h_1-1}_m \alpha^{h_2-1}_n } \nonu \\
& \times & \sqrt{\frac{(2h_1+2h_2-2r-1)!}{(2h_1-2)!(2h_2-2)!}}
\,N_{r}^{h_1, h_2}(m,n)\,,
\label{NfandClebsch}
\eea
where the relations Appendices
(\ref{alpha}) and (\ref{Delta}) are used.
This is nothing but $(2.18)$ of \cite{PRS} for even $r$.
In our case, the above relation holds for odd $r$.
This relation Appendix (\ref{NfandClebsch})  is only valid
for the range (physical range) where Clebsch-Gordan coefficients
satisfy the selection rules.
Because 
$ (1- \epsilon^{jj'j''})  =  (1-(-1)^{r-1}) $
in the function $F^{h_1-1, h_2-1, h_1+h_2-r-1}_{kk'k''}(\mu)$, we can consider only even $r$ case.

Then we can arrange the nontrivial relation between the generalized
hypergeometric function $\phi$ (\ref{spf}) and
$F^{jj'j''}_{kk'k''}$ symbol in Appendix (\ref{Fsym}).
For  $k=k'=k''=0$, 
we have
\begin{align}
F^{h_1-1, h_2-1,h_1+h_2-2r-1}_{0 0 0}(\mu)
&
=
-\frac{2\,\mi\,(-1)^r\,(2h_1-1)(2h_2-1)t_1[h_1]t_1[h_2]}{\Delta^{h_1-1,h_2-1,h_1+h_2-2r-1}t_1[h_1+h_2-2r](2h_1+2h_2-4r-1)}
\nonu\\
&
\times 
\sqrt{\frac{(2h_1-2)!(2h_2-2)!}{(2h_1+2h_2-4r-1)!}}
\,
\phi_{2r}^{h_1 ,h_2}(-\mu,1)
\,,
\label{relHyFf}
\end{align}
where we introduce
\bea
t_1[h]& \equiv&
\frac{2(-1)^{2h+1}(\frac{\mi}{2})^{h+1}}{(2h-1)}\sqrt{\frac{[h-\frac{3}{2}]_{h-1}}{[h-1]_{h-1}}}\,.
\label{t1h}
\eea
Here the falling Pochhammer symbol is used. 
We checked that Appendix (\ref{relHyFf}) holds for odd $r$.
Other four cases of $F$ symbols will be described later
in Appendices (\ref{Fconst}) and (\ref{Fconst1}).
According to Appendix (\ref{Fsym}), the power of $\mu$ is given by
$(j+j'-j'')=(2r-1)$. We observe that for the large values of $r$
closing to the sum of $h_1$ and $h_2$, there exists a factorized form
(where each factor is linear in $\mu$)
of $\mu$ in the left hand side of Appendix (\ref{relHyFf})
together with
Appendix (\ref{Fsym}).
On the other hand, for the small values of
$r$, there are some polynomials in $\mu$ which cannot be written in
terms of factorized forms.
In the region of middle values of $r$, there are
some factorized forms multipled by some polynomials.
In other words, there are two products in Appendix (\ref{Fsym})
specified by dummy variables $p$ and $q$. If they have 
same expression
with apprpriate possible values of $t$, then we will have
former behavior where there exists only factorized form.
If they are different from each other,
then we will have latter behavior which shows some polynomials.

On the other hand, we have seen that as we increase the
$r$ values which are the lower indices of
generalized hypergeometric functions,
the power of $\mu$ increases in the discussion of
(\ref{avalue}). Then how the $F$ symbols are related to the
generalized hypergeometric functions?
First of all, we have checked that the $\mu$ dependent parts of
the former and the latter are exactly the same for the range of
$h_1$ and $h_2$ we consider in this paper. The nontrivial part is
to determine the $\mu$ independent numerical values appearing in an
overall factors. They do depend on $h_1$,$h_2$ and $r$ as in
Appendix (\ref{relHyFf}). We have checked the relative coefficients
explicitly (Around both
the total spin, $h_1+h_2$, and the spin $r$ near $10$,
they are determined completely)
and it turns out that we have the relation between
the $F$ symbols and the generalized hypergeometric functions
\footnote{For example, for the $h_1=h_2=100$  case,
  we have several values for $F$ symbols as follows. When
  $r=3$, the $F$ contains $(-\nu^2-2 \nu+15329555)$ which appears in
  $\phi_{3}^{100,100}(-\nu,1)$. For $r=10$,
  the corresponding polynomial is given by $
  (\nu^8+8 \nu^7-18824252 \nu^6-112945624 \nu^5+80986952555566 \nu^4+323948563193240 \nu^3-112329283433337991596 \nu^2-224659214764103558088 \nu+46136714396921922707472825)$ which is common in
  $\phi_{10}^{100,100}(-\nu,1)$.
  For the $r=199$, we have
  $(-197+\nu)(-195+\nu) \cdots (-1+\nu)(3+\nu)(5+\nu) \cdots
  (199+\nu)$. Also we observe that  $\phi_{199}^{100,100}(-\nu,1)$
  contains this factorized form. Although we do not write down
  the relative coefficients (the number of decimal digits
  are order of several hundreds) here, they satisfy the relations in
  Appendix
  (\ref{relHyFf}). Therefore, we conclude that the nontrivial
  relation
Appendix (\ref{relHyFf}) holds for any $h_1$ and $h_2$.}.

\subsubsection{The first commutator of Appendix (\ref{shsRacah})}

We would like to construct the ${\cal N}=4$ higher spin generators
which provide the above ${\cal N}=2$ higher spin algebra
$shs[\mu]$.
In general, the spin $(j+1)$ generator
consists of
operator $K$ independent oscillator part
and the operator $K$ dependent oscillator part.
The number of oscillators is given by $2j$ with undetermined two
coefficients. 
Then we substitute this ansatz into the first of Appendix (\ref{shsRacah})
and use the previous relation between the oscillators
(\ref{IdEvenEven}) and other ones associated with the footnote
\ref{othercasesforcase1}.
We have found that
the mode dependent $SL(2)$ Clebsch-Gordan coefficient
is given by Appendix (\ref{NfandClebsch}) and symbol $F$ 
is given by Appendix (\ref{relHyFf}).
Then the above two coefficients can be fixed 
and by using (\ref{HSbasis}) we can write down
the bosonic spin $(j+1)$ generator in ${\cal N}=2$
higher spin algebra $shs[\mu]$ in terms of
the linear combination of two ${\cal N}=4$ higher spin generators.

It turns out that 
 the ${\cal N}=2$ higher spin generator 
can be expressed in terms of the ones of ${\cal N}=4$ higher spin
generators
\bea
\alpha^j_m \,T^j_m 
&=&
\underbrace{
\hat{y}_{(1}\ldots\,\hat{y}_{1}
}_{j+m} \,
\underbrace{
\hat{y}_{2}\ldots\,\hat{y}_{2)}
}_{j-m}
\Big(
\!
K \,t_1[j+1]\,(2j+1)+t_1[j+1]\,\nu-t_2[j+1]\,(2j-3)
\Big)
\nonu\\
&=& t_1[j+1]\,\Phi^{(j+1)}_{0,\,m}  +t_2[j+1]\, \tilde{\Phi}^{(j-1)}_{2,\,m} \,,
\label{Tbasis}
\eea
where the previous relations in (\ref{HSbasis}) and Appendix (\ref{t1h})
are used and 
\bea
t_2[h] \equiv \frac{(-2h+1+\nu)}{(2h-5)}\,t_1[h]\,.
\label{t2h}
\eea
There is also a linear $\nu$ dependence in the coefficient of
Appendix (\ref{Tbasis}) or in Appendix (\ref{t2h}). 
%

Therefore, we have the explicit realization of oscillators in
Appendix (\ref{Tbasis})
which
generates the first commutator of Appendix (\ref{shsRacah}).

\subsubsection{ Other (anti)commutators of Appendix (\ref{shsRacah})}

Other type of bosonic generator in the ${\cal N}=2$
higher spin algebra can be constructed similarly.
As done in previous example in Appendix (\ref{Tbasis}),
we make a linear combination
of oscillators with two undetermined coefficients.
One of them has an explicit operator $K$ dependence.
After substituting this ansatz into the second commutator of
Appendix (\ref{shsRacah}) and using the previous relations
Appendices (\ref{NfandClebsch}) and  (\ref{relHyFf}), we can determine
the above two coefficients.
For the quantities
\bea
u_1[h]& \equiv & -t_1[h]\,,
\qquad
u_2[h] \equiv \frac{(-2h+1-\nu)}{(2h-5)}\,t_1[h]\,,
\label{u12h}
\eea
the corresponding ${\cal N}=2$ higher spin generator
is given by either the oscillators or the ones of
${\cal N}=4$ higher spin generators
\bea
\alpha^j_m \,U^j_m 
&=&
\underbrace{
\hat{y}_{(1}\ldots\,\hat{y}_{1}
}_{j+m} \,
\underbrace{
\hat{y}_{2}\ldots\,\hat{y}_{2)}
}_{j-m}
\Big(
\!
K \,u_1[j+1]\,(2j+1)+u_1[j+1]\,\nu-u_2[j+1]\,(2j-3)
\Big)
\nonu\\
&=& u_1[j+1]\,\Phi^{(j+1)}_{0,\,m}  +u_2[j+1]\, \tilde{\Phi}^{(j-1)}_{2,\,m} \,.
\label{Ubasis}
\eea
There is also a linear $\nu$ dependence in the coefficient of
Appendix (\ref{Ubasis}) or in Appendix (\ref{u12h}). 
Therefore, the explicit realization of oscillators is given by
Appendix (\ref{Ubasis}) 
which
generates the second commutator of Appendix (\ref{shsRacah}).

Similarly,
together with
\bea
\psi_1[h]& \equiv & \frac{\mi^{h-\frac{1}{2}}}{2^{h+1}}\,
\sqrt{\frac{[h-2]_{h-\frac{3}{2}}}{(h-1)\,[h-\frac{1}{2}]_{h-\frac{3}{2}}}},
\qquad
\psi_2[h] \equiv \mi\, \frac{(h-1)}{(h-2)}\,\psi_1[h]\,,
\nonu \\
\bar{\psi}_1[h] &=& \psi_1[h],
\qquad
\bar{\psi}_2[h] \equiv -\psi_2[h]\,,
\label{psipsietal}
\eea
the remaining ${\cal N}=2$ higher spin generators
\bea
\alpha^j_\rho \,\Psi^j_\rho 
&=&
-2 \,e^{\frac{\mi \pi}{4}}
\underbrace{
\hat{y}_{(1}\ldots\,\hat{y}_{1}
}_{j+\rho} \,
\underbrace{
\hat{y}_{2}\ldots\,\hat{y}_{2)}
}_{j-\rho}
\Big(
K \,\mi \,\psi_1[j+1] \,(j ) + \psi_2[j+1]\,(j-1)
\Big)
\nonu\\
&=&
\psi_1[j+1]\,\Phi^{(j+\frac{1}{2})}_{\frac{1}{2},\,\rho}+\psi_2[j+1]\,\tilde{\Phi}^{(j-\frac{1}{2})}_{\frac{3}{2},\,\rho},
\nonu\\
\alpha^j_\rho \,\bar{\Psi}^j_\rho 
&=&
-2 \,e^{\frac{\mi \pi}{4}}
\underbrace{
\hat{y}_{(1}\ldots\,\hat{y}_{1}
}_{j+\rho} \,
\underbrace{
\hat{y}_{2}\ldots\,\hat{y}_{2)}
}_{j-\rho}
\Big(
    K \,\mi \,\bar{\psi}_1[j+1] \,(j ) +\bar{ \psi}_2[j+1]\,(j-1)
\Big)
\nonu\\
&=&
\bar{\psi}_1[j+1]\,\Phi^{(j+\frac{1}{2})}_{\frac{1}{2},\,\rho}+\bar{\psi}_2[j+1]\,\tilde{\Phi}^{(j-\frac{1}{2})}_{\frac{3}{2},\,\rho}
\,
\label{psipsibar}
\eea
provide the remaining five (anti)commutators of Appendix (\ref{shsRacah}).

Then the four ${\cal N}=2$ higher spin generators
are given by Appendices (\ref{Tbasis}), (\ref{Ubasis}) and (\ref{psipsibar})
in terms of oscillators or the ones in
${\cal N}=4$ higher spin generators which contain the
$2\times 2$ identity matrix in (\ref{HSbasis}).

\subsection{The (anti)commutators of oscillators
  by using the structure constants
in $shs[\mu]$}

Note that we can present the three cases (\ref{IdEvenEven}),
(\ref{IdEvenOdd-1}), (\ref{IdEvenOdd}), (\ref{IdOddOdd-1}) and
(\ref{IdOddOdd})
studied in previous sections
by using the structure constants in the ${\cal N}=2$ higher spin
algebra $shs[\mu]$ of (\ref{shsRacah}).
For example,
by using Appendices (\ref{Tbasis}) and  (\ref{Ubasis}) in order to remove
the operator $K$ dependent oscillator term and using the first two
relations of Appendix (\ref{shsRacah}),
we rewrite the left side of (\ref{IdEvenEven}) as follows:  
\bea
&&\big[ 
  \underbrace{
\hat{y}_{(1}\ldots\,\hat{y}_{1}
}_{h_1-1+m} 
\underbrace{
\hat{y}_{2}\ldots\,\hat{y}_{2)}
}_{h_1-1-m},\,
 \underbrace{
\hat{y}_{(1}\ldots\,\hat{y}_{1}
}_{h_2-1+n} 
\underbrace{
\hat{y}_{2}\ldots\,\hat{y}_{2)}
}_{h_2-1-n}
\big]
\nonu\\
&&=
\sum^{h_1+h_2-1}_{h=|h_1-h_2|+1} \,
\frac{\alpha^{h_1-1}_m \alpha^{h_2-1}_n}{4(2h_1-1)(2h_2-1)t_1[h_1]t_1[h_2]}
\,
\frac{(2h-1)t_1[h]}{\alpha^{h-1}_{m+n}}
\, C^{h_1-1, h_2-1, h-1}_{m, n, m+n}
\nonu\\
&&
\times
\Bigg[
(
f^{h_1-1,h_2-1,h-1}_{TTT}
+f^{h_1-1,h_2-1,h-1}_{UUU}
) 
\,  \underbrace{
\hat{y}_{(1}\ldots\ldots\hat{y}_{1}
}_{h-1+m+n} \,
\underbrace{
\hat{y}_{2}\ldots\ldots\hat{y}_{2)}
}_{h-1-m-n}
\nonumber\\
&&
+(
f^{h_1-1,h_2-1,h-1}_{TTT}
-f^{h_1-1,h_2-1,h-1}_{UUU}
) 
\,  \underbrace{
\hat{y}_{(1}\ldots\ldots\hat{y}_{1}
}_{h-1+m+n} \,
\underbrace{
\hat{y}_{2}\ldots\ldots\hat{y}_{2)}
}_{h-1-m-n} K
\Bigg]
\,,
\label{reWIdEvenEven}
\eea
where
Appendices
(\ref{alpha}), (\ref{t1h}), (\ref{CG}), (\ref{fs}) and (\ref{Fsym})
are used.
Other remaining (anti)commutators are presented in Appendix $G$. 

\subsubsection{The  structure constants
in the $shs[\mu]$ by using the generalized hypergeometric functions}

There are two different expressions for the same quantity.
For example, the right hand side of Appendix (\ref{reWIdEvenEven})
written in terms of the structure constants in $shs[\mu]$
should be equal to the right hand side of (\ref{IdEvenEven})
written in terms of the ones in $shs_2[\mu]$.
This leads to the previous relation in Appendix (\ref{relHyFf}).
There are also other nontrivial relations
which can be obtained from
(\ref{IdEvenOdd-1}), (\ref{IdEvenOdd}), (\ref{IdOddOdd-1}) and
(\ref{IdOddOdd})
and the corresponding ones in Appendix $G$ as follows.
Four different types of $F$ symbols are
\footnote{
As done before, for the $h_1=h_2=100$  case,
we have several values for the left hand side with
plus sign of Appendix (\ref{Fconst}) as follows. When
$r=3$, the linear combination of $F$
contains $(\nu^2-15485185)$ which appears in
the linear combination of generalized hypergeometric functions in
Appendix (\ref{Fconst}).
For $r=10$,
  the corresponding polynomial is given by $
  (\nu^8-19021364 \nu^6+82683010446934 \nu^4-
  115857948630733805076 \nu^2+48069142225065249313021425)  $
  which appears also in the corresponding generalized hypergeometric
  functions.
  For the $r=199$, we have
  $(-100+\nu)(-99+\nu) \cdots (-2+\nu)(1+\nu)(2+\nu) \cdots
  (98+\nu)(99+\nu)$. Also we observe that
   the corresponding generalized hypergeometric
  functions
  contain this factorized form.
  Therefore, we conclude that the nontrivial relation
Appendix  (\ref{Fconst}) holds for any $h_1$ and $h_2$.
  Similarly, if we take the plus sign of Appendix (\ref{Fconst1}),
  the left hand sides produce the following polynomials,
  $\nu$, $(3 \nu^8-57959612 \nu^6+255874861516642 \nu^4-
  364108448995236058268 \nu^2+153402332060210496539980275)$
 and $(-199+\nu)(-197+\nu) \cdots (-5+\nu)\nu(5+\nu)(7+\nu) \cdots
 (197+\nu)(199+\nu)$ respectively. After obtaining the nontrivial
 relative coefficients by varying $h_1,h_2$ and $r$,
 the nontrivial relation
 Appendix (\ref{Fconst1}) is satisfied for any $h_1$ and $h_2$.}
\bea
&&F^{h_1-1, h_2-\frac{1}{2},h_1+h_2-r-\frac{1}{2}}_{0 \,\frac{1}{2}\, \frac{1}{2}}(-\mu)
\pm F^{h_1-1, h_2-\frac{1}{2},h_1+h_2-r-\frac{1}{2}}_{0\, -\frac{1}{2} \,-\frac{1}{2}}(1-\mu)
\nonu\\
&&
=
-\frac{\mi^{r+1}\,(2h_1-1)(2h_2-1)t_1[h_1]\psi_1[h_2+\frac{1}{2}]}{\Delta^{h_1-1,h_2-\frac{1}{2},h_1+h_2-r-\frac{1}{2}}
\psi_1[h_1+h_2-r+\frac{1}{2}]\,(h_1+h_2-r-\frac{1}{2})}
\,
\sqrt{\frac{(2h_1-2)!(2h_2-1)!}{(2h_1+2h_2-2r)!}}
\nonu\\
&&
\times \frac{(h_1-\frac{1}{2}) \mp (h_1-r-\frac{1}{2})}{r}
\Bigg[
\phi_{r+1}^{h_1 ,h_2+1}(\mu,\tfrac{3\pm 1}{2})  \pm  \phi_{r+1}^{h_1 ,h_2+1}(1-\mu,\tfrac{3\pm 1}{2})
\Bigg]
\,,
\label{Fconst}
\eea
and
\bea
&&F^{h_1-\frac{1}{2}, h_2-\frac{1}{2},h_1+h_2-r}_{-\frac{1}{2}\, \frac{1}{2}\,0}(-\mu)
\pm (-1)^r\, F^{h_1-\frac{1}{2}, h_2-\frac{1}{2},h_1+h_2-r}_{\frac{1}{2} \,-\frac{1}{2}\,0}(1-\mu)
=
\nonu \\
&&
-\frac{2\,\mi^r \,(2h_1-1)(2h_2-1)\psi_1[h_1+\frac{1}{2}]\psi_1[h_2+\frac{1}{2}]}{\Delta^{h_1-\frac{1}{2},h_2-\frac{1}{2},h_1+h_2-r}
t_1[h_1+h_2-r+1]\,(2h_1+2h_2-2r+1)}
\,
\sqrt{\frac{(2h_1-1)!(2h_2-1)!}{(2h_1+2h_2-2r+1)!}}
\nonu\\
&&
\times
\frac{(h_1+h_2+\frac{1}{2})\mp(-1)^r (h_1+h_2+\frac{1}{2}-r)}{r}
\,
\nonu \\
&& \times
\Bigg[
\phi_{r+1}^{h_1+1 ,h_2+1}(\mu,\tfrac{3\pm (-1)^r}{2})  \pm(-1)^r  \phi_{r+1}^{h_1+1 ,h_2+1}(1-\mu,\tfrac{3\pm (-1)^r}{2})
\Bigg]
\,,
\label{Fconst1}
\eea
where the relations
Appendices
(\ref{t1h}), (\ref{psipsietal}), (\ref{Delta})
and (\ref{spf}) are used.
Each $F$ symbols is a linear combination of generalized hypergeometric
functions because the relative coefficients are different in
Appendices (\ref{Fconst}) and (\ref{Fconst1}),
compared to the one Appendix (\ref{relHyFf}).

Therefore, we have determined the structure constants,
Appendices (\ref{relHyFf}), (\ref{Fconst}) and (\ref{Fconst1}),
in the ${\cal N}=2$
higher spin algebra $shs[\mu]$ in terms of those in
 the ${\cal N}=4$
higher spin algebra $shs_2[\mu]$. 

\subsubsection{The relations between 
  the structure constants
in the $shs_2[\mu]$ and those in the $shs[\mu]$}

For convenience, we present 
the structure constants (\ref{3struct})
in terms of the ones in ${\cal N}=2$ higher spin algebra 
as follows:
\bea
&& \mathrm{BB}^{h_1,h_2}_{r,\,\pm}(m,n; \mu)
=
\nonu \\
&& 
(-\mi)^{r+1} \frac{(h_1+h_2-r-\frac{1}{2}) t_1[h_1+h_2-r] \alpha_m^{h_1-1}\alpha_n^{h_2-1}}{(2h_1-1)(2h_2-1) t_1[h_1]t_1[h_2] \alpha_{m+n}^{h_1+h_2-r-1}}
\nonu\\
&& \times
\Bigg[
F^{h_1-1, h_2-1,h_1+h_2-r-1}_{0 0 0}(-\mu)  \pm F^{h_1-1, h_2-1,h_1+h_2-r-1}_{0 0 0}(1-\mu)
\Bigg]
C^{h_1-1, h_2-1, h_1+h_2-r-1}_{m, n, m+n}
\,,
\nonu \\
&&
\mathrm{BF}^{h_1,h_2+\frac{1}{2}}_{r,\,\pm}(m,\rho; \mu) 
=
\nonu\\
&&
(-\mi)^{r+1} \frac{r\,(h_1+h_2-r-\frac{1}{2}) \psi_1[h_1+h_2-r+\frac{1}{2}] \alpha_m^{h_1-1}\alpha_{\rho}^{h_2-\frac{1}{2}}}{(2h_1-1)(2h_2-1) t_1[h_1]\psi_1[h_2+\frac{1}{2}] \alpha_{m+\rho}^{h_1+h_2-r-\frac{1}{2}}}
\, \frac{1}{\Big((h_1-\tfrac{1}{2})\mp (h_1-r-\tfrac{1}{2})\Big)}
\nonu\\
&&\times
\Bigg[
F^{h_1-1, h_2-\frac{1}{2},h_1+h_2-r-\frac{1}{2}}_{0 \, \frac{1}{2}\, \frac{1}{2}}(-\mu)  \pm F^{h_1-1, h_2-\frac{1}{2},h_1+h_2-r-\frac{1}{2}}_{0 \, -\frac{1}{2} \, -\frac{1}{2}}(1-\mu)
\Bigg]
C^{h_1-1, h_2-\frac{1}{2}, h_1+h_2-r-\frac{1}{2}}_{m, \rho, m+\rho}
\,,
\nonu \\
&&
\mathrm{FF}^{h_1+\frac{1}{2},h_2+\frac{1}{2}}_{r,\,\pm(-1)^r}(\rho,\omega; \mu)= 
\nonu\\
&&
(-\mi)^{r} \frac{r\,(h_1+h_2-r+\frac{1}{2}) t_1[h_1+h_2-r+1] \alpha_{\rho}^{h_1-\frac{1}{2}}\alpha_{\omega}^{h_2-\frac{1}{2}}}{(2h_1-1)(2h_2-1) \psi_1[h_1+\frac{1}{2}]\psi_1[h_2+\frac{1}{2}] \alpha_{\rho+\omega}^{h_1+h_2-r}
  }
\nonu\\
&& \times
\frac{1}{\Big((h_1+h_2+\tfrac{1}{2})\mp(-1)^r (h_1+h_2+\tfrac{1}{2}-r)
  \Big)}
\label{bbbfffintermsof} \\
&&\times
\Bigg[
F^{h_1-\frac{1}{2}, h_2-\frac{1}{2},h_1+h_2-r}_{- \frac{1}{2}\, \frac{1}{2}\,0}(-\mu)
\pm (-1)^r F^{h_1-\frac{1}{2}, h_2-\frac{1}{2},h_1+h_2-r}_{ \frac{1}{2} \, -\frac{1}{2}\, 0}(1-\mu)
\Bigg]
C^{h_1-\frac{1}{2}, h_2-\frac{1}{2}, h_1+h_2-r}_{\rho, \omega,\rho+\omega}
\,,
\nonu
\eea
where the inverse relations of Appendices
(\ref{relHyFf}), (\ref{Fconst})
and (\ref{Fconst1}) together with Appendix (\ref{NfandClebsch}) and
(\ref{3struct})
are used in Appendix
(\ref{bbbfffintermsof}) \footnote{The three upper indices of
  $F$ stand for the spins of the generators in the left hand side
  and the right hand side of the (anti)commutators respectively.
  In other words, the first two indices correspond to those of
  structure constants of (\ref{3struct}) and the third one
  corresponds to the difference between the upper indices and lower
  indices in the structure constants in (\ref{3struct}).
}.

This implies that we can write down the previous ${\cal N}=4$
higher spin algebra $shs_2[\mu]$ by using the
structure constants of  ${\cal N}=2$
higher spin algebra $shs[\mu]$ \footnote{Note that in
  Appendices (\ref{relHyFf}),
  (\ref{Fconst}),
  (\ref{Fconst1}) and (\ref{bbbfffintermsof}), we have checked them
  for $h_1, h_2
  \leq 100$. Moreover, all the computations in Appendices $F$ and $G$
  are done under this restriction.
}.

\section{The (anti)commutators of oscillators
  by using the structure constants
in ${\cal N}=2$ higher spin algebra $shs[\mu]$}

In this Appendix, we have considered the spins $h_1, h_2 \leq 100$. 
The left hand side  of (\ref{IdEvenOdd-1})
can be expressed as 
\bea
&&\big[ 
  \underbrace{
\hat{y}_{(1}\ldots\,\hat{y}_{1}
}_{h_1-1+m} 
\underbrace{
\hat{y}_{2}\ldots\,\hat{y}_{2)}
}_{h_1-1-m},\,
 \underbrace{
\hat{y}_{(1}\ldots\,\hat{y}_{1}
}_{h_2-\frac{1}{2}+\rho} 
\underbrace{
\hat{y}_{2}\ldots\,\hat{y}_{2)}
}_{h_2-\frac{1}{2}-\rho}
\big]
\label{f1}
\\
&&=
-\mi\,\!\!\!\!
\sum^{h_1+h_2-\frac{3}{2}}_{h=|h_1-h_2-\frac{1}{2}|} \,
\frac{2h\,\psi_1[h+1]\,\alpha^{h_1-1}_{m} \alpha^{h_2-\frac{1}{2}}_{\rho}}{2(2h_1-5)(2h_2-3)(t_2[h_1]+u_2[h_1])\psi_2[h_2+\frac{1}{2}]\,\alpha^{h}_{m+\rho}} 
\, C^{h_1-1, h_2-\frac{1}{2}, h}_{m, \rho, m+\rho}
\nonumber\\
&&
\times
\Bigg[
(
f^{h_1-1,h_2-\frac{1}{2},h}_{T\Psi\Psi}
+f^{h_1-1,h_2-\frac{1}{2},h}_{U\Psi\Psi}
+f^{h_1-1,h_2-\frac{1}{2},h}_{T\bar{\Psi}\bar{\Psi}}
+f^{h_1-1,h_2-\frac{1}{2},h}_{U\bar{\Psi}\bar{\Psi}})
\,  \underbrace{
\hat{y}_{(1}\,\ldots\,\hat{y}_{1}
}_{h+m+\rho} \,
\underbrace{
\hat{y}_{2}\,\ldots\,\hat{y}_{2)}
}_{h-m-\rho}
\nonumber\\
&&
+(f^{h_1-1,h_2-\frac{1}{2},h}_{T\Psi\Psi}
+f^{h_1-1,h_2-\frac{1}{2},h}_{U\Psi\Psi}
-f^{h_1-1,h_2-\frac{1}{2},h}_{T\bar{\Psi}\bar{\Psi}}
-f^{h_1-1,h_2-\frac{1}{2},h}_{U\bar{\Psi}\bar{\Psi}})
\,  \underbrace{
\hat{y}_{(1}\,\ldots\,\hat{y}_{1}
}_{h+m+\rho} \,
\underbrace{
\hat{y}_{2}\,\ldots\,\hat{y}_{2)}
}_{h-m-\rho} K
\Bigg],
\nonu
\eea
where the relations
Appendices (\ref{fs}), (\ref{CG}), (\ref{alpha}), (\ref{t1h}),
(\ref{t2h}), (\ref{u12h}), and (\ref{psipsietal}) are used.

The left hand side of (\ref{IdEvenOdd})
can be described by
\bea
&&\big[ 
  \underbrace{
\hat{y}_{(1}\ldots\,\hat{y}_{1}
}_{h_1-1+m} 
\underbrace{
\hat{y}_{2}\ldots\,\hat{y}_{2)}
}_{h_1-1-m}  K,\,
 \underbrace{
\hat{y}_{(1}\ldots\,\hat{y}_{1}
}_{h_2-\frac{1}{2}+\rho} 
\underbrace{
\hat{y}_{2}\ldots\,\hat{y}_{2)}
}_{h_2-\frac{1}{2}-\rho}
\big]
\label{f2}
\\
&&=
\mi\,\!\!\!\!
\sum^{h_1+h_2-\frac{3}{2}}_{h=|h_1-h_2-\frac{1}{2}|} \,
\frac{ 2h\,\psi_1[h+1]\,\alpha^{h_1-1}_{m} \alpha^{h_2-\frac{1}{2}}_{\rho}}{2(2h_1-1)(2h_2-3)(t_1[h_1]-u_1[h_1])\psi_2[h_2+\frac{1}{2}]\,\alpha^{h}_{m+\rho}} 
\, C^{h_1-1, h_2-\frac{1}{2}, h}_{m, \rho, m+\rho}
\nonumber\\
&&
\times
\Bigg[
(
f^{h_1-1,h_2-\frac{1}{2},h}_{T\Psi\Psi}
-f^{h_1-1,h_2-\frac{1}{2},h}_{U\Psi\Psi}
+f^{h_1-1,h_2-\frac{1}{2},h}_{T\bar{\Psi}\bar{\Psi}}
-f^{h_1-1,h_2-\frac{1}{2},h}_{U\bar{\Psi}\bar{\Psi}})
\,  \underbrace{
\hat{y}_{(1}\,\ldots\,\hat{y}_{1}
}_{h+m+\rho} \,
\underbrace{
\hat{y}_{2}\,\ldots\,\hat{y}_{2)}
}_{h-m-\rho}
\nonumber\\
&&
+(f^{h_1-1,h_2-\frac{1}{2},h}_{T\Psi\Psi}
-f^{h_1-1,h_2-\frac{1}{2},h}_{U\Psi\Psi}
-f^{h_1-1,h_2-\frac{1}{2},h}_{T\bar{\Psi}\bar{\Psi}}
+f^{h_1-1,h_2-\frac{1}{2},h}_{U\bar{\Psi}\bar{\Psi}})
\,  \underbrace{
\hat{y}_{(1}\,\ldots\,\hat{y}_{1}
}_{h+m+\rho} \,
\underbrace{
\hat{y}_{2}\,\ldots\,\hat{y}_{2)}
}_{h-m-\rho} K
\Bigg].
\nonu
\eea

The left hand side of (\ref{IdOddOdd-1}) can be 
written in terms of 
\bea
&&
\big\{ 
  \underbrace{
\hat{y}_{(1}\ldots\,\hat{y}_{1}
}_{h_1-\frac{1}{2}+\rho} 
\underbrace{
\hat{y}_{2}\ldots\,\hat{y}_{2)}
}_{h_1-\frac{1}{2}-\rho},\,
 \underbrace{
\hat{y}_{(1}\ldots\,\hat{y}_{1}
}_{h_2-\frac{1}{2}+\omega} 
\underbrace{
\hat{y}_{2}\ldots\,\hat{y}_{2)}
}_{h_2-\frac{1}{2}-\omega}
\big\}
\nonumber\\
&&=
\mi\,\!\!\!\!\sum^{h_1+h_2-1}_{h=|h_1-h_2|} \,
\frac{(1-(-1)^{h_1+h_2+h})(2h+1)\,\alpha^{h_1-\frac{1}{2}}_{\rho} \alpha^{h_2-\frac{1}{2}}_{\omega}\, t_1[h+1]}{16 (h_1-\frac{3}{2})(h_2-\frac{3}{2})\psi_2[h_1+\frac{1}{2}]\psi_2[h_2+\frac{1}{2}]\,\alpha^h_{\rho+\omega}}
\, C^{h_1-\frac{1}{2}, h_2-\frac{1}{2}, h}_{\rho, \omega, \rho+\omega}
\nonumber\\
&&
\times
\Bigg[
(
f^{h_1-\frac{1}{2},h_2-\frac{1}{2},h}_{\Psi \bar{\Psi}T}
+f^{h_1-\frac{1}{2},h_2-\frac{1}{2},h}_{\Psi \bar{\Psi}U}
)
\,  \underbrace{
\hat{y}_{(1}\,\ldots\,\hat{y}_{1}
}_{h+\rho+\omega} \,
\underbrace{
\hat{y}_{2}\,\ldots\,\hat{y}_{2)}
}_{h-\rho-\omega}
\nonumber\\
&&
+
(
f^{h_1-\frac{1}{2},h_2-\frac{1}{2},h}_{\Psi \bar{\Psi}T}
-f^{h_1-\frac{1}{2},h_2-\frac{1}{2},h}_{\Psi \bar{\Psi}U}
)
\,  \underbrace{
\hat{y}_{(1}\,\ldots\,\hat{y}_{1}
}_{h+\rho+\omega} \,
\underbrace{
\hat{y}_{2}\,\ldots\,\hat{y}_{2)}
}_{h-\rho-\omega} K
\Bigg].
\label{f3}
\eea

Finally, the left hand side of (\ref{IdOddOdd})
can be written 
as
\bea
&&
\big\{ 
  \underbrace{
\hat{y}_{(1}\ldots\,\hat{y}_{1}
}_{h_1-\frac{1}{2}+\rho} 
\underbrace{
\hat{y}_{2}\ldots\,\hat{y}_{2)}
}_{h_1-\frac{1}{2}-\rho},\,
 \underbrace{
\hat{y}_{(1}\ldots\,\hat{y}_{1}
}_{h_2-\frac{1}{2}+\omega} 
\underbrace{
\hat{y}_{2}\ldots\,\hat{y}_{2)}
}_{h_2-\frac{1}{2}-\omega} K
\big\}
\nonumber\\
&&=
-\!\!\!\!\sum^{h_1+h_2-1}_{h=|h_1-h_2|} \,
\frac{(1+(-1)^{h_1+h_2+h})(2h+1)\,\alpha^{h_1-\frac{1}{2}}_{\rho} \alpha^{h_2-\frac{1}{2}}_{\omega}\, t_1[h+1]}{16 (h_1-\frac{3}{2})(h_2-\frac{1}{2})\psi_2[h_1+\frac{1}{2}]\psi_1[h_2+\frac{1}{2}]\,\alpha^h_{\rho+\omega}}
\, C^{h_1-\frac{1}{2}, h_2-\frac{1}{2}, h}_{\rho, \omega, \rho+\omega}
\nonumber\\
&&
\times
\Bigg[
(
f^{h_1-\frac{1}{2},h_2-\frac{1}{2},h}_{\Psi \bar{\Psi}T}
+f^{h_1-\frac{1}{2},h_2-\frac{1}{2},h}_{\Psi \bar{\Psi}U}
)
\,  \underbrace{
\hat{y}_{(1}\,\ldots\,\hat{y}_{1}
}_{h+\rho+\omega} \,
\underbrace{
\hat{y}_{2}\,\ldots\,\hat{y}_{2)}
}_{h-\rho-\omega}
\nonumber\\
&&
+
(
f^{h_1-\frac{1}{2},h_2-\frac{1}{2},h}_{\Psi \bar{\Psi}T}
-f^{h_1-\frac{1}{2},h_2-\frac{1}{2},h}_{\Psi \bar{\Psi}U}
)
\,  \underbrace{
\hat{y}_{(1}\,\ldots\,\hat{y}_{1}
}_{h+\rho+\omega} \,
\underbrace{
\hat{y}_{2}\,\ldots\,\hat{y}_{2)}
}_{h-\rho-\omega} K
\Bigg].
\label{f4}
\eea

Therefore, we have five relations, Appendix
(\ref{reWIdEvenEven}), and Appendices
(\ref{f1}), (\ref{f2}), (\ref{f3}), and (\ref{f4}).


\end{document}